\documentclass[a4paper]{ar-1col}
\usepackage{natbib}
\usepackage{hyperref}
\usepackage{url,multirow}
\newcommand{\nat}{Nature}
\newcommand{\apj}{ApJ}
\newcommand{\aj}{AJ}
\newcommand{\araa}{ARA\&A}
\newcommand{\apjl}{ApJ}
\newcommand{\apjs}{ApJS}
\newcommand{\aap}{A\&A}

\newcommand{\mnras}{MNRAS}

\newcommand{\aapr}{A\&A Rev}
\newcommand{\pasp}{PASP}
\newcommand{\pasj}{PASJ}

\newcommand{\gca}{Geochim.~Cosmochim.~Acta}
\newcommand{\ssr}{Space Science Reviews}
\newcommand{\herschel}{\textit{Herschel}}
\newcommand{\spitzer}{\textit{Spitzer}}
\newcommand{\jqsrt}{J.~Quant.~Spec.~Radiat.~Transf.}
\newcommand{\arcsec}{$''$}
\newcommand{\kms}{km~s$^{-1}$}

\jname{Annu. Rev. Astron. Astrophys.}
\jvol{58}
\jyear{2020}
\doi{10.1146/((please add article doi))}

\begin{document}

\markboth{J{\o}rgensen et al.}{Astrochemistry during the formation of stars}

\title{Astrochemistry during the formation of stars}

\author{Jes K. J{\o}rgensen,$^1$ Arnaud Belloche,$^2$ and Robin T. Garrod$^3$
\affil{$^1$Niels Bohr Institute, University of Copenhagen, Copenhagen, Denmark; email: jeskj@nbi.ku.dk}
\affil{$^2$Max-Planck-Institut f{\"u}r Radioastronomie, Bonn, Germany; email: belloche@mpifr-bonn.mpg.de}
\affil{$^3$Departments of Chemistry and Astronomy, University of Virginia, Charlottesville, Virginia, USA; email: rgarrod@virginia.edu} }

\begin{abstract}
Star-forming regions show a rich and varied chemistry, including the presence of complex organic molecules -- both in the cold gas distributed on large scales, and in the hot regions close to young stars where protoplanetary disks arise. Recent advances in observational techniques have opened new possibilities for studying this chemistry. In particular, the Atacama Large Millimeter/submillimeter Array (ALMA) has made it possible to study astrochemistry down to Solar System size scales, while also revealing molecules of increasing variety and complexity.
In this review, we discuss recent observations of the chemistry of star-forming environments, with a particular focus on complex organic molecules, taking context from the laboratory experiments and chemical models that they have stimulated. The key takeaway points are:
\begin{itemize}
    \item The physical evolution of individual sources plays a crucial role in \\ their inferred chemical signatures, and remains an important area \\ for observations and models to elucidate.
    \item Comparisons of the abundances measured toward different star- \\ 
    forming environments (high-mass versus low-mass, Galactic center  \\ versus  Galactic disk) reveal a remarkable similarity, an indication \\ that the  underlying chemistry is relatively independent of \\ variations in their physical conditions.
    \item Studies of molecular isotopologs in star-forming regions provide a \\ link  with measurements in our own Solar System, and thus may \\ shed light  on the chemical similarities and differences expected in \\ other planetary systems. 
\end{itemize}
\end{abstract}

\begin{keywords}
astrochemistry, complex molecules, interstellar medium, interstellar molecules, star formation, submillimeter astronomy
\end{keywords}
\maketitle

\tableofcontents

\section{INTRODUCTION}
\subsection{Motivation}
The environments in which young stars form show a rich and varied chemistry. In fact, most of the molecules detected in the interstellar medium (ISM) to date have first been found in these regions -- whether in the cold starless/prestellar cores or in the warm gas surrounding young stars of high or low masses. These species range all the way from simple di- and tri-atomic neutral molecules, molecular radicals and ions to complex molecules. The latter, some with ten atoms or more, include species containing long unsaturated chains of  carbon atoms as well as saturated organics that can be considered the starting points for eventual prebiotic chemistry. The chemical networks describing the formation and destruction paths for these different species are strongly dependent on the underlying physical evolution of the star formation processes, such as the changes in density, temperature, and spectral shape and intensity of irradiation. 

Although molecules of varying degrees of complexity have also been detected in other regions including the envelopes around evolved stars, photodissociation regions, well-developed protoplanetary disks around Class II young stellar objects/T~Tauri stars, and even in distant galaxies, the chemistry in prestellar cores of molecular clouds and embedded protostellar stages is critical. These stages provide key laboratories for molecular astrophysics: through the high column densities characteristic of these regions we have by far the most complete molecular inventories of those, including censuses of  low-abundance organics and their isotopologs. Also, these stages are likely pivotal for linking the birth environments of young stars and the initial conditions in the emerging protoplanetary disks in terms of both their physics and chemistry.

\begin{figure}[!htb]
    \centering
    \includegraphics[width=\textwidth]{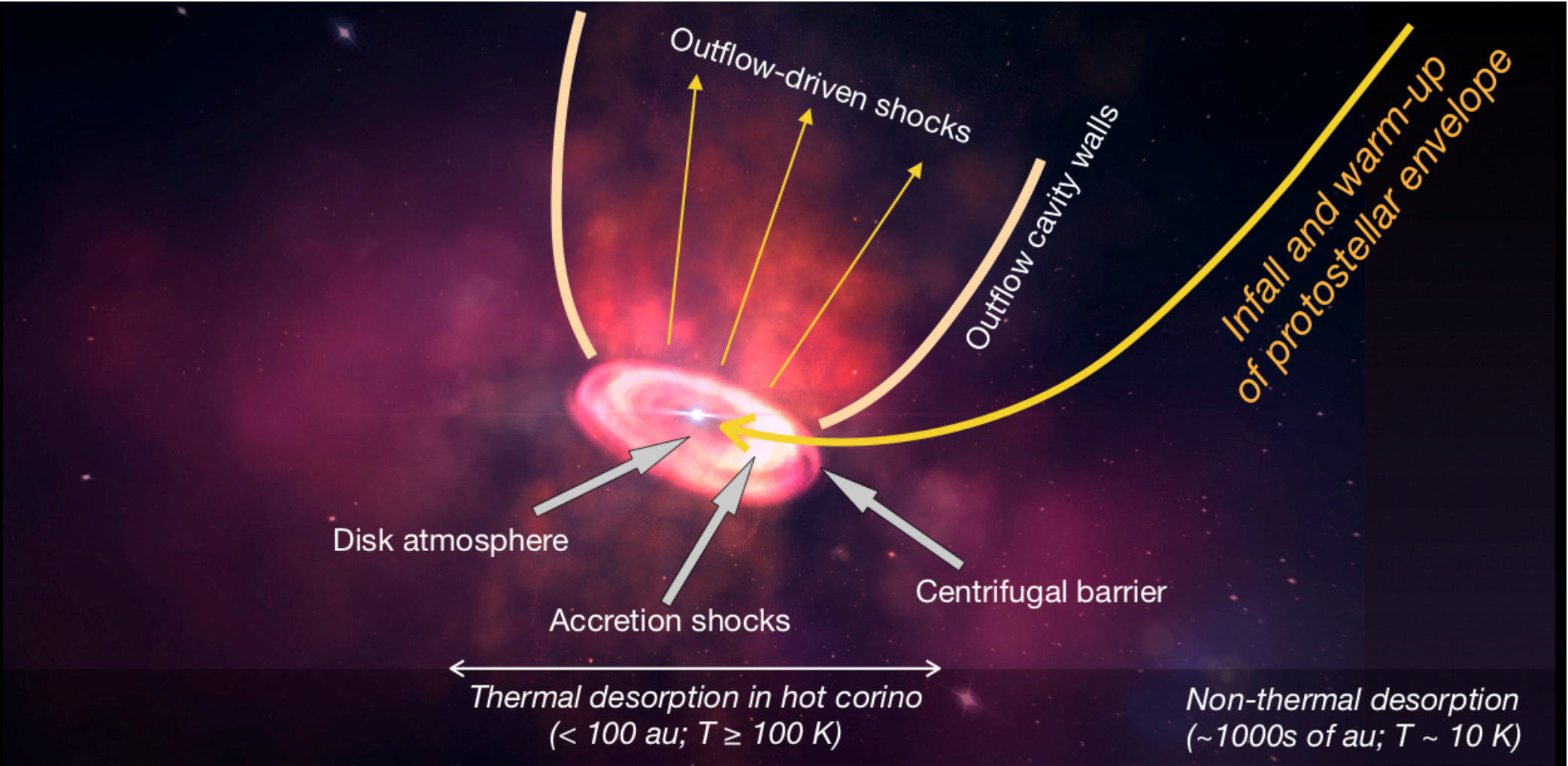}
    \caption{Schematic representation of a young solar-type protostar and its structural components that are key to its chemistry.Complex organic molecules are observed to be present in all of these components as the result of thermal and non-thermal desorption: on small and large scales within the protostellar envelope and disk, in shocks related to accretion at the disk surface and centrifugal barrier, and on larger scales associated with the protostellar outflow. UV irradiation close to the disk and outflow cavity are also potentially important in regulating the chemistry. The main focus of the previous review in this series by \cite{herbst09} was on the formation of complex organic species during the infall/warm-up phase (yellow thick arrow; see, i.p., their Fig.~14), but, as discussed in this review, in the recent years a much more complex picture has emerged. Background image by Per Bjerkeli.}
    \label{fig:cartoon}
\end{figure}

However, these sources also illustrate some of the major challenges in terms of understanding astrochemistry. In particular, recent observations with significant improvements in sensitivity and spatial resolution have revealed that complex chemistry is taking place in a wider range of the physical components of young protostars than considered previously (Fig.~\ref{fig:cartoon}). Understanding how the physical structure and evolution of young protostars influences the degree of molecular complexity that arises in their envelopes and disks, and how this may further influence chemical composition during the later planet-forming stages, remain some of the key challenges for astrochemistry.

\subsection{Star formation and astrochemistry}
The canonical scenario for the formation of a solar-type protostar starts at low temperatures of $\sim$~10~K at densities of $10^3-10^4$~cm$^{-3}$ with the formation of a dense prestellar core in a giant molecular cloud. In the denser parts of such cores, gaseous molecules collide with and stick to the surfaces of dust grains during their earliest stages and cold gas-phase chemistry will lead to abundance enhancements of, e.g., deuterium-containing molecules through fractionation. Star formation occurs when these cores collapse, leading to the formation of an opaque (second) hydrostatic core. Further infall leads to the release of gravitational potential energy heating the infalling envelope of dust and gas to temperatures of tens or hundreds of K, while the densities increase to $\sim 10^8-10^9$~cm$^{-3}$ in the inner $\sim$100~au regions around the central protostar. As the temperatures increase above 100~K, the water-rich ice mantles sublimate, injecting molecules into the gas phase -- giving rise to the so-called hot corino regions at high temperatures and densities, rich in saturated complex organics. These are also the scales at which protoplanetary disks are expected to arise owing to the conservation of angular momentum, a process that also leads to the launching of outflows and jets. While the overall physical evolution for high-mass stars through these stages is clearly more complex, many of the overall characteristics can be identified, including the extended hot cores with elevated temperatures where complex organics are present in some cases. Due to the high column densities of warm material, many of the first detections of complex organic molecules (COMs)\footnote{We keep the definition proposed by \citet{herbst09} and widely used in the community: a COM is a carbon-bearing molecule that has at least six atoms. Alternative names such as large astronomical molecule (LAM) or interstellar complex organic molecule (iCOM) have been proposed in the past decade but, as long as the context (astrochemistry, not biology or chemistry) is clear and the definition is stated, the term COM is adequate.} were made toward these high-mass regions.

The topic of astrochemistry and its link to star formation has been the subject of previous reviews in this journal.\footnote{In addition to the mentioned reviews from this journal, it is worth pointing out a number of reviews from the past decade: \cite{tielens13} described the physical and chemical processes governing the formation and evolution of molecules in the interstellar medium, \cite{caselli12review} and \cite{ceccarellippvi} addressed the link between astrochemistry in star forming regions and the Solar System, \cite{boogert15} discussed observations of ices, and \cite{vandishoeckppvi} investigated the water trail through the star formation process. A review on recent developments in mm/submm laboratory spectroscopy in support of observational astrochemistry was presented by \citet{widicus-weaver19}.}
 Van~Dishoeck~\&~Blake~\citeyear{vandishoeck98} focused on the overall chemical evolution of star forming regions that at the time had come within reach through advances in (sub)millimeter wavelength single-dish telescopes, space-borne infrared telescopes, and previous generations of millimeter-wavelength interferometers. These efforts underlined the importance of molecular astrophysics as a tracer of the physical changes taking place during the star formation process. This  includes the freeze-out of molecules on the surfaces of dust grains, the resulting grain-surface chemistry leading to more complex species, and eventually the release of those into the gas phase during the collapse due to thermal desorption close to young stars or in outflow driven shocks -- or their incorporation into protoplanetary disks.

A decade later, dedicated observational efforts, laboratory studies, and sophisticated gas-phase and grain-surface chemical models had shifted the focus from the relatively simple species to the formation of COMs -- the subject of the review by \cite{herbst09}. Extensive observations of hot cores had provided the first unbiased surveys providing complete censuses of the molecular line content of individual high-mass protostars covering wide spectral ranges of the windows where the atmosphere is mostly transparent, as well as systematic, more focused, inventories of networks of species toward groups of sources. Targeted observations of low-mass protostars had started revealing the rich chemistries of these sources as well -- including the detections of saturated complex organics in the inner envelopes of deeply embedded protostars as well as in shocks associated with their outflows. 

Today, yet another decade later, gigantic steps forward have been taken due to the systematic molecular studies at THz frequencies by the \emph{Herschel Space Observatory} (\emph{Herschel}), significant upgrades to many (sub)millimeter wavelength single-dish telescopes and interferometers especially in terms of the receivers and correlators, and, in particular, the advent of the \emph{Atacama Large Millimeter/submillimeter Array} (ALMA) that has pushed molecular astrophysics studies by orders of magnitude in sensitivity and spatial resolution.

\subsection{Outline of this paper}
In this review, we focus on the complex chemistry taking place from the point where star formation is initiated by the formation of dense (prestellar) cores, through their collapse to form young protostars and their circumstellar disks. We describe the opportunities and challenges encountered with recent advances in observations, modeling, and laboratory experiments (Sect.~\ref{s:advances}) and provide an overview of detections of complex molecules in different environments (Sect.~\ref{s:inventories}). This is followed by discussions of the importance of the physical conditions on the chemistry  reflecting both the non-homogeneous conditions in star forming environments (Sect.~\ref{s:environments}) and  the changes occurring during the formation and early evolution of stars (Sect.~\ref{s:physical_evolution}). The final two sections focus on constraints on the formation of complex organic molecules and the link between star forming environments and our own Solar System. Specifically, we describe the insights that can be obtained by studies of isotopic fractionation (Sect.~\ref{s:fractionation}) and by comparing systematic chemical inventories across samples of sources to measurements from our own Solar System and the predictions from models (Sect.~\ref{s:origin_complexity}).

\section{RECENT ADVANCES AND NEW CHALLENGES}
\label{s:advances}
\subsection{Advances in observational techniques}
Significant advances within astrochemistry have been made over the past decade thanks to new telescopes and improvements in instrumentation at existing facilities. The  key features offered by these facilities are \emph{(i)} the improvement in the large instantaneous bandwidths covered with high spectral resolution by individual instruments, \emph{(ii)} the sensitivity offered by large apertures and excellent observation sites, \emph{(iii)} improved spatial resolution with, in particular, combined array antennas, and \emph{(iv)} coverage of high frequency windows in the far-infrared with high spectral resolution using space-based telescopes such as \emph{Herschel} and the Stratospheric Observatory for Infrared Astronomy (SOFIA). Each of these aspects provides new opportunities as well as challenges.

\subsubsection{Increase of instantaneous bandwidth at many observational facilities}
\begin{figure}
    \centering
    \includegraphics[width=\hsize]{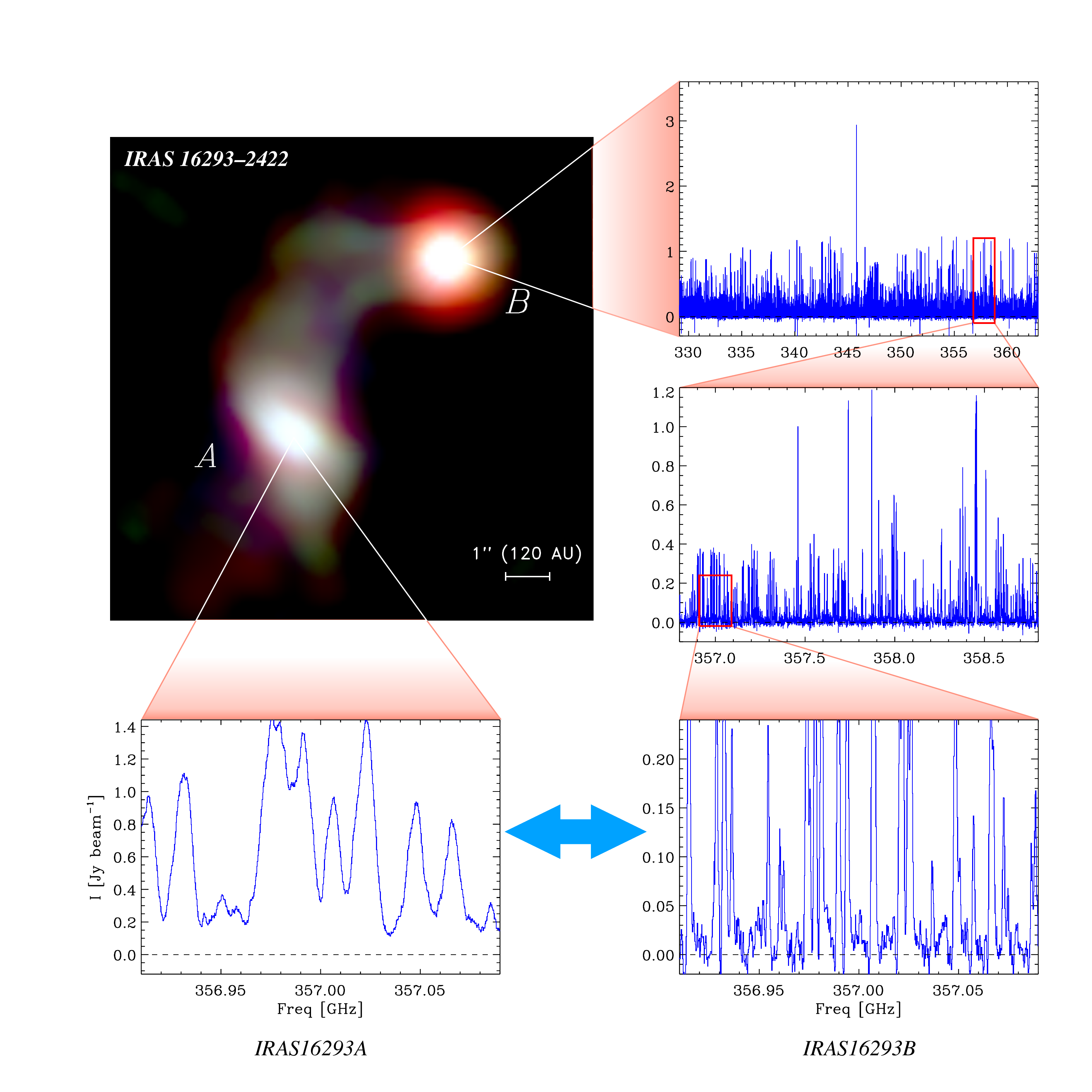}
    \caption{Example of spectra encountered from the ALMA/PILS survey \citep{jorgensen16}. The image in the top left corner shows the dust continuum radiation from the protostellar binary with the ``edge-on'' oriented IRAS16293A source and ``face-on'' oriented IRAS16293B source at wavelengths of 3.0~mm (red), 1.3~mm (green), and 0.8~mm (blue). The different orientations cause the line widths to differ significantly for the two sources ($\approx$1~\kms\ for IRAS16293B and $\approx$~5~\kms\ for IRAS16293A). The panels in the right-hand column show spectra progressively zooming in from the full range covered in the survey toward IRAS16293B. The most zoomed-in version is compared to the same spectrum for IRAS16293A illustrating the enhanced line confusion at those scales due to the wider lines of that source. For reference, the RMS noise levels in the spectra are 8--10~mJy~beam$^{-1}$, thus most of the features seen represent significant line emission.}
    \label{f:pils_overview}
\end{figure}

One of the key aspects of \emph{Herschel} and ALMA as well as upgrades
of receivers and correlators on facilities such as the Atacama Pathfinder EXperiment (APEX), the Institut de Radioastronomie Millim{\'e}trique (IRAM) 30~m
telescope, the NOrthern Extended Millimeter Array (NOEMA), and the Submillimeter Array (SMA), has been the increase in instantaneous
bandwidth obtainable while keeping a relatively high spectral
resolution. This is particularly important for performing unbiased spectral surveys covering large frequency ranges. For example, the
``Herschel observations of EXtra-Ordinary Sources (HEXOS)'' survey of Orion performed a spectral scan from 480 to 1907 GHz
(with two small gaps) with 1.1~MHz spectral resolution and identified more than 13,000 spectral lines, i.e., 10 lines per GHz \citep{crockett14}. From ALMA the ``Exploring Molecular Complexity with ALMA'' (EMoCA) survey of the high-mass star forming region Sgr~B2(N) \citep{belloche16} and the ``Protostellar Interferometric Line Survey'' (PILS) of the nearby low-mass protostar IRAS~16293--2422 \citep{jorgensen16} have been the main unbiased studies. The high angular resolution of EMoCA revealed that the secondary hot core Sgr~B2(N2), with about 6500 lines detected above 7$\sigma$ between 84 and 114~GHz (about 220 lines per GHz), has narrow lines ($\sim$~5~\kms) compared to the lines measured with single-dish telescopes toward Sgr~B2(N). This reduction in spectral confusion was decisive for the identification of new species. The main component of PILS was a systematic survey of the 329--363~GHz range of ALMA's Band~7 at 0.25~\kms\ spectral resolution (see Fig.~\ref{f:pils_overview}). With the narrow lines ($\sim$~1~\kms) at selected positions toward the protostellar system the line confusion is reached at low levels and more than 10,000 lines above 5$\sigma$ can be identified (about 300 lines per GHz).

\begin{textbox}
\section{New facilities}
The three main new facilities for studies of star forming regions that have started operations in the past decade are:
\begin{itemize}
\item {\bf ALMA:} 66 telescope array operating at submillimeter wavelengths. ALMA has demonstrated its metier for high resolution, high sensitivity imaging of the distribution of molecules in star-forming regions both near and far (2011--).
\item {\bf \emph{Herschel}:} 3.5~m space-based observatory operating at far-IR (THz) wavelengths. Key contributions concern the presence of water and organics in star forming regions (2009--2013).
\item {\bf SOFIA:} 2.5~m airborne telescope that gives access to far-IR windows not observable from the ground, and in particular to important cooling lines of the ISM (2010--).
\end{itemize}
\end{textbox}

The key advantage of the large bandwidth lies in the possibility for identifications of new species as well as accurate modeling of line emission leading to robust derivations of excitation temperatures and column densities. In particular, secure identifications of new species need a high number of well-isolated lines that can be assigned and modeled (see Sect.~\ref{ss:reporting}). Furthermore, with access to larger bandwidths, the range of energy levels and line strengths covered by individual species provides highly accurate constraints on their excitation. With of order 15--20 identified lines, the typical uncertainties on excitation temperatures and column densities become less than 10--20\% when local thermodynamic equilibrium (LTE) can be assumed \citep[e.g.,][]{jorgensen18}. Laboratory astrophysics has also benefited from these technical developments \citep[see][and references therein]{widicus-weaver19}. New spectrometers with large bandwidths are starting to be used in the laboratory to measure the rotational spectra of molecules -- in particular, complex ones \citep[e.g.,][]{wehres18a,cernicharo19}. 

The challenges with such wide spectral scans lie mainly in dealing with huge data volumes and in the need for new tools for their analysis. Previous more targeted, and lower sensitivity, studies typically only picked up (at most) some handfuls of lines of an individual species. In those cases classical rotation diagram methods worked well, because one could deal practically with separating and fitting individual lines. However, in the larger, typically more confused, spectra, it is often more useful to fit the entire observed spectrum using synthetic models (see Fig.~\ref{f:improvedspectro} for examples). Such models predict the full spectra under, e.g., the LTE assumption, taking into account optical depth of individual lines specifically -- and accounting for multiple molecular species at once, thus implicitly sorting out chance alignments, line blends and, also importantly, testing for non-detections (see also Sect.~\ref{ss:reporting}).

\subsubsection{Improvements in sensitivity}
\label{ss:sensitivity}
Another important aspect of the telescope advances with \emph{Herschel} and ALMA has been the significant improvements in sensitivity, both due to the increased collecting area compared to previous facilities as well as (for ALMA) the excellent conditions on the telescope site. This improvement enables more statistical approaches to astrochemical studies, whether in terms of the ranges of lines and species observed for individual sources or in terms of targeting larger samples of sources. In particular, the latter makes it possible to determine the degree to which variations in molecular emission signatures for different sources are caused by the influence of, e.g., physical conditions (the impact of the external radiation field, variations in cosmic-ray flux, degree of turbulence etc.) on the resulting chemistry.

Moving into this new sensitivity regime also brings new challenges. One example is that the line-confusion limit is now reached also for sources that were relatively line-poor in shallower surveys. A spectrum has reached the confusion limit when emission from spectral lines is detected in basically every single spectral channel \citep[see, e.g., the 1.3~mm single-dish spectra toward Sgr B2(N) of][]{belloche13}, implying that a longer integration time with the same telescope would not bring any new information. This issue becomes more severe at higher frequencies because the spectral line density of a mixture of COMs is expected to be relatively uniform in frequency space. The linewidth in frequency space is, however, increasing with frequency (in astronomical spectra the linewidth is expected to be constant in velocity space) and therefore the likelihood of line overlaps, and thereby confusion,  in line-rich sources increases with frequency as well.

Reaching the confusion limit raises some practical problems: in particular if the observed angular resolution is coarse compared to the kinematical structure of the source it becomes difficult, if not impossible, to estimate the underlying continuum emission (see, e.g., bottom left panel of Fig.~\ref{f:pils_overview}). Rather than subtracting the continuum based on line-free channels before performing Fourier transforms and cleaning of the interferometric data, one either needs to take the continuum into account in the synthetic spectrum modeling or define the continuum by fitting the  flux distributions pixel-by-pixel in the imaged datacubes (see \citealt[][]{sanchezmonge18} and \citealt{jorgensen16} for example methodologies). While the former has the advantage of most closely resembling the actual data with minimum tinkering, it is often impractical -- in particular, if the spatial distribution of molecular line emission is considered for line-rich sources. 

\begin{table}
\caption{List of molecules mentioned in this review with more than three atoms.}    \label{tab:molecules}
    \begin{center}
    \begin{tabular}{ll|ll}\hline
Species & Formula & Species & Formula \\ \hline
Acetaldehyde & CH$_3$CHO                & Glyoxal & HC(O)CHO                              \\
Acetamide & CH$_3$C(O)NH$_2$            & Hydroxylamine & NH$_2$OH                        \\
Acetic acid & CH$_3$COOH                & Isocyanic acid & HNCO                           \\      
Acetone & CH$_3$C(O)CH$_3$              & Methane & CH$_4$                                \\    
Ammonia & NH$_3$                        & Methanimine & CH$_2$NH                          \\  
Benzene & c-C$_6$H$_6$                  & Methanol & CH$_3$OH                             \\
Benzonitrile & c-C$_6$H$_5$CN           & Methoxymethanol & CH$_3$OCH$_2$OH               \\
Butyl cyanide & C$_4$H$_9$CN            & Methyl acetylene & CH$_3$CCH                    \\
Cyanoacetylene & HC$_3$N                & Methyl amine & CH$_3$NH$_2$                     \\
Cyanodiacetylene & HC$_5$N              & Methyl chloride & CH$_3$Cl                      \\
Cyanoformaldehyde & NCCHO               & Methyl cyanide & CH$_3$CN                       \\
Cyanomethanimine & NHCHCN               & Methyl formate & CH$_3$OCHO                     \\
Cyanomethyl radical & CH$_2$CN          & Nitrous acid & HONO                             \\
Cyclopropenone & c-H$_2$C$_3$O          & Propanal & C$_2$H$_5$CHO                        \\
Dimethyl ether & CH$_3$OCH$_3$          & Propanol & C$_3$H$_7$OH                         \\
Ethanimine & CH$_3$CHNH                 & Propenal & C$_2$H$_3$CHO                        \\
Ethanol & C$_2$H$_5$OH                  & Propyl cyanide & C$_3$H$_7$CN                   \\
Ethylene glycol & (CH$_2$OH)$_2$        & Propylene oxide & c-CH(CH$_3$)CH$_2$O           \\
Formaldehyde & H$_2$CO                  & Quinoline & C$_9$H$_7$N                         \\
Formamide & NH$_2$CHO                   & Thioformaldehyde & H$_2$CS                      \\
Formic acid & HCOOH                     & Vinyl cyanide & C$_2$H$_3$CN                    \\
Glycolaldehyde & CH$_2$(OH)CHO          & Urea & NH$_2$C(O)NH$_2$                         \\ 
Glycolonitrile & HOCH$_2$CN             &         &                \\ \hline
    \end{tabular}
    \end{center}
\end{table}

Several strategies can be adopted to beat the spectral confusion limit in order to detect species that have a low abundance or weak lines. Observations at higher angular resolution can reveal regions with smaller velocity dispersion or separate sources that have different systemic velocities and were blended in larger beams. This strategy led to significant advances for instance in Sgr~B2(N) \citep[][]{belloche16,belloche19} or IRAS~16293 \citep{jorgensen16}. 
Finding sources with intrinsically narrow linewidths is another promising avenue that is illustrated, for example, by the recent detection of methoxymethanol\footnote{See Table~\ref{tab:molecules} for a list of the names and formulae for molecules discussed in this review.} in the hot core MM1 of NGC~6334I \citep[][]{mcguire17}.

Finally, going to lower frequency where the spectral confusion is less severe is another option, provided that the emission lines are still strong enough to be detected. IRAM has started to explore frequencies below 80 GHz (down to 70 GHz with NOEMA and 73 GHz with the 30-m telescope), and Bands 1 and 2 of ALMA will be valuable in this respect in the near future. COMs have been detected at even lower frequencies, for instance propanal and benzonitrile with the GBT \citep[][]{hollis04a,mcguire18}.
While confusion is not (yet) an issue at these low frequencies, the difficulty lies in the excitation of the molecules that does not follow LTE in the environments probed by these observations. Collision rate coefficients are not available for many COMs, making a reliable estimate of their column densities under such conditions a challenge. However, recent progress has been made in this respect for, e.g., methyl formate and methanimine \citep[][]{faure14,faure18}.

\subsubsection{Increased spatial resolution}
\label{sss:spatial_resolution}
ALMA has clearly pushed observations at submillimeter wavelengths to a new regime with its high angular resolution and sensitivity providing images of dust and gas with a resolution of 0.01\arcsec, corresponding to few au scales in nearby star forming regions. Even in ALMA's intermediate baseline configurations, the achieved angular resolution of $\approx$0.1\arcsec\ represents an improvement by a factor 5--10 in angular resolution compared to what is typically achieved at other facilities. Besides helping with issues such as line confusion described above, the advantages of imaging at these spatial scales are obvious: it  makes it possible to look at the spatial coincidences and separations between different species and thereby reveal their chemical relations and their link to the underlying source physical structures.

One challenge encountered in high-resolution studies of low- and high-mass protostars comes from the optical thickness of the observed line and continuum emission. Many sources show unresolved continuum structures that become optically thick, in particular, in the higher-frequency ALMA bands. The continuum optical thickness may, for example, suppress the line emission on protostellar disk scales \citep[e.g.,][]{harsono18}. These effects made it necessary for the PILS and ReMoCA surveys to focus on positions offset from the main continuum peaks to derive reliable column densities.

\subsection{Spectroscopic identifications of new species}\label{ss:reporting}
There are a number of considerations to make when reporting detections of new species or
presenting derivations of their physical/chemical characteristics. This is becoming particularly important with the new sensitive observations, where line-rich spectra of individual sources are often revealed serendipitously. 

\subsubsection{Considerations for new detections}
\label{sss:newdet}
While it may be uncontroversial to report the detection of a common species toward sources belonging to a well-studied group, more care needs to be taken with an exotic claim (such as a completely new species or the detection of a species toward a type of region where it has not previously been seen). In the former case, it is often sufficient just to note the rest frequency of the species from common spectroscopic references, but in the latter case a range of transitions is required to be measured independently. The number of transitions needed for a secure claim strongly depends on the spectral line density \citep[see][]{neill12,halfen06}.

For new detections, a number of other criteria should be fulfilled as well: the line widths and local-standard-of-rest (LSR) velocities of all transitions of a given species have to be consistent or, if they vary (for example, as a function of energy level), an explanation needs to be provided (obviously an interesting scientific result in its own right). Likewise it is critical to check that the column density and/or excitation temperature derived based on the measured transitions do not predict other transitions to be present across the observed spectral ranges where they are not seen, e.g., from lines of higher intrinsic strengths or more favorable energy levels. Chance alignments from transitions of other (more common) species should also be checked. In particular, in the ALMA era it is often found that even fairly high energy levels of common organic molecules can be populated (including rotational levels in vibrationally or torsionally excited states) that can lead to serious false identifications. To fulfill all these criteria, especially in the case of hot cores or corinos, a recommended approach is to fit a complete spectral survey with a synthetic spectrum that accounts for all identified molecules rather than rely on simple independent Gaussian fits of individual lines of a given species. The formalism behind such synthetic spectra is described in \citet{moller17} and a number of publicly available tools are available to calculate them (e.g., \href{http://cassis.irap.omp.eu}{\emph{\underline{CASSIS}}}, \href{https://xclass.astro.uni-koeln.de}{\emph{\underline{XCLASS}}},  \href{https://www.iram.fr/IRAMFR/GILDAS/doc/html/weeds-html/weeds.html}{\emph{\underline{Weeds}}}, and \href{http://cab.inta-csic.es/madcuba/MADCUBA_IMAGEJ/ImageJMadcuba.html}{\emph{\underline{MADCUBA}}}). Details about basic radiative transfer equations and the derivation of column densities can be found in, e.g., \citet{goldsmith99} and \citet{mangum15}.

An often overlooked aspect is the impact of continuum emission on small scales: the presence of high column densities of dust on small scales can introduce frequency-dependent continuum opacity significantly altering the intensities of observed molecular transitions if split over multiple bands. The strength of the continuum emission may also be such that the corresponding ``background'' temperature terms in the equation of radiative transfer cannot be neglected \citep[see, e.g., Sect.~4.4 of][]{belloche19}. Also, these effects strongly affect the use of traditional methods such as population diagrams that typically assume a negligible amount of background radiation \citep[see][for details about population diagrams]{goldsmith99}.

Finally, the spectroscopic reference needs to be evaluated in terms of the accuracy of the listed frequencies and any extrapolations from laboratory measurements (see Fig.~\ref{f:improvedspectro} and Sect.~\ref{ss:labspectro}). We encourage astronomers to inspect the documentation for individual species in the spectroscopic databases such as the \href{https://cdms.astro.uni-koeln.de}{\emph{\underline{Cologne Database for Molecular Spectroscopy}}} (CDMS; \citealt{cdms1,cdms2}) and \href{https://spec.jpl.nasa.gov}{\emph{\underline{Jet Propulsion Laboratory's database for molecular spectroscopy}} (JPL; \citealt{jpl})} and cite the relevant spectroscopic studies, and spectroscopists to make their measurements and associated predictions as well as partition functions available in such databases in order to increase the impact of their work in the astrochemical community.

\subsubsection{Reporting physical properties}\label{sss:reportingproperties}
Both in the context of new detections, and when reporting properties from line emission for other species, there is also a range of issues that need to be considered and reported. Exact positions, beam sizes, and, in the case of interferometric observations of extended structures, spatial sensitivity -- in terms of $(u,v)$-coverage -- obviously belong to such critical information. For constraints on the column densities of the emission, it is important to consider the source size in comparison to the angular resolution of the observations. In mapping observations, this involves a comment about the distribution of the material (i.e. whether it is Gaussian or homogeneous with respect to the beam) and whether the column density is estimated from a single position or pixel (where the measured flux density in janskys per beam can be translated into an effective radiation temperature in kelvins) or whether an integration over a larger area is adopted (where the beam-to-pixel size ratio also needs to be taken into account).

As mentioned above, when constructing a model for the excitation of a given molecule, the important physical parameters besides the column density and excitation temperature are the systemic velocity and line width. In particular, the latter -- together with the assumed extent -- is important when considering possible optical depth effects. On the scales routinely probed with ALMA, many lines of the main isotopologs \citep[and in some cases even less abundant isotopologs;][]{jorgensen18} become optically thick. In those cases, it is often possible to obtain apparently good fits in population diagrams, but if those lines in reality are optically thick then no constraints are obtained on the actual column densities. For such species it is therefore critical to seek additional constraints to confirm which lines are optically thin -- either by looking at rarer variants or studying transitions with lower line strengths -- when constraining their column densities.

It is also important to consider whether or not local thermodynamic equilibrium (LTE) is a reasonable approximation, i.e., whether the densities of the main collision partners (typically H$_2$) in the region are sufficiently high that the  excitation of a molecule is collision-dominated. In hot corinos and cores where the densities exceed $10^{7}$--$10^{8}$~cm$^{-3}$, LTE is often achieved for most species and transitions of interest, but in more tenuous environments (e.g., prestellar and translucent cores, outflow regions) this is not necessarily the case. It is also worth keeping in mind that while LTE may work well at high frequencies (e.g., at submillimeter wavelengths) for a given source, this may not be the case for the same source at lower frequencies (e.g., centimeter wavelengths) if the observations trace more extended emission. In the case of non-LTE excitation it becomes even more problematic to trust the assignments of lines and thus claims of detections and properties of a given species. This issue may outweigh the advantage that the line confusion at longer wavelengths may be lower.

For species excited to temperatures of 100--150~K or above it is important to note whether the partition function associated with a given spectroscopic entry includes the vibrational contribution or whether it is purely rotational. In the latter case, vibrational corrections need to be applied to derive the full column density of the molecule (see, e.g., Sect.~5 of \citealt{margules17}).

A final consideration is whether column densities or abundances are reported for the modeled species. Abundances are very often referenced relative to the column density of H$_2$, as estimated from the dust continuum or CO gas-phase lines. However, such estimates may be problematic (e.g., due to different spatial distributions of CO and the targeted species) or highly model-dependent (e.g., the dust temperature and assumed dust opacity law). Often it is helpful to focus on the relation between a given species and its possible chemical precursor, but again here care must be taken that the choice of reference species does not introduce spurious correlations, e.g., due to assumptions about the  excitation or the line optical thickness of the reference species.

\subsubsection{The importance of laboratory spectroscopy}
\label{ss:labspectro}
The identification of interstellar COMs rely heavily on the accuracy of frequencies and line strengths derived from the analysis of laboratory measurements \citep[see review by][]{widicus-weaver19}. The complexity of the COM Hamiltonians due to, e.g., internal rotation, makes extrapolations beyond the range of frequencies measured in the laboratory uncertain or even unreliable. With the new era of sensitive spectral broadband datasets at submillimeter wavelengths, laboratory measurements at high frequencies are becoming even more crucial. The spectroscopy community has already begun to extend the spectral characterization of a number of COMs into the submm domain \citep[e.g.,][]{wehres18b,kolesnikova18,motiyenko19}. But even the extension of the spectroscopy of COMs from the cm range, investigated several decades ago, to the mm range already represents significant progress \citep[e.g.,][]{cernicharo16,alonso16,martinDrumel19}.

\begin{figure}[ht]
\includegraphics[width=0.5\paperwidth]{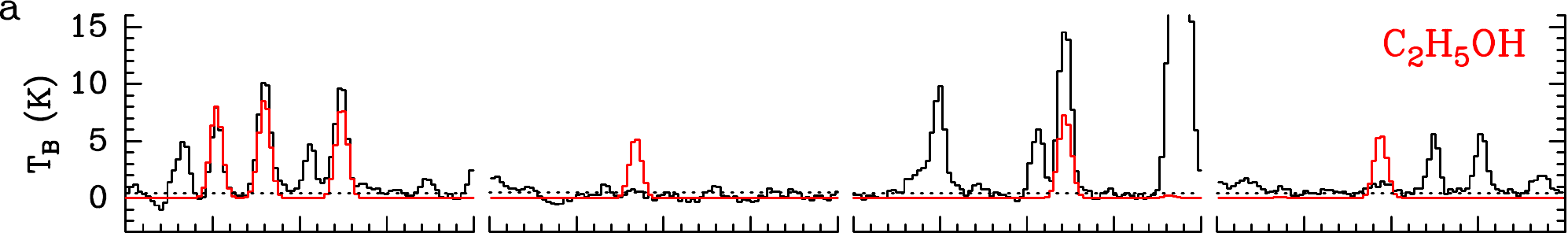}
\includegraphics[width=0.5\paperwidth]{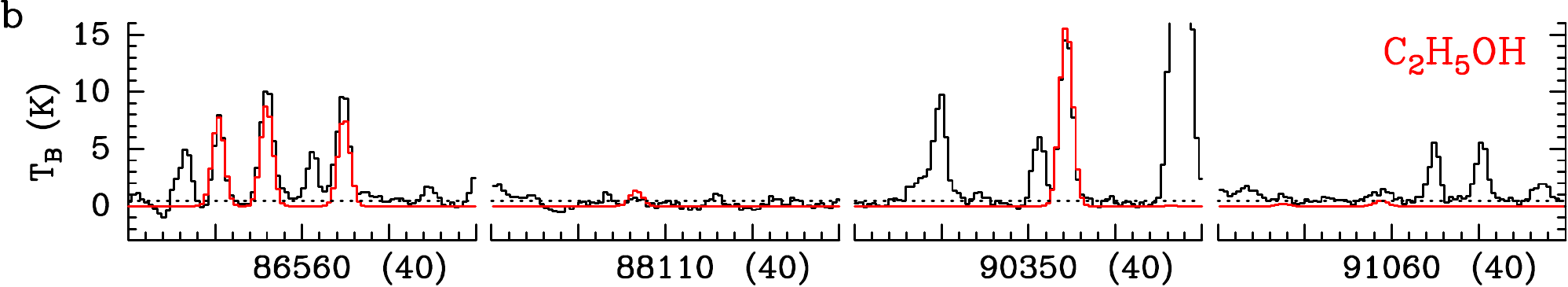}
\includegraphics[width=0.5\paperwidth]{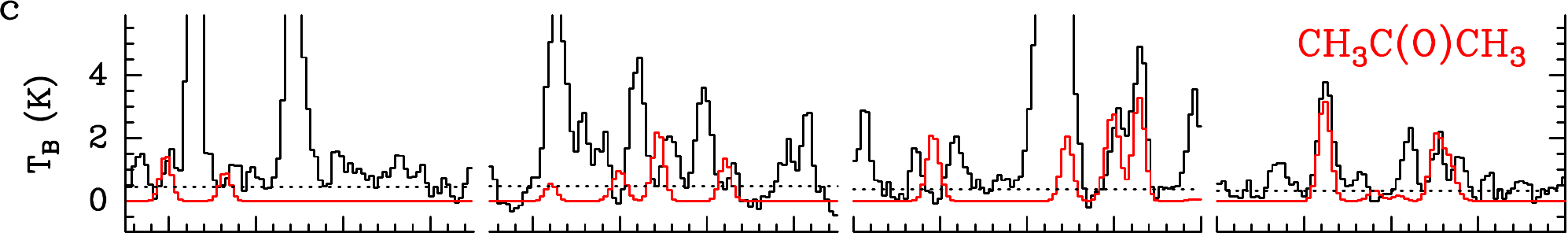}
\includegraphics[width=0.5\paperwidth]{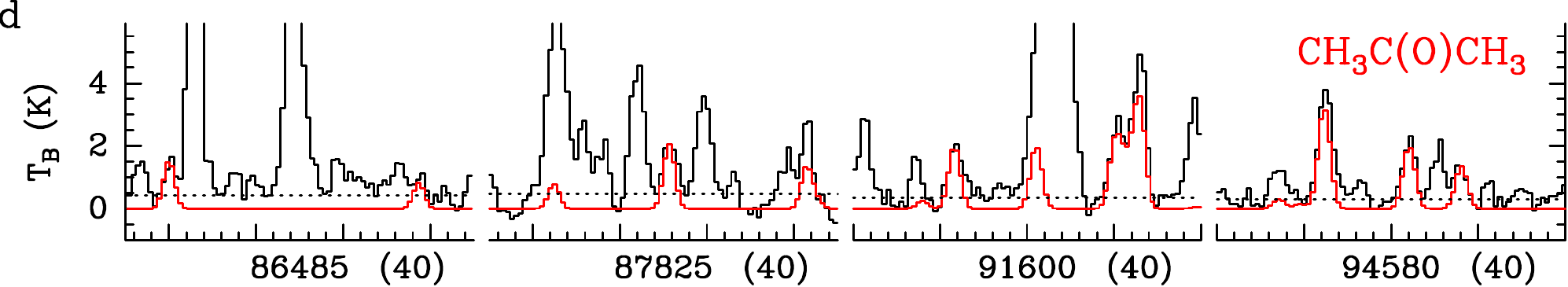}
\caption{Synthetic spectra (red) of ethanol (a, b) and acetone (c, d) computed assuming LTE, overlaid on spectra (black) of the hot molecular core Sgr~B2(N2) observed with ALMA in the frame of the EMoCA project \citep[][]{belloche16}. The synthetic spectra of panels b and d were computed with recent spectroscopic predictions produced by \citet{mueller16a} and \citet{ordu19}, respectively. They agree much better with the observed spectra than the synthetic spectra computed with older predictions from the JPL database (panels a and c). The central frequency and frequency range (in parentheses) are indicated in MHz below each sub-panel.
}
\label{f:improvedspectro}
\end{figure}

A more accurate characterization of the Hamiltonian of COMs with advanced modeling codes is also critical: the identification of new interstellar COMs with low abundance requires a good knowledge of 
the complete spectra of more abundant, known, COMs -- that is, not only their strongest lines, but also the weaker ones. The recent progress made with, for example, C$_2$H$_5$OH \citep[e.g.,][]{mueller16a} or CH$_3$C(O)CH$_3$ \citep[e.g.,][]{ordu19} illustrates how critical this is (see Fig.~\ref{f:improvedspectro}). There is also a strong need to characterize the rotational spectrum of COMs in their vibrationally excited states \citep[e.g.,][]{mueller16b,degliesposti17}, and of their isotopologs \citep[e.g.,][]{margules16,zakharenko19}, to advance the line identification of astronomical spectra.

\subsection{Linking observations, chemical modeling, and laboratory experiments}
\label{ss:modeling}
Our understanding of the chemistry of star-forming regions is constantly informed by ongoing efforts to simulate it both in the laboratory and through computational means. The latter includes both the calculation of rates and parameters of individual processes and full-scale chemical kinetic simulations of astronomical objects. However, observational discoveries of new molecules have often been the driving factor in the choice of species to study.

Over the past decade or so there has been an increased interest in determining the importance, or otherwise, of dust grain-surface chemistry to the production of complex organic molecules, including through thermal or energetic processing of mixed molecular ices, as well as their relative importance compared to, e.g., gas-phase processes. While the field has largely moved beyond the facile ``either/or'' argument, there is nevertheless still disagreement as to the precise origins of even some of the most commonly observed molecules in star-forming regions, such as formamide (NH$_2$CHO). Quantum rate calculations for the gas-phase reaction of formaldehyde (H$_2$CO) with the radical NH$_2$, leading to formamide, have suggested variously that this mechanism is very {\em efficient} \citep{barone15,skouteris17} or that it is prohibitively {\em inefficient} \citep{song16}. Meanwhile, chemical models indicate \citep[e.g.,][]{garrod13,quenard18} that with warm ($\sim$20~K) dust grains, formamide can be produced through reactions within/upon the ice mantles. Even for molecules such as dimethyl ether, with plausible production mechanisms in both the gas phase and on grains (see Fig.~\ref{fig:DME}), the relative contributions may also be dependent on the interaction of large-scale physical conditions like temperature with molecule-specific microscopic quantities such as binding energies. Thus, it is likely that there is no blanket description that applies equally well to all COMs, even within a single astronomical source. The chemistry of star-forming regions may perhaps be best described as an intrinsically well-coupled gas-grain chemical system. The recent push to explore the limits of both gas and grain-surface/ice chemistry has led to the investigation of a number of new types of reaction mechanisms, which may become standard parts of the toolbox for astrochemical simulations of all kinds.

\begin{marginnote}[120pt]
\entry{Radical}{A molecular species that has an unpaired valence-shell electron, making it typically very reactive.}
\end{marginnote}

\subsubsection{Chemical modeling}
Chemical kinetic models of star-forming regions calculate the time-evolution of chemical abundances, as determined by a network of reactions and processes and their rates. The main advances made in this field over the past decade involve either the improvement or expansion of chemical networks, the detailed treatment of grain-related chemistry, or the improvement of astrophysical inputs such as density, temperature, and radiation fields.

Models of complex molecule production in star-forming regions have frequently concentrated on high-mass star-forming regions, i.e. hot cores. Since early models of hot-core chemistry, such as that of \cite{brown88}, some form of two-stage physical model has been assumed. In those, the first stage involves the low-temperature evolution and/or collapse of a molecular cloud core, with the second treating the hot-core stage proper, with higher densities and temperatures. A major feature of the first stage involves the build-up of molecular ice mantles on dust grains, composed primarily of simple hydrides such as H$_2$O, NH$_3$, and CH$_4$, as well as CO, CO$_2$, H$_2$CO, and CH$_3$OH. In the second stage, these are ejected into the gas phase -- either immediately, or gradually, according to a time-dependent release as temperatures are ramped up to observed values \citep{viti99}. The thermal ejection of grain-mantle molecules is a major driver of gas-phase chemistry, through both ion-molecule and neutral-neutral reactions \citep{charnley92}.

\subsubsection{Grain-surface chemical treatments}
Models with some explicit treatment of grain-surface chemistry are now fairly standard, due to the need to consider ice evolution. The use of three-phase models (ice mantle, surface, and gas phase), based on the treatment by \cite{hasegawa93}, has also become more common. These models treat the upper ice layer as a separate phase from the underlying bulk, which in some cases is treated as being chemically active \citep{garrod13}. Other models have considered multiple layers within the bulk \citep{taquet14}. In general, gas-grain models calculate grain-surface abundances using the same so-called ``rate equation'' approach as is used for pure gas-phase chemistry. This method ignores the inherently stochastic nature of the grain-surface chemistry, and can lead to divergence from numerically-exact treatments. \cite{vasyunin13} successfully constructed a three-phase model that treats both gas and grain chemistry accurately using a macroscopic Monte Carlo method, followed more recently by \citet{lu18}. Rate-equation models can also be adapted to replicate stochastic surface chemistry reasonably well, using the modified-rate method \citep{garrod08a, garrod09}, although the method has not been widely adopted \citep{furuya15}. 

\begin{marginnote}[120pt]
\entry{Photolysis}{The process of photodissociation of molecules, and subsequent associated mechanisms, caused by photons typically at UV/Visible wavelengths.}
\entry{Radiolysis}{The ionization, and subsequent associated processing, of a target material caused by the impingement of an ionizing particle.}
\end{marginnote}

Much modeling work over the past decade has involved the development of chemical networks for the grain-surface production of COMs. Early grain chemistry networks \citep{tielens82} included mechanisms for the build-up of certain COMs by repetitive atomic addition, as well as through reactions between radicals \citep{allen77}. \cite{garrod06} presented a model of hot-core chemistry in which the surface mobility, and thus the reactivity, of radicals on grain surfaces would be enhanced by including the gradual warm-up of the gas and dust, which had previously been considered only in the context of the {\em desorption} of grain-surface molecules \citep{viti99, viti04}. In these models, radicals are produced largely as the result of photodissociation (i.e. photolysis) of stable, solid-phase species such as methanol (CH$_3$OH), caused by cosmic ray-induced UV photons. 
Reactions between the radical products of methanol dissociation become prominent at temperatures of 20--40~K resulting in the subsequent formation of  many COMs. Many of those linger on grain surfaces until temperatures upward of 100~K are reached where they sublimate and become observable in the gas phase. The models were expanded to include a larger network of radical reactions \citep{garrod08b} that was quite successful in providing a basic framework for the production of many COMs, which has been extended over subsequent years to include newly-detected species \citep{garrod17, mueller16a, belloche17}, especially toward Sgr B2(N).

\begin{figure}[!htb]
    \centering
    \includegraphics[width=\textwidth]{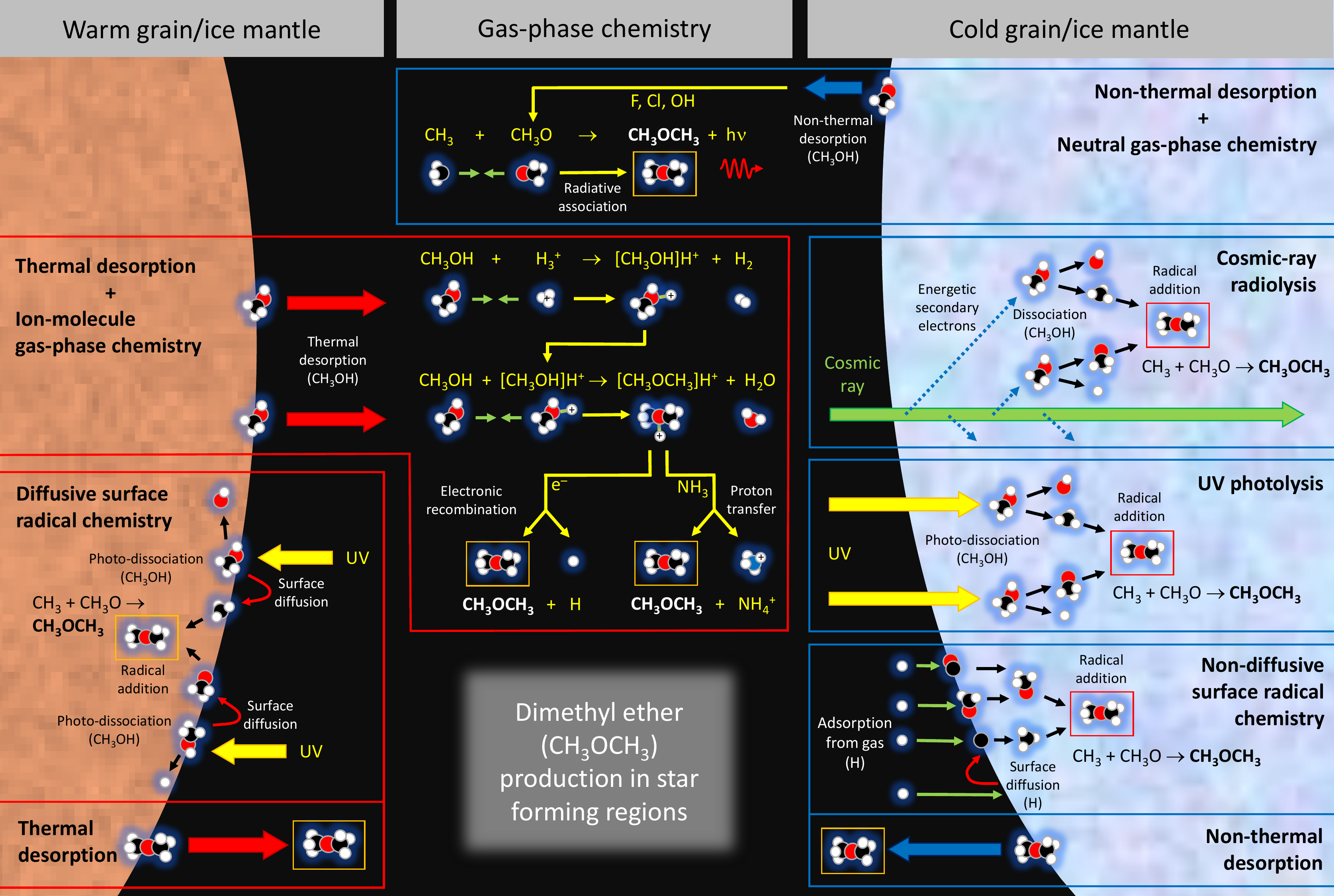}
    \caption{An illustrative selection of proposed pathways to the formation of the representative COM dimethyl ether, CH$_3$OCH$_3$, in star-forming regions, under either warm or cold conditions. Not all possible pathways are shown. Each of the processes occurring on cold grains or within cold ice mantles could also occur at higher temperatures, although with arguably lower significance. The warm grains should be assumed to inherit much of the ice-mantle composition of the earlier, cold dust grains. The cold UV-induced photolysis process implicitly assumes minimal thermal diffusion of dissociation products. The cosmic-ray radiolysis process could also involve ``suprathermal'' dissociation products (not shown). UV photolysis and cosmic ray-induced radiolysis may occur either within the ice mantles or on their surfaces. On the warm grains, surface diffusion of many radicals is expected to be rapid, but diffusion rates within the ice mantles remain uncertain, and could also depend on the presence of ice porosity. Gas-phase protonation of, e.g., methanol may occur by reaction with multiple different molecular ions and not only H$_{3}^{+}$. Electronic recombination of protonated dimethyl ether, [CH$_3$OCH$_3$]H$^+$, is expected to produce dimethyl ether in no more than 7\% of cases (see text), while proton transfer to ammonia (NH$_3$) may be very efficient.}
    \label{fig:DME}
\end{figure}

Importantly for gas-grain models, uncertainties exist in the desorption energies (i.e. binding energies) and diffusion barriers of many chemical species. The former control the temperatures at which atoms or molecules may thermally desorb from the dust grains and into the gas, while the latter determine the rates at which reactive species may migrate on the grain surfaces through diffusion, allowing them to react \citep[see also][]{cuppen17}. In the case of radicals, both values are often poorly defined by experiment, due to these reactive species' short lifetimes. Some grain-surface reactions require a chemical activation energy barrier to be overcome. Below $\sim$100~K, many such processes important to astrochemistry involve quantum tunneling of H atoms, however, the rates for barrier-mediated reactions used in chemical networks are often poorly defined. Hydrogen addition to unsaturated COMs, which often involves an activation barrier, could be especially important to interstellar abundances \citep{alvarez-barcia18,krim18}. 

The degree of diffusion that can occur within the bulk ice is also uncertain and could indeed be minimal for species other than H and H$_2$. Although such diffusion is technically required in the models to allow radical photo-products to react within the bulk, new methods are emerging to treat those reactions without mediation by diffusion, allowing COMs to be formed even at very low temperatures \citep{garrod19}, in line with experimental evidence \citep[e.g.,][]{henderson15}.

The strength of the radical-radical reaction process for COMs is affected both by the warm-up timescale of the core, and by the rates of ice photodissociation, which scale with the cosmic-ray ionization rate. These two physical inputs are somewhat degenerate in their chemical effects. Recent models \citep[e.g.,][]{barger19} indicate that short warm-up timescales (on the order of 10$^4$ yr) and high cosmic-ray ionization rates ($\sim$10$^{-16}$ s$^{-1}$) may provide the best agreement with observations. Dedicated models of chemistry in individual cores within Sgr B2(N) \citep{bonfand19, willis19} support the latter finding. Photodissociation rates of individual solid-phase molecules are also not well constrained by laboratory data, although they are expected to be lower than equivalent gas-phase processes by a factor of a few \citep{kalvans18}.

\cite{shingledecker18} presented a gas-grain model of cosmic ray-induced radiolysis of interstellar ice mantles. In the case of cosmic rays, the impinging particle (which may be an electron, proton, or a heavier atomic nucleus) has far more energy than is required to produce a single molecular ionization, and the resultant electron will also be capable of producing further ionization events.
The cascade of electrons produced by a single cosmic ray results in dissociation, excitation, ionization, and recombination of species within the ice. \citeauthor{shingledecker18} considered this process as a means of forming COMs in quiescent interstellar clouds (mostly through reactions between neutral product species, including radicals), but the mechanism could also plausibly influence COM abundances in actively star-forming cores (see also Sect.~\ref{sss:labexperiments}).

\subsubsection{Gas-phase mechanisms}
Gas-phase production mechanisms for COMs have long been a part of the chemical networks, and the production of some relatively complex species through ion-molecule gas-phase chemistry remains uncontroversial. For example, for dimethyl ether, CH$_3$OCH$_3$ (Fig.~\ref{fig:DME}), the gas-phase reaction of methanol with protonated methanol (CH$_3$OH$_2^+$) can lead to the formation of protonated dimethyl ether. This species can recombine with an electron to produce dimethyl ether in up to 7\% of cases, while the remainder of recombinations should result in the fragmentation of the underlying C--O bonds \citep{hamberg10}. Chemical models indicate that the large abundances of gas-phase methanol, following grain-mantle ejection, could make this process effective enough to explain dimethyl ether observations in star-forming regions. However, for other molecules, the inefficient conversion of protonated molecules to a stable, neutral COM via electronic recombination presents a problem. The main gas-phase {\em destruction} mechanism for many COMs will also be their protonation by small molecular ions, followed by destructive electronic recombination.

\cite{taquet16} proposed a solution to this problem, citing the reaction of protonated COMs with gas-phase NH$_3$ (following its  desorption) as a way to remove a proton without destroying the underlying COM structure. They also suggest that gas-phase mechanisms could produce a selection of COMs through ion-molecule chemistry, with proton-transfer to ammonia being the terminal step. These authors further proposed that the heating of grains as the result of episodic accretion events in low-mass sources (see Sect.~\ref{ss:episodic_accretion}) could induce thermal desorption of molecules that would promote gas-phase chemistry. The formation of methyl formate would proceed via the reaction of protonated methanol with formic acid, leading to the {\em trans}-conformer of protonated methyl formate \citep{neill11}. Further conversion to produce the commonly-observed {\em cis}-methyl formate would be required, but the efficiency of the overall process under interstellar conditions is not well constrained.

There has recently been renewed interest in neutral gas-phase reaction mechanisms for COM production (prompted by the detection of cold COMs in prestellar cores), especially those involving radicals. Such mechanisms include radiative association, a process by which two species react to form a single product that is stabilized by the emission of a photon. These reactions are inefficient for small-molecule production, but for larger products -- including COMs -- the efficiency could be much greater. Calculations by \cite{vuitton12} suggest radiative association is rapid for large hydrocarbon radicals. \cite{vasyunin13b} found the radiative association reaction CH$_3$ + CH$_3$O $\rightarrow$ CH$_3$OCH$_3$ + $h\nu$ to be effective under prestellar core conditions. Rates for radiative association are generally difficult to constrain in the laboratory, so detailed calculations will be required to confirm the production efficiencies for specific COMs. \cite{balucani15} noted further that the abstraction of a hydrogen atom from dimethyl ether by Cl and, to a lesser extent, F atoms -- followed by reaction of the resultant radical with an O atom -- could lead to the production of methyl formate itself. \cite{skouteris18} recently considered similar processes beginning with ethanol 
to form glycolaldehyde, and acetic and formic acid. Many of the above-mentioned reactive processes have been tested with chemical models under only a limited set of physical conditions (centered mainly around cold, prestellar core conditions) and are dependent on the efficient gas-phase production of radicals. However, this neutral and radical chemistry provides an important new dimension to  networks for COM production and further investigation will be highly valuable.

\subsubsection{Laboratory experiments}\label{sss:labexperiments}
The production of COMs through the irradiation of mixed molecular ices has long been an important area of astrochemical research. The relative influence of cosmic ray-induced ultraviolet/visible irradiation (i.e. photolysis) versus direct cosmic-ray impingement (i.e. radiolysis) in interstellar COM production continues to be debated. Complex organics typical of star-forming regions can be produced through either scheme \citep{bennett07,bennett11,oberg09,oberg16}. However, the action of ``supra-thermal'' (i.e. excited or kinetically energetic) species may play a larger role in radiolysis, while COM production through photolysis would be closer to the radical-addition mechanisms considered in many astrochemical models. \cite{oberg09} demonstrated that a large range of observed COMs could be formed through photolysis of solid methanol, with product abundances varying with temperature (20--70~K). More recently, \cite{butscher16} showed that HCO radicals produced by formaldehyde photodissociation could be stored in the ice at low temperatures, becoming more mobile and reactive as temperatures increase. The relative orientations of the radicals during reaction may also be important to the product ratios and reaction efficiencies \citep{bergner16,enrique-romero16,butscher17}.

A major area of new ice-chemistry research involves the production of COMs at very low temperatures without the need for diffusion of large radicals (thermally or otherwise), nor with any initiating energetic process. \cite{fedoseev15} found that co-deposition of H and CO at low temperatures could produce COMs, which they suggest occurs through the production of glyoxal (HCOCHO) via HCO radical addition, which is subsequently hydrogenated to produce glycolaldehyde and ethylene glycol. In this scheme, the low-temperature chemistry that produces methanol by hydrogen addition would allow more complex species to be formed in tandem. Reactions between HCO radicals would be the result not of diffusive meetings, but rather their occasional production in close enough proximity that little or no diffusion would be required. The HCO radical could also plausibly react with CO itself, mediated by an activation energy barrier but with little need to diffuse. Subsequent studies have highlighted cold pathways to other COMs, through deposition of additional species \citep{fedoseev16,qasim19,dulieu19}. In this review, we refer to such processes generically as {\em non-diffusive}, although they may indeed be initiated by the diffusion of H atoms, which are mobile even at very low temperatures. It seems likely that non-diffusive mechanisms could make a significant contribution to the COM content of star-forming regions. However, methyl formate does not seem to be produced in sufficient quantities to explain its interstellar abundance and photolysis may therefore still be required \citep{chuang16} if it originates in the solid phase. Astrochemical models have treated non-diffusive reactions in the past on a case-by-case basis \citep{garrod11}, but new models are beginning to include more general treatments for this cold chemistry \citep[e.g.,][]{jin20}.


\section{CHEMICAL INVENTORIES IN REGIONS OF STAR FORMATION}\label{s:inventories}
\subsection{Recent salient results about molecular complexity}\label{ss:complexity}
Between 2010 and 2019, 58 new species were detected in the ISM, at a rate exceeding the approximately 4 new detections per year on average since 1968 \citep{mcguire18}. About 36\% of those new detections were COMs, of which 71\% were detected toward dark clouds or star-forming regions, with the majority being highly saturated species. Since 2014, 6 out of 16, i.e. $\sim$40\% of the new COM detections, were made with ALMA -- a clear demonstration of its capabilities for astrochemical studies. But large single-dish telescopes still have a role to play, as illustrated by the detections of 9 new COMs with the GBT and IRAM 30~m telescopes over the same period (4 and 5, respectively, for these telescopes).

Three of the new COM detections reported over the past five years have revealed the wider complexity of the molecular structures produced by chemistry in star-forming regions. The ALMA detection of the branched alkyl molecule iso-propyl cyanide toward the hot core Sgr~B2(N2) with an abundance only 2.5 times lower than that of its straight-chain structural isomer normal-propyl cyanide has shown that branched molecules can easily form in the ISM \citep[][]{belloche14}. Astrochemical simulations subsequently showed that at the next stage of complexity in the alkyl cyanide family (C$_4$H$_9$CN), some of the branched isomers should even dominate over the straight-chain one \citep[][]{garrod17}. This remains to be verified observationally. The detection of a chiral molecule has also enlarged our inventory of the types of molecular structures present in the ISM: \citet{mcguire16} reported the detection of propylene oxide in the extended, low-density region of the protocluster Sgr~B2(N) with the GBT and the Parkes telescope. These observations do not tell us if propylene oxide has an enantiomeric excess in Sgr~B2, as was found for some chiral molecules in meteorites \citep[see, e.g.,][]{pizzarello11}, but \citet{mcguire16} discuss the prospects for a measurement of this.. Finally, the identification of an aromatic COM in the ISM was reported by \citet{mcguire18} who detected benzonitrile toward the cold dense core TMC-1 with the GBT. Apart from the fullerenes, this molecule is, with 13 atoms, the largest molecule identified in the ISM so far.  As those authors' chemical models indicate, its presence is likely connected to the chemistry of benzene. However, it is less clear whether benzene itself should be formed predominantly through a ``bottom-up'' mechanism starting from simpler species, or whether it  represents the smaller end of a spectrum of PAHs that are relatively long-lived and photo-stable.

Attempts to detect larger molecules such as butyl cyanide (15 atoms) or the heterocyclic aromatic molecule quinoline (C$_9$H$_7$N, 17 atoms) have not been successful so far \citep[][]{ordu12,garrod17,cordiner17}. \citet{garrod17} predicted an abundance of butyl cyanide relative to propyl cyanide in Sgr B2(N) just below the detection limit of the EMoCA survey, but a preliminary search for this molecule in the three times more sensitive survey ReMoCA on the same source \citep[][]{belloche19} has not succeeded yet. Likewise, no secure detection of propanol (12 atoms) has been achieved so far \citep[][]{mueller16a,tercero15,qasim19}.

\subsection{Detections in the warm gas of hot cores and corinos}\label{s:hotcorinos}
Through efforts with ALMA, \emph{Herschel}, and other millimeter/submillimeter facilities, significant advances have been made toward the collection of systematic molecular inventories for a larger number of sources, including those that are generally less line-rich. For example, many of the complex species previously only detected toward high-mass star forming regions are now also seen toward solar-type protostars. These include molecules containing three carbon atoms, such as acetone and propanal \citep{lykke17}, and species of prebiotic interest such as glycolaldehyde \citep{jorgensen12,coutens15,taquet15} and formamide \citep{kahane13,coutens16}.

At the current level there is no evidence that the degree of molecular complexity is any less for low-mass hot corinos than their high-mass counterparts, although the former typically have fainter lines. This is not to say that there are no differences between low- and high-mass star forming regions. There are significant non-detections in some regions: for example, species such as methylamine, CH$_3$NH$_2$, and hydroxylamine, NH$_2$OH, have been sought due to their potential roles as precursors for amino acids. However, so far only CH$_3$NH$_2$ has been detected in the ISM and only toward a few select regions such as Sgr~B2(N) \citep{kaifu74}, Orion~KL \citep{pagani17}, G10.47+0.03 \citep{ohishi19}, and NGC~6334 \citep{bogelund19methylamine}, but not in general toward high-mass star forming regions or low-mass protostars \citep{ligterink15,ligterink18amines}. No secure detection of NH$_2$OH has been reported so far despite dedicated searches toward high- \citep{pulliam12} and low-mass star forming regions \citep{mcguire15,ligterink18amines}. An analysis of the PILS data  shows that the CH$_3$NH$_2$ upper limit, as well as the detection of methanimine (CH$_2$NH), a precursor species of CH$_3$NH$_2$, imply abundances with respect to methanol and formamide more than an order of magnitude lower in IRAS16293B compared to Sgr~B2(N) \citep{ligterink18amines}.

Conversely, the targeted studies of solar-type protostars have revealed the presence of some species not previously seen toward high-mass star forming regions -- for example, the simple halogen-bearing organic methyl chloride \citep{fayolle17}, and nitrous acid \citep{coutens19hono}. Due to high degrees of line confusion typically seen toward high-mass star forming regions, it is not clear whether those species differ in abundance compared to high-mass star forming regions, as the increased line confusion toward Sgr~B2(N2) in the EMoCA survey produces upper limits on the abundance of these two species comparable to the inferred values in IRAS~16293B. Another recent example of a ``first discovery'' in low-mass protostars is that of glycolonitrile. This species was detected in IRAS~16293B \citep[][]{zeng19} but searches for it toward Sgr~B2(N) have been unsuccessful so far, both in single-dish surveys \citep[][]{margules17} and in the EMoCA and ReMoCA interferometric surveys. 

\subsection{Low-density environments and starless/prestellar cores}\label{ss:prestellar_coms}
Some particularly interesting observations of complex molecules and related species involve detections toward cold starless or prestellar cores. While unsaturated carbon-chains have long been known to be present toward such cores (TMC-1 and L134N being the most well-studied examples), recent studies have also yielded detections of a number of saturated COMs toward dense, low-mass prestellar cores such as L1689B in the Ophiuchus molecular cloud region \citep{bacmann12}, L1544 in Taurus \citep{vastel14,jimenezserra16} and Barnard~5 in Perseus \citep{taquet17}, as well as their likely counterparts for high-mass stars such as the infrared dark cloud IRDC028.34+0.06 \citep{vasyunina14} and the prestellar core candidate W43-MM1/6 \citep{molet19}. 
Detected molecules include acetaldehyde, methyl formate, and dimethyl ether. Maps of the spatial distributions of these species toward some of the cores \citep{bizzocchi14,soma15} suggest that they are prevalent in regions offset from the center of the core as traced by the submillimeter continuum radiation or infrared extinction. 

The detections of COMs in absorption toward Sgr~B2(N) also provide insight into the chemistry of cold clouds. Some of these detections correspond to the large-scale envelope of Sgr~B2(N) itself, but a few COMs (CH$_3$OH, CH$_3$CN, CH$_3$CHO, NH$_2$CHO) were also detected at velocities corresponding to translucent clouds along the line of sight, both in the Scutum arm and in the Galactic center region \citep[][]{corby16,thiel17,thiel19}. The abundances of CH$_3$CN, CH$_3$CHO, and NH$_2$CHO relative to methanol were found by \citet{thiel17} to be similar to those derived for the same molecules detected in absorption in the $z = 0.89$ spiral galaxy located in front of the quasar PKS~1830-211 \citep[][]{muller11,muller13}. This suggests that the processes leading to chemical complexity in the translucent parts of molecular clouds have remained similar since the universe was half its current age.

The detections of COMs in these cold regions emphasize two points: firstly, the formation of these highly-saturated organics cannot be dependent on the same warm, diffusive, grain-surface chemistry that many models rely on to explain COMs in hot-core regions. Secondly, if they originate on the grains, they must be released into the gas through some non-thermal mechanism. Plausible theories include cosmic-ray sputtering of the ices, photo-desorption by UV photons, and so-called ``reactive desorption'', in which excited, newly-formed molecules spontaneously desorb. However, neither of the latter two mechanisms are found to be efficient for methanol in the laboratory \citep[e.g.,][]{bertin16,martin-domenech16}. The presence of COMs in translucent regions, in which photo-destruction of those molecules by ambient interstellar UV photons would be more significant, presents a greater problem. Formation of the COMs within transient or unresolved
structures \citep[e.g.,][]{garrod05} could provide a possible explanation. An alternative idea is that reactions of H and O atoms with the bare surfaces of carbonaceous grains could be a source of COMs. Experiments have for example shown that formaldehyde may be produced in this way \citep{potapov17}.

These new detections have reset the clock for COM formation to an even earlier stage than previously thought. An important task for the future will be to determine to what degree hot cores and corinos have inherited chemistry from their earlier evolutionary stages.

\subsection{Outflows}
\label{ss:outflows}
The first discovery of COMs toward an outflow-related shock was made toward the chemically active outflow L1157-B1 \citep{arce08}. This region has since then been the target of 
unbiased line surveys using the Nobeyama 45~m telescope  \citep[][]{yamaguchi12}, the IRAM 30~m telescope \citep[ASAI,][]{lefloch17}, and \textit{Herschel}/HIFI \citep[CHESS, e.g.,][]{codella10chess,codella12,benedettini12}. These surveys have led to a number of new detections in such shocked regions, including of phosphorhus mononitride \citep[PN;][]{yamaguchi11}, as well as inventories of COMs there \citep{sugimura11,lefloch17}.  Beyond the L1157 outflow, the systematic inventories of complex organics beyond CH$_3$OH and CH$_3$CN in outflow regions remain relatively sparse. \cite{oberg10,oberg11serpens} reported the detections of COMs, including CH$_3$OCHO, CH$_3$CHO, and CH$_3$OCH$_3$, toward outflows in the B1-b and SMM4-W regions with abundances of up to a few percent with respect to methanol. With their relatively rich spectra, the shocks in the L1157 outflow have also been the targets of a number of interferometric studies aimed at searching for chemical differentiation between the various species either between the different shocks or within individual ones \citep{fontani14,burkhardt16,codella17}. Such studies, as well as ones for high-mass protostellar outflows \citep[e.g.,][]{palau17}, show great potential for observations of chemistry in the \emph{time domain} that can provide much needed constraints, but naturally require good, independent constraints on the time progression of the shocks as well as the physical changes they induce  \citep[e.g.,][]{burkhardt19}.

\subsection{External galaxies}
COMs have also been detected in external galaxies. In Sect.~\ref{ss:prestellar_coms} we already mentioned the detections reported in the $z=0.89$ spiral galaxy in front of PKS~1830-211, with more than 42 simple and complex molecules identified with ATCA and ALMA \citep[][]{muller14}. Extragalactic detections of methanol, methyl cyanide, and methyl acetylene were already reported three decades ago, in particular in the central regions of the starburst galaxy NGC~253, with abundances similar to abundances in our Galaxy \citep[][]{henkel87,mauersberger91}. More recently, hot cores have been reported in the Large Magellanic Cloud (LMC). \citet{shimonishi16} identified a hot core toward the high-mass young stellar object ST11 with ALMA on the basis of its compact size, high density, and high temperature, but no COM was detected, with upper limits implying that methanol is at least one order of magnitude less abundant than in Galactic hot cores. Further progress was made by \citet{sewilo18} who detected not only methanol but also dimethyl ether and methyl formate toward two hot cores of N113, one of the most prominent star-forming regions in the LMC. The abundances of those species were found to be similar to those found in Galatic hot cores after accounting for the difference in metallicity. Methanol has been detected at even lower metallicity, in  a cold core of the Small Magellanic Cloud, with an abundance, not corrected for metallicity, comparable to similar galactic cold sources, which may suggest an enhanced production of methanol at low metallicity \citep[][]{shimonishi18} although the  implication is somewhat dependent on the assumptions about the dust properties and temperatures used to derive the H$_2$ column density (Sect.~\ref{sss:reportingproperties}).

A limited number of gas-grain chemical models of Magellanic Cloud chemistry have been published: \cite{acharyya18} present an example of a model of hot-core chemistry in these environments. They found that the effects of varied physical conditions, particularly the minimum dust temperature achieved, could be more influential than the metallicity. Unfortunately, the relatively few detections of COMs so far in the Magellanic Clouds do not place strong constraints on the chemical and physical processes at work.

\section{CHEMICAL DIFFERENTIATION IN THE ENVIRONMENTS OF STAR FORMATION}\label{s:environments}
The environment of a star forming region can have a strong impact on its chemistry: the temperature, cosmic-ray ionization rate, radiation field, and timescale of evolution all influence the rates at which molecules are created or destroyed, and possibly introduce chemical differentiation. In turn, the chemical content of a source can be used to infer its physical properties and history. Good examples of these links are the two favorite hot-core regions, the protocluster Sgr~B2(N) in the Galactic Center (GC) region and the 20-times more nearby Orion~KL region. Both have been targets of broad spectral line surveys over the past four decades to probe their COM content, but their physical characteristics may in fact be very different. Furthermore, other regions of low- and high-mass star formation are starting to show evidence of chemical differentiation through the increased sensitivity, spectral coverage, and angular resolution of current observational facilities (see Sect.~\ref{s:advances}). In this section we therefore focus on the observational relationships between the physical structure in star forming environments and chemical differentiation.

\subsection{Sgr~B2(N): the impact of cosmic rays and an extended reservoir of COMs}
\label{ss:sgrb2}

The GC giant molecular cloud Sgr~B2 shows a rich chemical diversity, related both to compact hot cores and to more extended material where complex molecules are also present. The fact that high-mass star formation is happening in Sgr B2 has been known for several decades, with the detection of several dozen ultracompact HII regions indicating the presence of a protocluster of young O and B stars \citep[e.g.,][]{gaume95}. Recently, a large population of several hundred compact, high-density sources in a likely earlier stage of evolution was unveiled with ALMA through their dust continuum emission, with a large fraction of them being interpreted as tracing high-mass young stellar objects \citep[][]{sanchezmonge17,ginsburg18}. In Sgr B2(N) in particular, three new hot cores have been identified through their molecular line forests with ALMA \citep[][]{bonfand17}; all coincide with previously known Class II methanol masers \citep[][]{caswell96}, which are exclusively associated with young high-mass stars \citep[][]{minier03,xu08}.

While the global chemical composition of Sgr~B2's main sites of star formation, Sgr~B2(N) and Sgr~B2(M), could be derived from single-dish observations \citep[e.g.,][]{belloche13,neill14}, significant progress has been made thanks to ALMA's high angular resolution and sensitivity that allowed the chemical composition of the individual hot cores of Sgr B2(N) to be determined and accurate column densities to be derived \citep[][]{belloche16,bonfand17}. Through models of the observed COM abundances of Sgr~B2(N)'s hot cores, \citet{bonfand19} found evidence for a cosmic-ray ionization rate a factor 50 higher than the canonical value of $1.3 \times 10^{-17}$~s$^{-1}$ usually assumed for dense gas in the galactic disk. This is in qualitative agreement with the higher cosmic-ray ionization rate values found for the diffuse medium of the GC region compared to the diffuse medium in the galactic disk \citep[][]{indriolo15,lepetit16}.

Another interesting discussion concerns the thermal history of the region and its impact on the formation of COMs. Dust temperatures measured with \textit{Herschel} are higher in the GC region compared to the galactic disk. Temperatures of 20--27~K were derived for Sgr B2 \citep[][]{guzman15} but they may be affected by the feedback of its star formation. A better proxy for the physical conditions that were prevalent during the earlier prestellar phase in Sgr~B2 may be G0.253+0.016, a GC molecular cloud with little on-going star formation. Dust temperatures of 19--27~K were measured in this cloud \citep[][]{longmore12}. \cite{bonfand19} showed that minimum temperatures above 20~K during the collapse history of Sgr~B2(N)'s hot cores would not be consistent with their current COM abundances. Either this discrepancy implies that the \textit{Herschel} maps are not sensitive to the lowest temperatures at high densities or that our understanding of the formation of COMs, and indeed ices of all kinds, in a region like Sgr~B2 would have to be revised. Specifically, the retention of CO ice on grains much above 20~K or so is unlikely, rendering even methanol production a challenge, as it has no efficient gas-phase formation mechanism \citep{garrod07}. This discussion may also have implications for more local (low-mass) star forming regions where the presence of nearby high-mass stars may cause the temperature floor to be elevated \citep[e.g.,][]{jorgensen06}.

A feature of Sgr~B2 that has been known for nearly two decades but has become more prominent in the past one is the presence of COMs distributed on large scales in the cloud, in addition to the COMs present on small (arcsec) scales in the hot cores. GBT detections of CH$_2$(OH)CHO, C$_2$H$_3$CHO, C$_2$H$_5$CHO, (CH$_2$OH)$_2$, c-H$_2$C$_3$O, NH$_2$CHO, CH$_3$C(O)NH$_2$, NCCHO, CH$_3$CHNH, and NHCHCN in absorption and/or emission with low excitation temperatures is an indirect indication that these molecules are present at low densities and distributed over arcminute scales in Sgr~B2 \citep[][]{hollis04a,hollis04b,hollis05,hollis06a,hollis06b,remijan08,loomis13,zaleski13}. Mapping observations confirm this with, for example, emission of CH$_3$OH, CH$_3$CN, CH$_3$CCH, NH$_2$CHO, CH$_3$CHO, and HC$_5$N extending over several arcminutes in maps obtained with the Mopra telescope at 3 and 7~mm \citep[][]{jones08,jones11} and absorption of CH$_2$CN, CH$_3$CHO, and CH$_3$CHNH detected with ATCA at 7~mm over the extent of the background free-free continuum emission \citep[$\sim$10$''$,][]{corby15}.

The connection between this low-density, colder emission and the hot cores remains unclear. Recently, \citet{li17} reported a change in the relative abundances of CH$_2$(OH)CHO and (CH$_2$OH)$_2$ over scales of 15$'$ in Sgr~B2 with the Shanghai Tianma 65\,m radio telescope with the [(CH$_2$OH)$_2$]/[CH$_2$(OH)CHO] abundance ratio decreasing from the extended, low-density region to the dense regions Sgr~B2(N) and Sgr~B2(M) where star formation occurs. The recent detections of COMs in cold cores (Sect.~\ref{ss:prestellar_coms}) suggest that a similar set of physical conditions could be producing such molecules in both environments, although -- if formed on dust grains -- the mechanism of desorption could be quite different. The action of shocks in galactic center environments could provide a more substantial ejection of mantles into the gas \citep[see, e.g.,][]{requena-torres06}. Recently, ALMA observations revealed a network of filaments on scales of a few tenths of a parsec toward Sgr~B2(N) \citep{schwoerer19}. The kinematics of the filaments show evidence for mass flows toward a central hub where the main hot core, Sgr~B2(N1), is located and imply a timescale of 60--300 kyr for the formation of this central hub. Further investigations of the relation between these filaments and both the formation of the high-mass stars and the chemistry in the extended and compact environments may shed more light on these issues.

\subsection{Orion~KL: chemical impact of a past explosion}
\label{ss:orion}
Thanks to its short distance, Orion~KL can be studied in greater detail than Sgr~B2 and many signs of chemical differentiation have been seen within this source. For example, resolved CARMA observations showed that cyanides peak at different locations from other species such as CH$_3$OCH$_3$ and CH$_3$OCHO \citep{friedel08}. A differentiation between N-bearing COMs (cyanides and NH$_2$CHO) and O-bearing COMs (alcohols, CH$_3$OCH$_3$, and CH$_3$OCHO) has also been inferred from differences in their excitation temperatures from \textit{Herschel} observations by \citet{crockett15}, who found that the former trace hotter gas (200--300~K) than the latter (100--150~K). However, the CH$_3$C(O)CH$_3$ emission detected with PdBI resembles more the cyanide emission than that of CH$_3$OCH$_3$ and CH$_3$OCHO \citep[][]{peng13,feng15}, two COMs that were found to be strongly spatially correlated \citep[][]{brouillet13}. To complicate the situation even further, \cite{favre17} demonstrated with ALMA that CH$_3$COOH and (CH$_2$OH)$_2$ trace a more compact region located closer to the hot core than other O-bearing COMs. Analyzing a larger sample of O-bearing COMs detected with ALMA, \citet{tercero18} confirmed this result and showed that there is in fact a more general spatial segregation between the COMs containing a C-O-C structure and the COMs containing a C-OH bond, the former tracing the compact ridge and the latter the hot core.  \cite{tercero18} interpret this segregation as resulting from the chemistry being dominated by different radicals in these two regions, methoxy (CH$_3$O$\cdot$) in the former and hydroxymethyl ($\cdot$CH$_2$OH) in the latter. One caveat in these discussions is that integrated intensity maps intended to trace a particular transition of a given COM may be contaminated by emission from other species, due to the complex velocity structure of the Orion~KL region \citep[][]{pagani17}. A more reliable method would be to construct column density maps from LTE fits of the spectrum of each pixel (e.g., using the VINE method; \citealt{calcutt18a}).

As already noted, some species show indications of temperatures across the region ranging from about 100~K up to 300--450~K toward the cyanide peak at the northeastern part of the hot core \citep[][]{bell14}. Among the COM detections recently reported in Orion~KL, the detection of propyl cyanide is particularly interesting. Both the straight chain (\textit{n}) and the branched (\textit{i}) isomers were detected with ALMA \citep[][]{pagani17}. The authors derived an abundance ratio of the \textit{n}- and \textit{i}-isomers that is similar to the one obtained by \citet{belloche14} in Sgr~B2, but varying by a factor of 3 (from 2 to 6) across the region. These variations may reveal a non-uniform thermal history across the Orion~KL region \citep[][]{pagani17}.

Still, significant evidence has emerged that mechanical rather than thermal processes may strongly influence the chemistry of Orion KL. It has long been known that the region shows a prominent wide-angle outflow with a peculiar structure of finger-like filaments in H$_2$ emission \citep[][]{taylor84} possibly the result of an explosive event \citep[][]{allen93}. A recent explanation for this explosion is that embedded radio and infrared sources within the cloud were originally part of a multiple young stellar system that violently disintegrated $\sim$500~yr ago as a result of a close dynamical interaction, creating at the same time the wide-angle outflow \citep[e.g.,][]{bally05,gomez05,zapata09,bally17,luhman17}. The absence of a self-luminous submillimeter, radio, or infrared source embedded in the hot molecular gas and the fact that there are no outflow filamentary structures in the shadow of the hot core pointing away from the explosion center led \citet{zapata11} to suggest that Orion KL is not a typical hot core with internal heating by a nascent star but a pre-existing dense structure that was heated up from the outside by the gas that was accelerated by the explosion. An alternative explanation \citep{goddi11,bell14} is that the hot core is externally heated due to interaction with the compact SiO outflow driven by source \emph{I} \citep{plambeck09}. In either case, the chemistry of the Orion~KL ``hot core'' could thus be dominated by shocks rather than thermal heating by embedded stars, which corroborates earlier suggestions that the hot core is externally heated \citep[e.g.,][]{blake96}. The context of the explosion may also shed new light on the structure of the Orion~KL region: \citet{pagani17} argue that the compact ridge, with its narrow velocity dispersion, is located in front of or behind the rest of the region and has not been affected by the explosion yet.

\begin{figure*}[t!]
 \resizebox{\textwidth}{!}{\includegraphics[scale = 0.75,trim=0 20 49 0,clip]{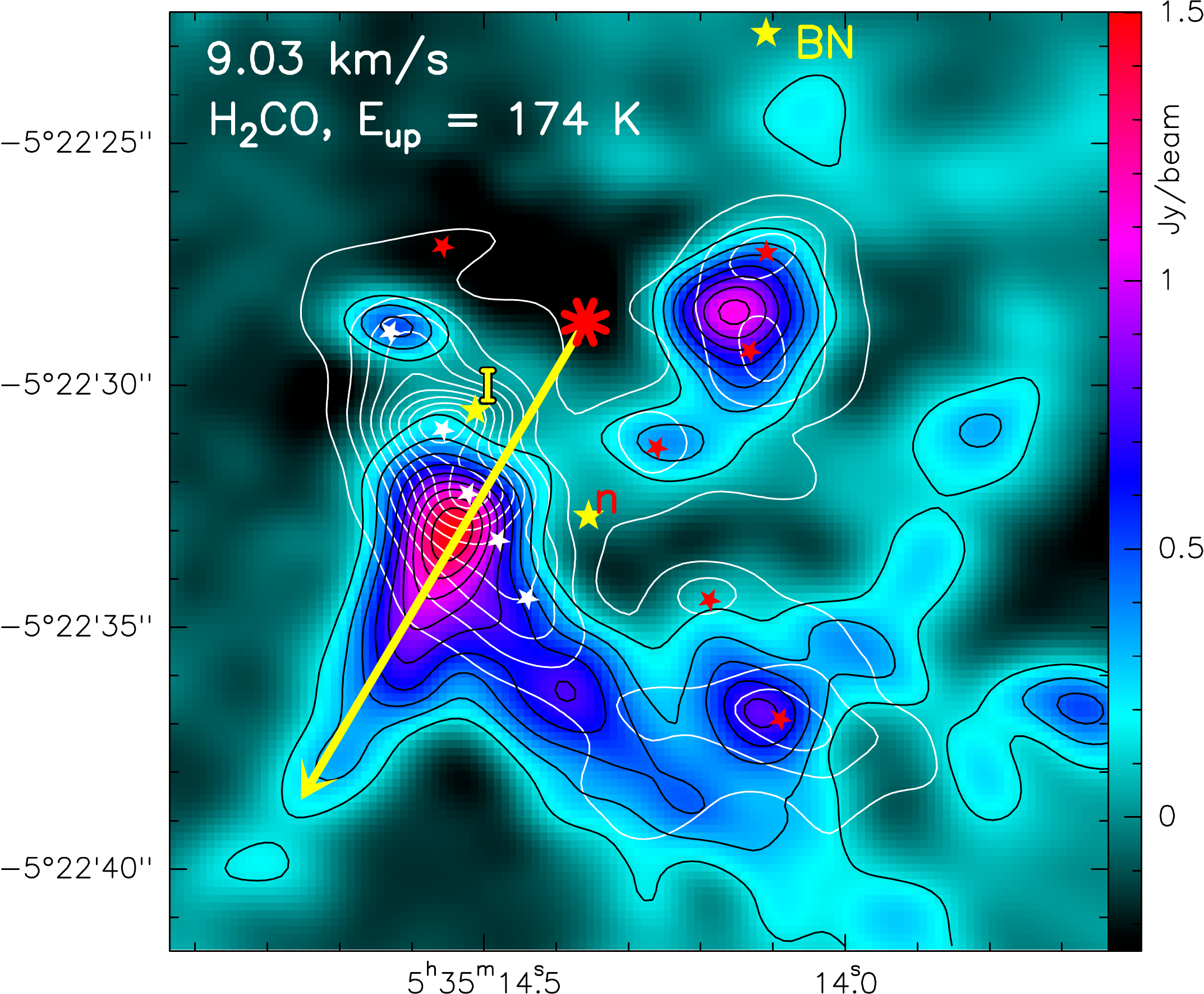}
 \includegraphics[scale = 0.753, trim=85 20 49 0,clip]{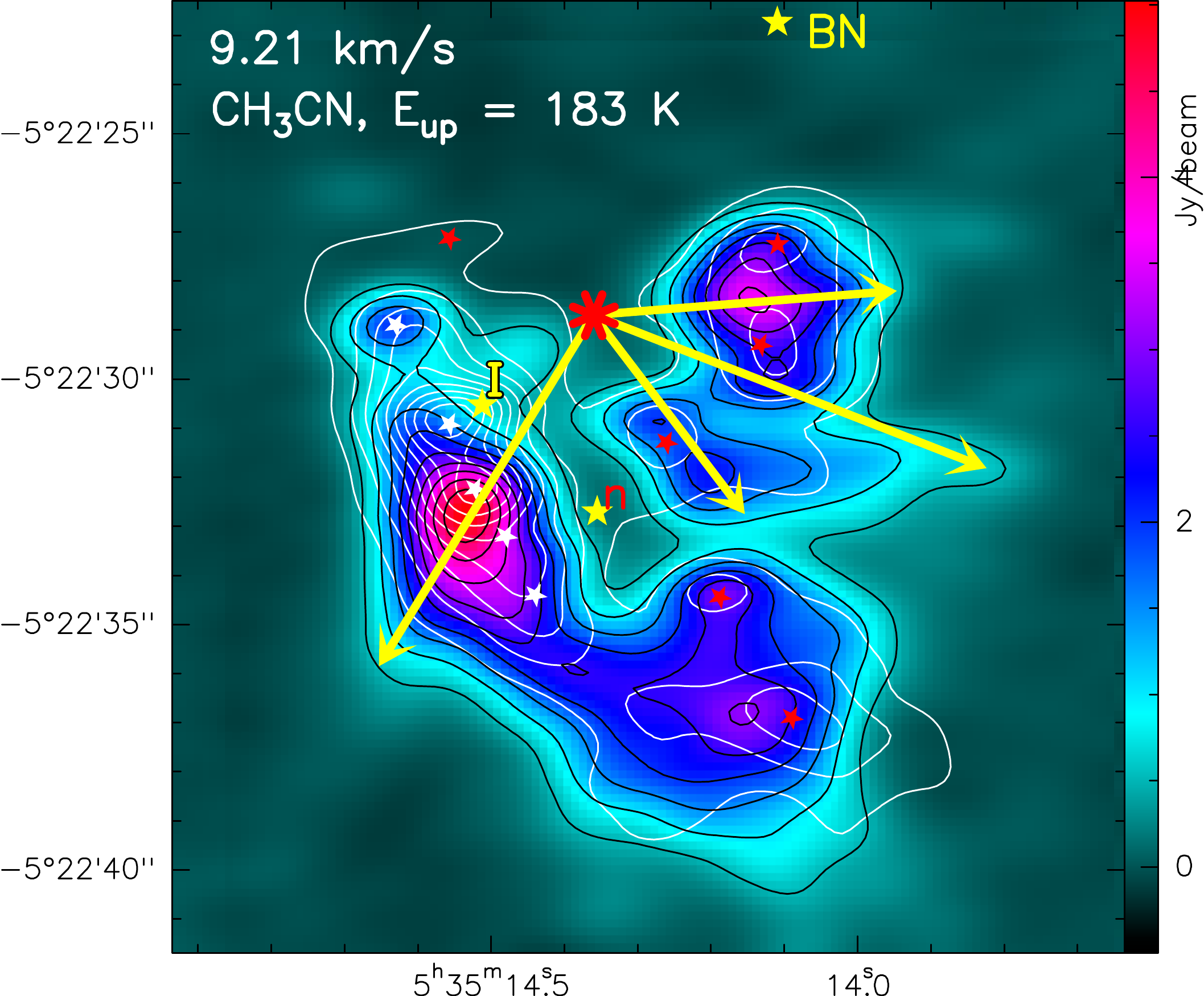}
 \includegraphics[scale = 0.753, trim=85 20 49 0,clip]{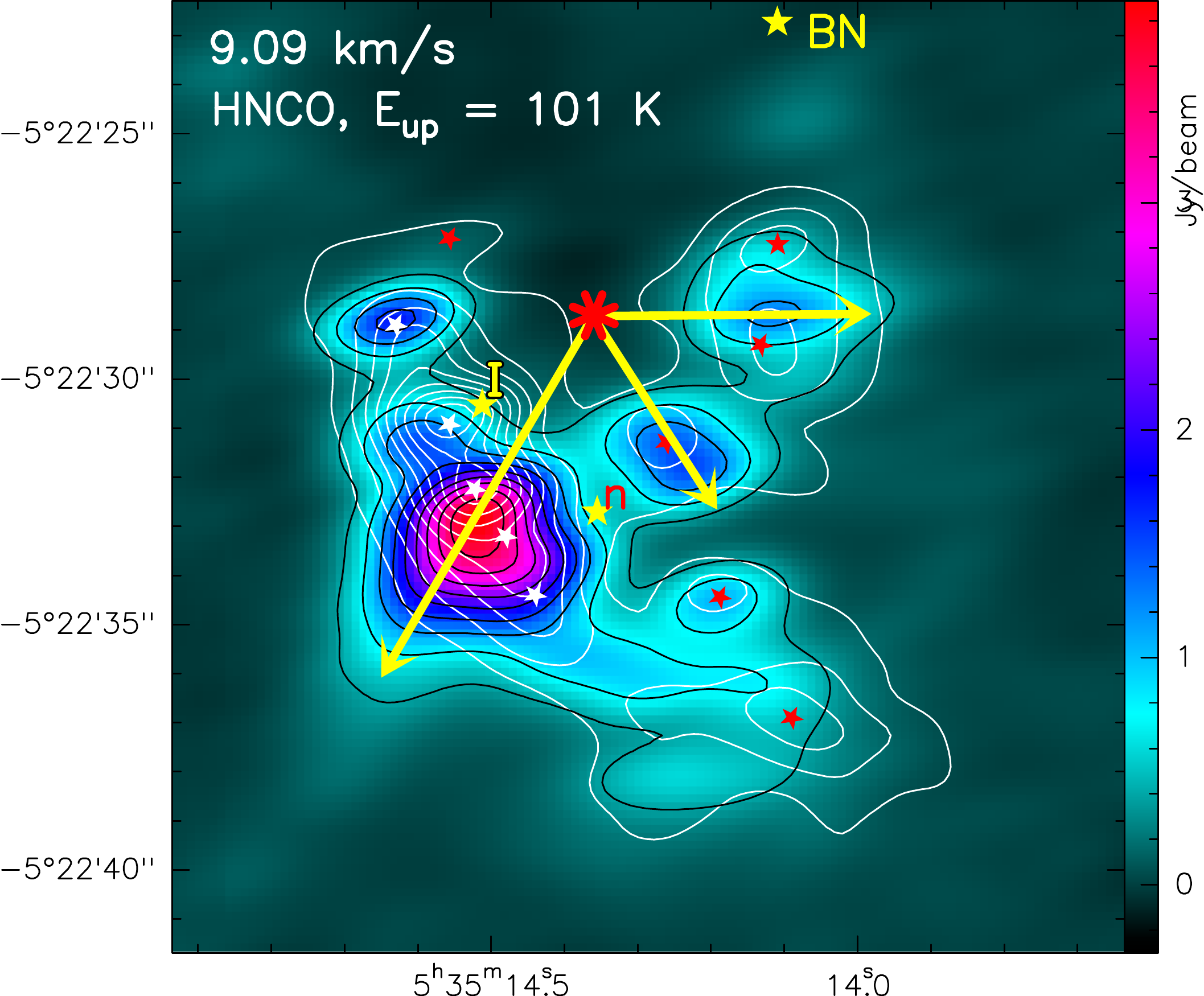}}
 \resizebox{\textwidth}{!}{\includegraphics[scale = 0.75,trim=0 0 49 0,clip]{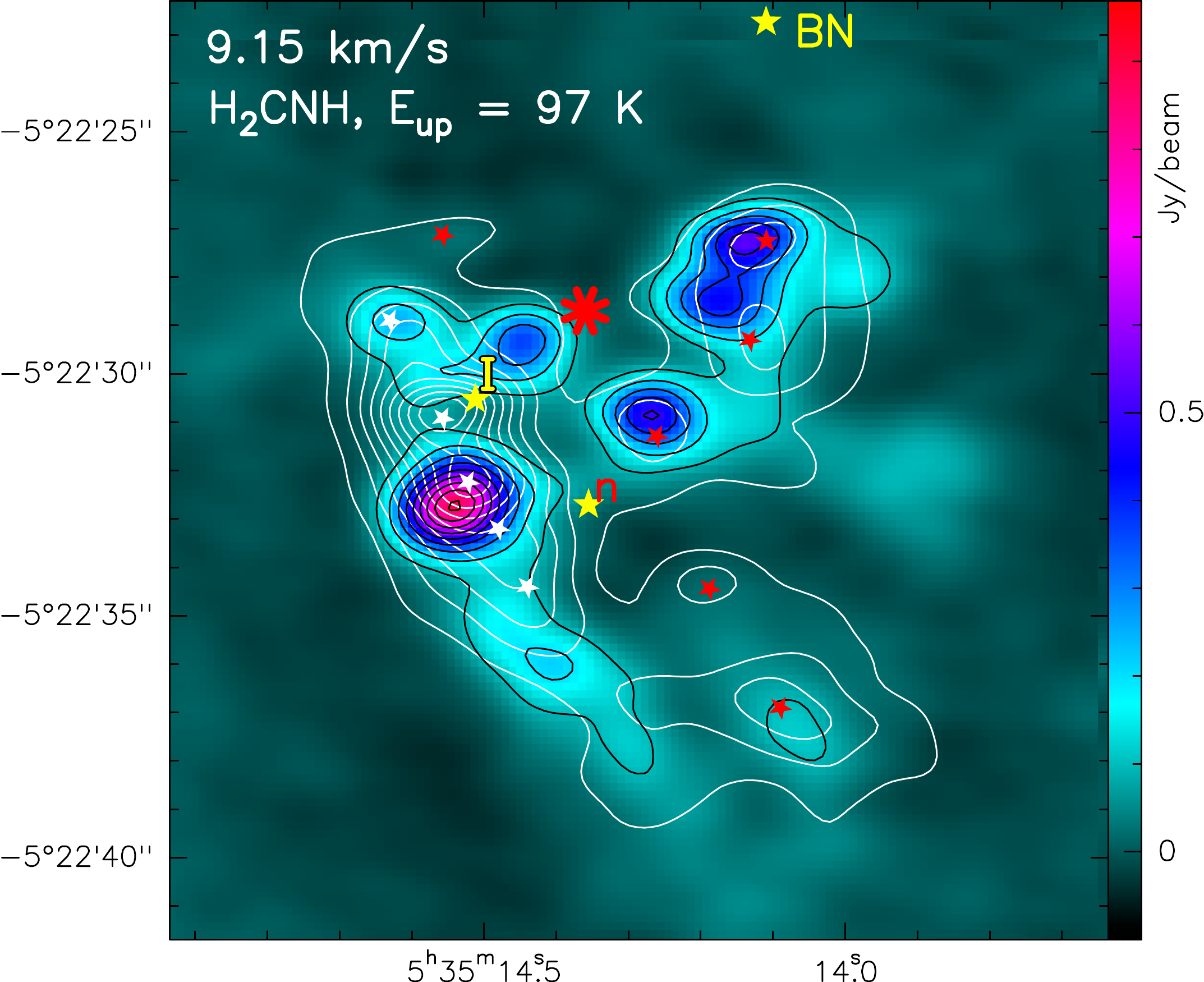}
 \includegraphics[scale = 0.743, trim=85 0 49 0,clip]{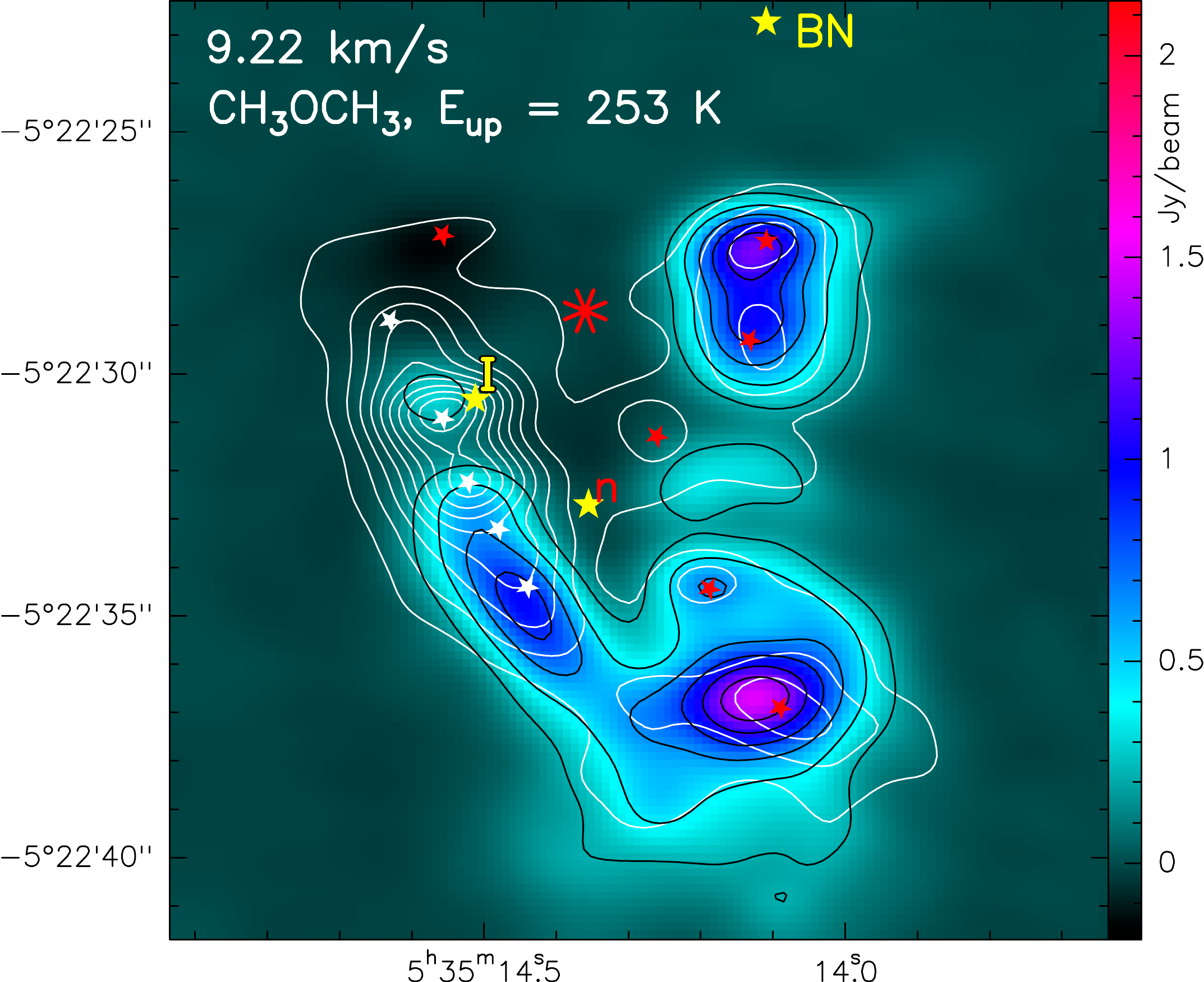}
 \includegraphics[scale = 0.743, trim=85 0 49 0,clip]{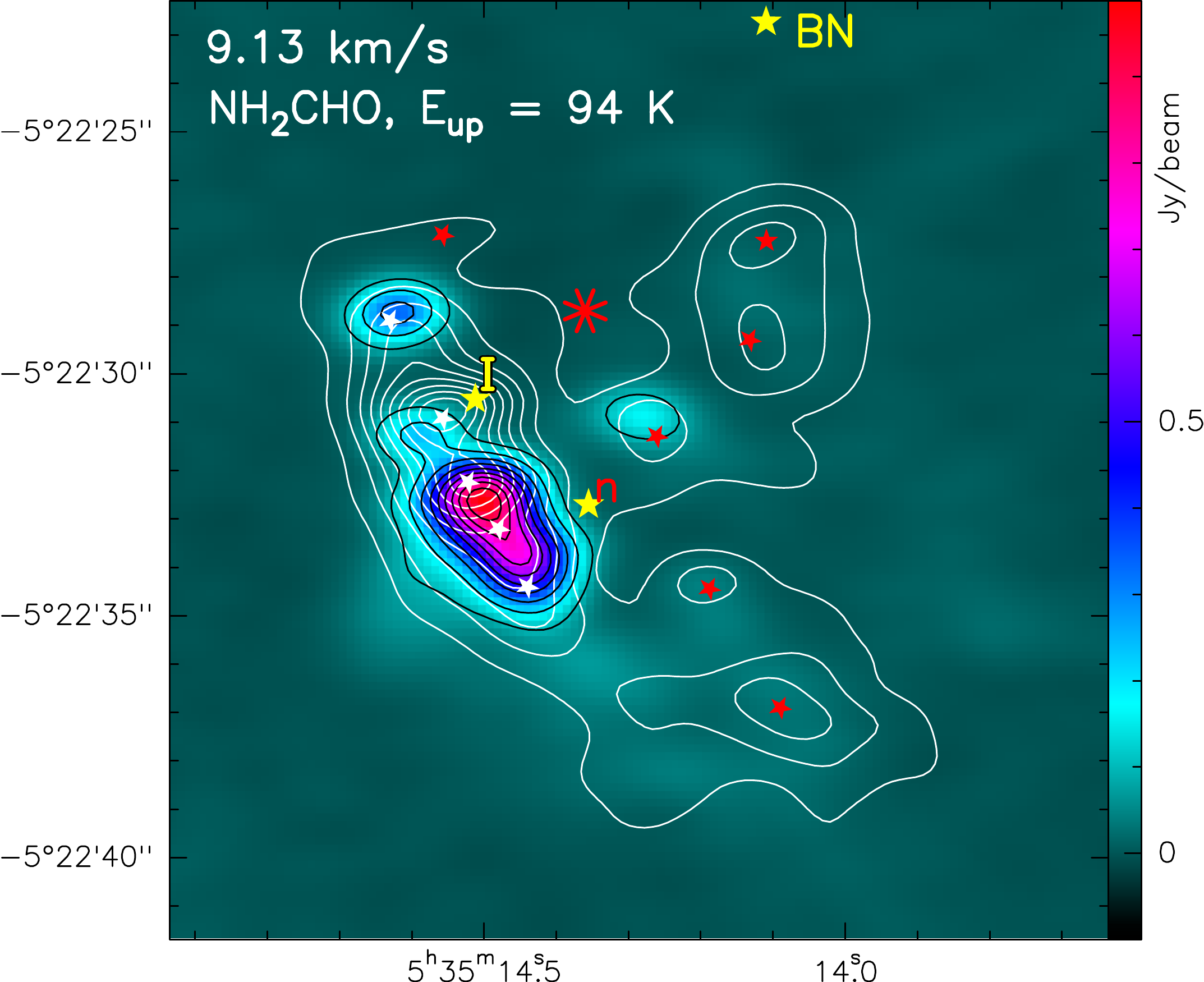}}
 \caption{A selection of species seen at 9 km\,s$^{-1}$ (color and black contours), compared to the 1.2~mm continuum emission (white contours) towards the Orion KL region. The stellar objects BN, I, and n are indicated. Small red or white 5-pointed stars mark infrared peaks. The yellow arrows starting from the explosion center (red eight-pointed star) suggest possible displacement of gas linked to the explosive event, which occurred $\sim$500 years ago. The first row shows species with conspicuous displacements (from left to right: H$_2$CO, CH$_3$CN, HNCO) while the second row shows species close to their production sites (from left to right: CH$_2$NH, CH$_3$OCH$_3$, NH$_2$CHO) (adapted from Pagani et al. 2019).}
\label{f:orion_alma}
\end{figure*}
A recent ALMA study by \citet{pagani19} presents an interesting case of probing chemistry in a new way in relation to the explosive event. In addition to kinematic evidence pointed out in their previous work \citep[][]{pagani17}, elongations found in the channel maps of various simple and complex molecules, which point back toward the explosion center, are interpreted by the authors as the result of
these molecules being dragged away by the expanding gas that was pushed out by the explosion (see Fig.~\ref{f:orion_alma}). The authors suggest that molecules that do not show such elongations are quickly destroyed in the gas phase after their release from the dust ice mantles, while those showing such elongations survive longer. If this interpretation holds true, then the Orion~KL region would turn out to be a kind of time-of-flight experiment that would allow us to explore interstellar chemistry in the time domain. This would be a major observational breakthrough. 

\subsection{Chemical differentiation in other hot cores}\label{ss:otherhotcores}
Another avenue to improve our understanding of the formation of COMs in high-mass star forming regions, complementary to the detailed studies in Sgr~B2 and Orion~KL mentioned above, is to target large samples of sources and compare their chemical compositions. \citet{bisschop07} found that the species (COMs or otherwise) that they detected with the JCMT toward seven high-mass YSOs were divided into two groups, one tracing hot gas ($>$100~K) and the other one cold gas ($<$100~K). Further studies of small ($<$10 sources) samples of high-mass star forming regions targetting several COMs have been performed \citep[e.g.,][]{calcutt14,suzuki18,taniguchi18}, but the lack of angular resolution, the inhomogeneity of the spectral coverage, or the small number of investigated COMs have prevented the emergence of a clear chemical pattern from these studies. \citet{hernandez-hernandez14} targeted 17 hot cores in CH$_3$CN with the SMA and found a correlation between the CH$_3$CN abundance and its rotational temperature, which they interpreted as evidence for a gas-phase formation of CH$_3$CN. \citet{oberg14} argued that the observed increase of [CH$_3$CN]/[CH$_3$OH] with temperature could make this ratio an evolutionary tracer of massive YSOs.

\begin{figure}[!htb]
    \centering
    \includegraphics[width=\textwidth]{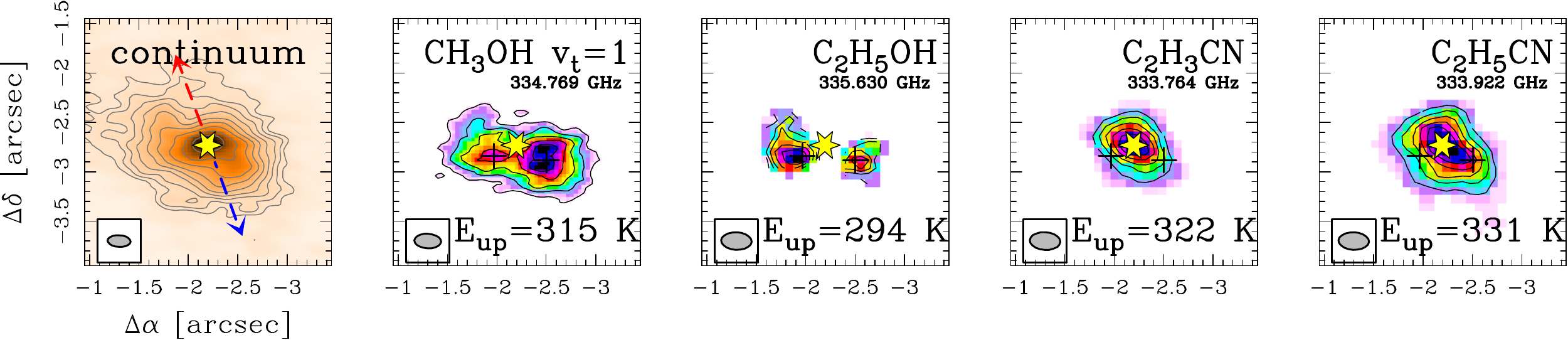}
    \caption{Continuum emission (left panel) and integrated intensity maps of transitions of several COMs (other panels) toward the young high-mass protostar G328.2551-0.5321 obtained with ALMA. Methanol and ethanol trace two spots (black crosses) offset from the protostar (yellow star) that have velocities consistent with rotation around the outflow axis (blue and red arrows). In contrast, the emission of vinyl and ethyl cyanide peaks on the protostar. Adapted from \citet{csengeri19}.}
    \label{f:g328}
\end{figure}

In addition to Orion~KL, chemical differentiation between N- and O-bearing COMs has previously been reported in W3, W75N, W51e1/e2, and G19.61-0.23 \citep[e.g.,][]{wyrowski99,qin15,remijan04,kalenskii10,qin10}.
More recently, ALMA observations revealed a chemical differentiation on small scales (a few 100 au) toward the young high-mass YSO G328.2551--0.5321, with the N-bearing COMs peaking toward the protostar and the O-bearing COMs toward two spots offset from the protostar \citep[][]{csengeri18,csengeri19}. The authors interpreted these two spots, which have velocities consistent with a rotation pattern, as tracing accretion shocks onto a disk (see Fig.~\ref{f:g328} and Sect.~\ref{ss:transition2disk}). Chemical differentiation on scales smaller than 1000~au was also reported by \citet{allen17} among the four continuum peaks detected with ALMA toward G35.20-0.74N, but a scenario to explain this differentiation is still lacking. 

Determining spatially-resolved abundance profiles of COMs in protostellar envelopes is a means to test the predictions of chemical models in order to identify their formation routes. For instance, the spatially-resolved analysis of the emission of CH$_3$OH, CH$_3$CN, and CH$_3$CCH toward the massive YSO NGC7538~IRS9 revealed a change in COM chemistry at temperatures above 25~K, likely reflecting the onset of an efficient ice chemistry above this temperature \citep[][]{oberg13}. Such spatially-resolved studies of COMs should be generalized toward samples of YSOs in the future. Another strategy is to study families of molecules, such as isomers, in targeted observations of large sample of sources. For instance in the case of the C$_2$H$_4$O$_2$ family, \citet{elabd19} found a bi-modal column density distribution of glycolaldehyde (GA) with respect to methyl formate (MF), while acetic acid correlates uniformly well with the latter. The sources with a high MF/GA ratio are exclusively high-mass objects while both high-mass and low-mass objects populate the region with low MF/GA ratios \citep{elabd19}, similar to comets and quiescent clouds \citep[][]{rivilla17}. The reason for this bi-modal distribution is not currently understood: a difference in the balance between UV-driven ice processing, versus other mechanisms such as cold, non-diffusive, surface chemistry, may provide one plausible explanation (see Sect.~\ref{sss:labexperiments}). Further observational and astrochemical modeling studies are required to investigate this discrepancy and the underlying physical mechanisms.

\subsection{Spatial differentiation in IRAS~16293-2422}\label{ss:iras16293_spatial}
Chemical differentiation can also be resolved in solar-type regions. Imaging observations of IRAS~16293-2422 indicate some spatial differentiation between N- and O-bearing complex organics \citep[see][]{bisschop08}. However, also within these groups of different complex organics, systematic differences can be seen toward one of the two components in the system, IRAS16293A: toward this source a number of species, including HNCO, H$_2$CO, CH$_2$(OH)CHO, (CH$_2$OH)$_2$, NH$_2$CHO, CH$_3$CHO, and C$_2$H$_5$OH show  compact emission (independent of excitation level) while others show more extended emission \citep{manigand19iras16293A}. This spatial differentiation seems to be related to the derived excitation temperature of the species ranging from above 90~K to 180~K with the ones with higher excitation temperatures showing more compact emission. Toward the other, lower luminosity, component of the system, IRAS16293B, no significant spatial differentiation is resolved at 0.5$''$ angular resolution. However, variations in excitation temperatures are also seen on smaller scales stretching over a somewhat larger range from about 100 to 300~K \citep{jorgensen16} with a few species (e.g., CH$_3$OH, CH$_3$OCHO) showing evidence for both an optically thick foreground layer with temperatures of about 100 K, with the bulk of the optically thin emission originating from molecules at the higher temperatures. A possible explanation for the spatial differentiation and variations in excitation temperatures may lie in the binding energies of the individual species in the ices causing this spatial stratification \citep{manigand19iras16293A} as also suggested in connection with gas-grain chemical modeling of hot cores \citep[e.g.,][]{garrod13}.

\section{LINKING THE PHYSICAL AND CHEMICAL EVOLUTION OF PROTOSTARS}
\label{s:physical_evolution}
Over the last decade significant work has been done on searching for and characterizing the properties of disks around embedded protostars in their Class~0 and I stages. In particular, with the advances in sensitivity and angular resolution by ALMA and other millimeter-wave interferometers, it is found that Keplerian disks exist around a large number of Class~I protostars \citep[e.g.,][]{harsono14,yen15sample,artur19sample} and at least some Class~0 protostars \citep{tobin12l1527,murillo13,lindberg14alma,yen17}, while counter-examples where no signs of Keplerian rotation are seen down to $\sim$10~au scales certainly also exist \citep[e.g.,][]{yen15b335,jacobsen19}. 

These high resolution observations have brought forward the close link between the physical structure and evolution of young stars and their disks and their astrochemistry. In particular, the youngest circumstellar disks serve as an interesting conduit between the COMs observed in the warm gas (Sect.~\ref{s:hotcorinos}) and the initial conditions for its later, protoplanetary, evolution. However, on the scales of the protostellar envelopes where the temperatures increase above 100~K due to heating by the central protostar and complex organics sublimate from ices, the physics due to the presence of the disks become highly non-trivial (see also Fig.~\ref{fig:cartoon}). Also, the degree by which the chemistry may be altered, or even completely reset, as material is accreted to the disk, remains unclear. In this section, we discuss recent results concerning the link between the physical and chemical evolution of embedded protostars with particular focus on the formation of disks and accretion.

\subsection{Chemical changes in the transitions from envelopes to disks}
\label{ss:transition2disk}
From a theoretical physical point of view the formation of a circumstellar disk is thought to be associated with the presence of accretion shocks where material enters the newly formed disk \citep[e.g.,][]{cassen81,neufeld94}. Besides their potential for resetting the chemistry, such shocks may also affect the cosmic-ray ionization rate both on small scales around individual protostars and on larger scales in more clustered environments \citep{gaches18}.

Based on \emph{Spitzer Space Telescope} observations of water lines toward the Class~0 protostar NGC~1333-IRAS4B, \cite{watson07} suggested that the water vapor had its origin in the warm and dense gas possibly related to an accretion shock due to infall from the protostellar envelope onto a very young disk. However, images of the infrared lines from \herschel\ and \spitzer\ show that the  water line emission is in fact related to the outflow to the south \citep{herczeg12}. This is in contrast to narrow, i.e., non-shocked, line emission of the H$_2^{18}$O isotopolog that peaks toward the embedded protostar itself \citep{iras4b_h2o}.

Using ALMA, \cite{sakai14} found kinematical evidence of a distinct component, the so-called ``centrifugal barrier'', in a region of about 100~au separated from  the Class~0 protostar L1527 and surrounding its Keplerian disk. \cite{sakai14} showed that SO and other sulfur-containing species were enhanced at this location and argued that this was a result of sublimation of grain mantles due to a weak accretion shock. Subsequent studies of other protostars show similar kinematical features, although in different molecular tracers  \citep[e.g.,][]{sakai16,oya16,oya17}. In an ALMA survey of a sample of protostars from the Ophiuchus star-forming region, \cite{artur18,artur19sample} detected warm ($E_{\rm up}\approx 200$~K) SO$_2$ lines toward five of the more luminous sources. The lines were found to be wide and, in some cases, showed velocity gradients perpendicular to the outflow direction, a possible indication that the emission is related to shocks toward the surfaces of the emerging disks -- although the compact nature of the emission made it impossible to rule out, for example, an outflow/jet origin.

Indications of accretion shocks at the envelope-disk interface have also been suggested for high-mass protostars. In high angular resolution ALMA observations of a massive protostar associated with the massive clump G328.2551--0.5321, \cite{csengeri18} found evidence for CH$_3$OH emission spots spatially offset from the central unresolved continuum emission associated with the protostar (see Sect.~\ref{ss:otherhotcores} and Fig.~\ref{f:g328}). Based on the distinct kinematics of the CH$_3$OH emission compared to the outflow traced by lines of SiO and SO$_2$, \citeauthor{csengeri18} argued that the CH$_3$OH peaks are associated with accretion shocks at the centrifugal barrier  while a transition of HC$_3$N in its first vibrationally excited state shows more compact emission, possibly tracing the properties of the accretion disks around more massive protostars.

However, significant work still needs to be done to establish an internally consistent picture of the different kinematical and chemical components of protostars as illustrated by different interpretations of similar observations of individual sources. For example, \cite{oya16} argued that CH$_3$OH and other COMs were present at the centrifugal barrier toward IRAS16293A as a result of either a weak accretion shock or protostellar heating, in contrast to H$_2$CS that also traces the disk on smaller scales. Van~'t~Hoff~et al.~\citeyear{vanthoff19} on the other hand found that multiple spatially-resolved H$_2$CS transitions were consistent with the temperature in a passively heated envelope down to small scales overlapping with the emission from some of the complex organics. 

Another example is offered by the Class 0 protostar, L483, for which  unsaturated carbon-chain species are present on large, single-dish, scales \citep[e.g.,][]{sakai09,agundez08,agundez19}, while saturated COMs are present on small, interferometric, scales \citep{oya17,jacobsen19}. Based on the kinematical model of CS emission, \cite{oya17} argued that the centrifugal barrier in this source is at radii of 30--200~au with COM and SO emission possibly tracing an unresolved Keplerian disk within this. However, \cite{jacobsen19} demonstrated that the emission from CS and the COMs both show similar kinematical signatures consistent with material infalling under the conservation of angular momentum down to radii of 10--15~au rather than Keplerian rotation.

\subsection{The physical/chemical structure of embedded disks}
For protostars with established Keplerian disks, other interesting questions arise concerning their physical structures: does the distribution of molecules in the gas and ices in these disks reflect those around more evolved (T~Tauri) stars where significant freeze-out appears in the mid-plane, or does the active accretion change the physical structures, for example by increasing the disk temperatures that are critical for the resulting chemistry? Through radiative transfer modeling of gas-phase lines observed with the SMA toward two Class~I protostars with well-established disks, \cite{brinch13} found the former to be the case: relatively simple models for T~Tauri disks could be applied to those sources with simple chemistry dictated by the amount of shielding, in turn determining in what fraction of the disk  CO is frozen out. A counter-example is offered by the disk around the more deeply embedded protostar L1527: by comparing ALMA data to detailed radiative transfer models, \cite{vanthoff18l1527} showed that even in the mid-plane of this disk, the temperature is above 20~K out to radii of at least 75~au, thereby causing CO to be in the gas phase in the bulk of the disk.

The exact distribution and location of complex organics around Class~I protostellar sources also remains puzzling. Toward the protostars in the ALMA Ophiuchus survey by \cite{artur19sample}, no significant emission from lines in the prominent CH$_3$OH $7_k$--$6_k$ branch was seen toward the locations of the protostars. These results demonstrate that the warm regions of the disk where gas-phase CH$_3$OH can be present (or where weak shocks could act) must be relatively limited in extent \emph{and} that the envelopes surrounding these disks must have relatively low column densities of material in their inner warm regions. In the standard picture of rotating collapse of a protostellar core and disk formation, the latter would be a natural consequence with the envelope density profile flattening out on small scales where the disk emerges. A simple prediction would consequently be that the presence of complex organics in the envelopes on small scales and the presence of extended Keplerian disks should be anti-correlated.

Recently \cite{lee17coms,lee19hh212} presented an in-depth analysis of the HH~212 protostellar system in the L1630 cloud of Orion. The system shows a characteristic morphology with an edge-on dust lane with emission from COMs distributed parallel to the dust lane -- but offset above and below it. The complex organics show signs of a velocity gradient in the elongated direction, indicative of rotating motions (the region of the disk where Keplerian motions would be present is only marginally resolved). These features led \cite{lee17coms} to suggest that these species are present in the ``atmosphere'' of the disk and either formed in the cold regions at large scales and sublimated at high temperatures in the disk atmospheres -- or alternatively formed through rapid gas-phase reactions there. 

Another example of differentiation between dust continuum and line emission is seen toward the Class~I protostar TMC1A. Through ALMA long-baseline observations of that source,  \cite{harsono18} demonstrated that the emission of CO isotopologs ($^{13}$CO and C$^{18}$O) was clearly offset from the central location of the protostar and resolved dust continuum emission. Through detailed line radiative transfer modeling, \citeauthor{harsono18} show that this differentiation in TMC1A  best can be explained by the existence of large (millimeter-sized) grains causing the dust to become optically thick and thus blocking the emission of the gas lines. Due to the high column densities traced by ALMA in its long baseline configurations, some care must be taken when interpreting the spatial differentiation between species and dust emission on these scales due to the high continuum optical depth even in the absence of large grains such as those inferred by \citeauthor{harsono18} (see Sect.~\ref{sss:spatial_resolution}).

\subsection{Episodic accretion and snowlines in protostellar envelopes}
\label{ss:episodic_accretion}
The emergence of disks around embedded protostars may also have profound implications on the overall physical evolution of the infalling material from the protostellar envelopes. If disks around the embedded protostars are comparable in mass to those around more evolved T~Tauri stars as is suggested by dust continuum observations \citep[e.g.,][]{tychoniec18,jorgensen09}, these disks may easily undergo periods of instability causing the accretion onto the central stars, and consequently their luminosities, to be non-steady (episodic) -- thereby also causing the temperatures in the ambient disk and envelope heated by the central protostar to vary. Observationally, there is  photometric evidence, both statistically (see \citealt{audardppvi},  \citealt{dunhamppvi} and references therein) as well as for individual protostars \citep[e.g.,][]{safron15}, that such luminosity variations indeed take place.

\begin{figure}[ht]
\includegraphics[width=\textwidth]{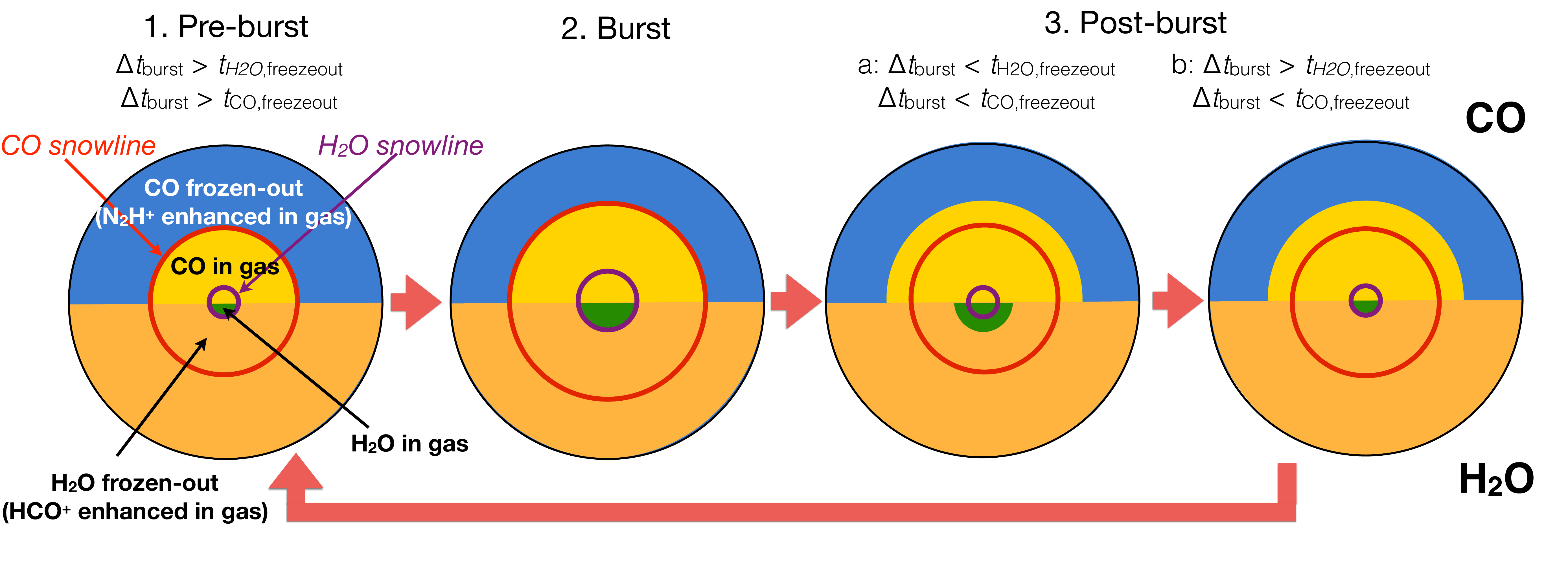}
\caption{Changes of the chemical signatures of an embedded protostellar envelope before, during, and after a burst in accretion or temporal increase in luminosity. Panel 1: before the burst, molecules are frozen-out on grains beyond the snowlines of the individual species (CO and H$_2$O illustrated at the top and bottom, respectively, as examples) and in the gas phase within these. Species destroyed through gas-phase chemical reactions with those species are enhanced in the regions where they are frozen-out (e.g., N$_2$H$^+$ and HCO$^+$ enhanced where CO and H$_2$O, respectively, are frozen-out). Panel 2: During the burst, the luminosity of the protostar, and thus the temperatures in the envelope, increase. As a consequence the snowlines are shifted to larger radii where the species are sublimated instantaneously. Panel 3: In the quiescent phase after the burst, the snowlines shift inwards again, but due to the density-dependence of the freeze-out process, molecules remain in the gas phase for a longer period of time. The timescale for each species will depend on the density where its snowline is located. Eventually the freeze-out will bring the chemistry down to its original state, again assuming that the time interval between bursts, $\Delta t_{\rm burst}$, is longer than the molecular freeze-out time-scales. Figure based on \cite{jorgensen13,jorgensen15}.}
\label{f:episodicAccretion}
\end{figure}

From a chemical point of view, temporal changes in the temperatures are naturally very important. For example, an (episodic) increase in luminosity for a given protostar will cause the regions where specific molecules  sublimate from the surfaces of grains  to become more extended compared to sources in a more quiescent phase of steady accretion \citep[e.g.,][and Fig.~\ref{f:episodicAccretion}]{lee07,visser12,visser15,rab17}. As the time-scale for the subsequent freeze-out may be long (inversely proportional to the density) compared to the duration of such bursts, those extended regions of sublimated molecules may persist for significant periods of time even after the protostar has returned to its quiescent phase. Thus, observations of the molecular signatures of embedded protostars may provide an ``archeological'' view of their recent evolution. Episodic accretion events may also be reflected in the observed ice and dust features of the protostars. For example, the existence of pure CO$_2$ ices observed in the mid-infrared spectra of protostars is an indication that some thermal processing and subsequent freeze-out has taken place at some point during the preceding evolution \citep[e.g.,][]{ehrenfreund97,gerakines99}, possibly due to previous accretion bursts \citep[e.g.,][]{kim12,poteet13}. \cite{taquet16} demonstrated that the cyclic sublimation and freeze-out of the icy mantles, could lead to the build-up of appreciable amounts of complex organic species in the inner envelopes of protostars.

Observationally, these suggestions can be tested by measuring the extents of primary species directly sublimating from grains or secondary species either produced or destroyed subsequently in the gas phase (Fig.~\ref{f:episodicAccretion}). Studies of the images of the C$^{18}$O emission toward protostars observed with the SMA \citep{jorgensen15,frimann17} show evidence for emission extending beyond the predictions based on the current luminosities for 20--50\% of the sources. Given that the time-scale for C$^{18}$O to freeze out   is of order 10$^4$~yr at the densities where it is found in the gas phase for these protostars, the statistics suggest that each protostar typically undergoes an accretion burst once every 2--5$\times 10^4$~yrs, resulting in an increase in its luminosity by 1--2 orders of magnitude. A comparison to numerical predictions of large-scale MHD simulations demonstrates that these results are robust even when the possible complex geometry of the sources is taken into account \citep{frimann16b}. An alternative to targeting the CO isotopologs is to look for the absence of N$_2$H$^+$ emission toward the center of the protostar where CO sublimates. Spatially resolved images of the emission from N$_2$H$^+$ and CO show these species to be anti-correlated. For example, \cite{anderl16} presented observations and modeling of N$_2$H$^+$ toward four "non-bursting" protostars from the sample studied in C$^{18}$O by \cite{jorgensen15} and show that the N$_2$H$^+$ structure indeed was consistent with that seen in C$^{18}$O, as well as the current luminosities of the sources. \cite{hsieh18} used ALMA to observe seven very low luminosity protostars (``VeLLOs'') in C$^{18}$O and N$_2$H$^+$ and found that for five of those sources, the C$^{18}$O emission and related N$_2$H$^+$ suppression implied recent accretion bursts with a time interval between bursts of 1.2--1.4$\times 10^4$~yr.

Complementary constraints to those from CO and N$_2$H$^+$ observations can be obtained by observing species that sublimate at different radii due to their different binding energies in the ices. Because of the decrease in density with increasing radii in the protostellar envelopes, this would cause the freeze-out time-scale for such species to differ from those of CO, thus providing complementary information concerning the frequencies and magnitudes of the bursts. A particularly interesting example is that of H$_2$O, which sublimates at much smaller scales of the protostellar envelope where the temperature exceeds 90--100~K, compared to the 20--30~K of CO, and consequently freezes out much more quickly again. Water is difficult to image using millimeter wavelength interferometers due to  Earth's atmosphere, but species such as HCO$^+$ is suppressed by the presence of H$_2$O, in a similar way as N$_2$H$^+$ is by CO, and can thus be used as indirect tracers \citep[][and Fig.~\ref{f:episodicAccretion}]{visser15}. Indeed, images of the H$_2^{18}$O and H$^{13}$CO$^+$ isotopologs of these species toward the Class~0 protostars NGC~1333-IRAS2A \citep{vanthoff18snowline} and IRAS~15398-3359 \citep{jorgensen13,bjerkeli16water} show similar anti-correlations to those of CO and N$_2$H$^+$, but on much smaller scales. The extent of this HCO$^+$ suppression toward IRAS~15398-3359 \citep{jorgensen13} implies that the protostar has undergone a recent burst but only within the previous 100--1000~yr. High angular resolution images of the CO outflow from this source show the presence of bullet-like structures also supporting the interpretation of ejection events on $\sim$100~yr time-scales \citep{bjerkeli16outflow}. Another  example is the FU Orionis object V883 Ori for which an increased luminosity moved the water snowline outwards and led to large column densities of methanol and other COMs over extended regions where they are observable with ALMA \citep{vanthoff18v883ori,lee19v883ori}.

\cite{hsieh19} performed a statistical survey of N$_2$H$^+$ and HCO$^+$ using ALMA toward 39 Class~0 and I protostars from the Perseus molecular cloud. Almost all sources in their sample show evidence for post-burst signatures in N$_2$H$^+$ while the number of sources with post-burst signatures in HCO$^+$ decreases from the Class~0 to I stages. These differences suggest that the time intervals between bursts increase from 2400~yr in the Class~0 stage to 8000~yr in the Class~I stage. An explanation for this evolution could be that the accretion bursts are results of instabilities in disks that are more likely to occur in the earliest protostellar stages.

\subsection{Summary}
The above discussions raise a number of outstanding questions concerning the physical and chemical structure of embedded disks and their formation and evolution. Are the pictures of accretion shocks and episodic accretion generally applicable? To what degree do these cause the chemistry to be reset or lead to differing chemistries with new molecules formed? How are COMs distributed around embedded protostars -- and do they point to one physical origin, or perhaps several different ones connected to those shown in Fig.~\ref{fig:cartoon}?  What role does grain growth play in the chemical evolution as well as the interpretation of line emission and chemistry on small scales? These discussions are still suffering from a lack of unbiased datasets that are comparable in terms of the observed molecular line tracers. The obvious next step will therefore be to expand the observed number of sources and species systematically -- and establish generally applicable models that can account for the differences between low- and high-mass protostars and, importantly, make predictive statements. 


\section{FRACTIONATION}\label{s:fractionation}
One of the important tools to study the physics and chemistry of the ISM is to target less abundant isotopologs of specific species. This may be a useful way of finding transitions that, while optically thick for the main isotopolog, \emph{may} be optically thin for the rarer isotopolog due to its lower column density. However, from an astrochemical point of view, it is also well established that for many molecular species in the ISM, the isotope ratios (e.g., D/H, $^{12}$C/$^{13}$C, or $^{14}$N/$^{15}$N) measured as column density ratios between their isotopologs, differ significantly from the elemental abundances representative for the local conditions in their ambient interstellar medium. These variations, referred to as isotopic fractionation, can either be enhancements or depletions of the rarer isotopologs compared to the local ISM isotopic ratio and range numerically from a factor of few differences for the heavier elements, such as carbon and nitrogen, to enhancements of several orders of magnitude for deuterium. In the Solar System, significant variations are also seen in the isotopic compositions in different bodies and materials. As these variations are thought to reflect processes that took place during the earliest stages of the Solar System, a detailed understanding of the different physical and chemical processes behind isotopic fractionation may shed light on its origin and relation to star and planet formation as we observe it now (see sidebar title Isotopolog and Isotopomer).
\begin{textbox}[!htb]
\section{Isotopolog and Isotopomer}
{\emph{Isotopologs} refer to molecules that only differ in their isotopic composition. Isotopologs can have one or more atoms substituted by isotopes, occasionally referred to as singly-, doubly-, etc. substituted variants.\\[1.0ex] \emph{Isotopomers}, or isotopic isomers, are species that consist of the same number of each isotopic atom, but the positions of those atoms within the molecule differ.\\[1.0ex]As examples, CH$_2$DOH, CH$_3$OD, $^{13}$CH$_3$OH, and CH$_3^{18}$OH are  singly-substituted isotopologs of methanol, CH$_3$OH. Of these, CH$_2$DOH and CH$_3$OD are isotopomers where the D-isotope is substituted on either the CH$_3$ or OH functional group.}
\end{textbox}

A range of processes can lead to isotopic fractionation in the ISM. For common species such as CO and N$_2$, the high abundance of the main isotopologs, and thus opacity at the frequencies of their transitions, causes these species to be self-shielded against photodissociation, while their rarer isotopologs may be photodissociated over larger columns. As a result of this \emph{isotope-selective photodissociation} (see, e.g., \cite{visser09photodissociation} for the case of CO and \cite{heays14} for N$_2$), the molecular carrier becomes depleted in the rarer isotopic variant as the depth or extinction increases ($^{12}$CO becomes more abundant than $^{13}$CO compared to the $^{12}$C/$^{13}$C ratio, $^{14}$N$_2$ more abundant than $^{15}$N$^{14}$N compared to the $^{14}$N/$^{15}$N ratio etc.) while the atomic gas becomes enriched in the rarer isotope ($^{13}$C, $^{15}$N etc.). Depending on the exact chemical networks both the molecular and atomic isotopic signatures may be carried over to other gaseous species and possibly solids \citep[e.g.,][]{lyons05,furuya18}. The efficiency of these processes  naturally depends on the strength and spectral shape of the incoming radiation as well as physical conditions in the environment, e.g., the distribution of material and shielding by grains.

Deeper in the cloud, \emph{chemical fractionation} may play a more prominent role. A classical example is the balance between H and D through the isotope exchange reaction:
\begin{equation}
{\rm H}_3^+ + {\rm HD} \quad\leftrightarrow\quad {\rm H}_2{\rm D}^+ + {\rm
  H}_2 \label{e:DH_frac}.
\end{equation}
At the low temperatures of prestellar cores, the exothermicity of this reaction causes the deuterium of HD to be driven toward H$_2$D$^+$, which thus becomes increasingly abundant relative to H$_3^+$. As H$_3^+$ is one of the key ingredients for many gas-phase chemical networks in the ISM, its enhanced deuteration can then be carried to a number of other common species, such as HCO$^+$ and N$_2$H$^+$ enhancing the abundance of their deuterated isotopologs. The above reaction also has an important role in determining the D/H ratios for species formed on grain surfaces: the main source of atomic H or D in star forming regions is through dissociative recombination of H$_3^+$ or H$_2$D$^+$ with electrons. Thus, the over-abundance of H$_2$D$^+$ leads to an enhanced atomic D/H ratio in the gas at low temperatures, which can be transferred to molecules formed on grains through hydrogenation of neutral species \citep{tielens83}. The effectiveness of these deuteration processes depends on a number of factors, including the temperature, the freeze-out of species such as CO, that acts as the main destroyer of H$_3^+$ and H$_2$D$^+$, and the ortho-to-para ratios of the species in Eq.~\ref{e:DH_frac} \citep[e.g.,][]{sipilae15benchmark}. 

In this section, we discuss the origin of isotopic fractionation as it occurs in connection to the process of star formation. The intent is not to provide a comprehensive review of all observational and modeling efforts. Rather we highlight a few key points with particular emphasis on how measurements of the relative abundances of different isotopologs can be used to trace the link between complex species and water in star forming environments and the Solar System.

\subsection{Water}
Perhaps, the most well-known discussion of fractionation in regions of
star formation and the link to the Solar System concerns the origin of
Earth's water and its deuterium isotopic composition compared to
Solar System bodies and star forming regions. It has long been known that the D/H ratio in Earth's oceans of $1.5\times 10^{-4}$ \citep[e.g.,][]{robert00} is enhanced compared to the cosmic D/H ratio of $1.5-2.0\times 10^{-5}$ \citep[e.g.,][]{linsky03,prodanovic10}. Furthermore, a number of other Solar System bodies including comets \citep[e.g.,][]{lis13} and meteorites \citep[e.g.,][]{alexander12} are also known to show enhanced D/H ratios. While a lot of emphasis has been put on understanding the delivery of water to Earth through comparisons of the D/H ratios between these bodies, e.g., whether water was brought to Earth through cometary or meteoritic impacts, another relevant question from a star-formation astrochemical point of view is when water was formed and how much processing occurred before it ended up as ices in the protoplanetary disk. Through detailed models of the water deuterium chemistry, \cite{cleeves14} and \cite{furuya17} found that the formation of deuterated water is inefficient in the disk itself, but rather inherited without significant alteration from the earlier prestellar stages. These results highlight the importance of the prestellar chemistry for the origin of water -- both in our own Solar System as well as other planetary systems.

Knowledge about the water deuteration in star forming regions comes largely from gas-phase observations at far-infrared and (sub)millimeter wavelengths. In general, the column densities of water ice are large toward pre- and protostellar cores and the ice solid-state features can readily be studied at infrared wavelengths. However, the corresponding solid-state features of deuterated water, HDO, are difficult to disentangle due to sensitivity and overlap with more prominent species \citep[e.g.,][]{dartois03} and only tentative detections have been reported toward solar-type protostars \citep{aikawa12}.

With, in particular, its HIFI instrument, \emph{Herschel} provided significant insight into the distribution of gaseous water around solar-type protostars. Those observations demonstrated that small amounts of water are present in the gas phase at large scales of protostellar envelopes \citep{coutens12,mottram13} as well as cold prestellar cores \citep{caselli10}, a result of cosmic rays causing the water to photodesorb off the grains \citep{caselli12water}. The abundance of water in the gas phase remains low, but still the column densities are high enough that HDO emission from the same regions could be estimated for a few sources. In those regions, the HDO/H$_2$O ratio is determined by gas-phase chemistry at low temperatures and was found to range from $\sim$1\% to 20$\%$ \citep{coutens12,coutens13,liu11}.

Deriving the D/H ratios of water ices sublimating in the warm regions close to the central protostar using single-dish and \textit{Herschel} measurements is more problematic: these regions are heavily diluted in the large beams of such observations and the interpretation therefore heavily depends on radiative transfer models of the emission, which in turn are complicated due to the separation of the multiple (unconstrained) physical components within the beam and the fact that both the dust and some prominent lines become optically thick on small scales \citep[e.g.,][]{visser13}. Alternatively, interferometric studies of lines of less abundant isotopologs make it possible to zoom in and constrain the column densities of water on comparable scales where the emitting regions are relatively homogeneous in terms of their physics and the densities high enough that LTE is a reasonable approximation \citep[see discussions in][]{taquet13,persson14}. In this manner, estimates of the HDO/H$_2$O ratios have been derived for the typical Class~0 hot corinos with values of order 0.1\% \citep[e.g.,][]{iras4b_hdo,persson13,persson14,taquet13}, i.e., more than an order of magnitude below those found on larger scales and more in line with the highest numbers found for Oort family comets in our own Solar System.

The differences between the D/H ratios in the cold gas in the outer envelope and in the warm gas on small scales can be successfully modeled in different manners. Either the increased D/H ratio on large scales is caused by photodesorption, predominantly acting on the outermost layers of ice mantles that are formed latest and therefore most deuterium rich \citep{taquet14}, or by continued gas-phase deuteration of water once desorbed \citep{furuya16}.

The same models must also explain measured differences in the deuteration between the singly and doubly-deuterated variants of water, i.e., the HDO/H$_2$O and D$_2$O/HDO abundance ratios specifically. Statistically, the former ratio should be four times higher than the latter, since there are two indistinguishable variants of singly-deuterated water depending on which H-atom is substituted. However, direct interferometric measurements show that the D$_2$O/HDO ratio is in fact higher than the HDO/H$_2$O ratio by a factor of 7 \citep{coutens14} -- i.e., D$_2$O is more than an order of magnitude more abundant than HDO with respect to what should be expected from the HDO/H$_2$O ratio and the statistics mentioned above. In the models by \cite{furuya16}, the higher D$_2$O/HDO ratio can be explained if water ice forms throughout the evolution of the dense prestellar cores, from the earliest more tenuous stages where the D/H ratio is low to the later stages where CO freeze-out and a drop in the ortho-to-para ratio of H$_2$ causes the deuteration to be more efficient. In this scenario, the bulk of the (non-deuterated) water forms early in this evolution, while the deuterated species (HDO and D$_2$O) are formed in the dense stages and their relative abundances thus reflect the higher D/H ratios there. 

\begin{figure}
    \centering
    \includegraphics[width=\textwidth]{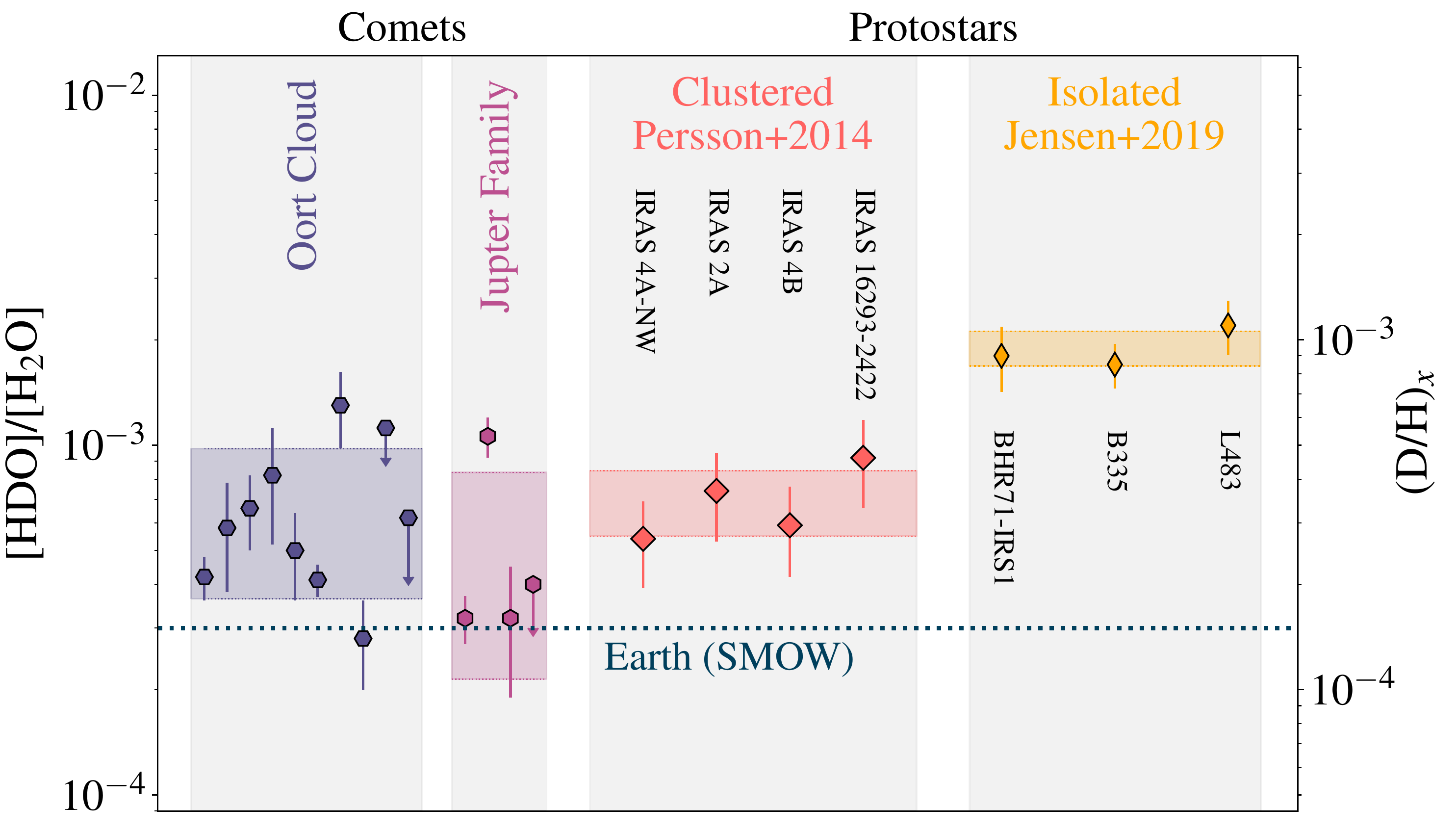}
    \caption{Measurements of the HDO/H$_2$O abundance (left axis) and D/H ratio corrected for statistics (i.e. when taking into account the number of equivalent hydrogen atoms; right axis) toward Oort Cloud and Jupiter Family comets as well as protostars formed in clustered and more isolated regions. The two groups of protostars show a difference in the HDO/H$_2$O ratios by a factor 2--4 with the ratios for the protostars from the more clustered regions closer to those of the comets. These differences may reflect the conditions in the parental environments of the protostars and provide a diagnostic for the environments in which the Sun was formed. Figure based on \cite{jensen19} and references therein.}
    \label{fig:HDOH2O}
\end{figure}

While studies of water's D/H ratios show interesting potential for studies of its formation under different conditions, it should be cautioned that the samples are still relatively small. A few studies have started addressing, e.g., the temporal and environmental variations. For example, toward the Class~I protostar SVS13b \cite{codella16svs13} derive limits to the HDO/H$_2$O comparable to that of the Class~0 protostars, which again could be taken as an argument in favor of the water formation and deuteration during the prestellar phases and little processing happening during the evolution of the protostars themselves. An example of the potential environmental dependence is seen in recent measurements by \cite{jensen19} of the HDO/H$_2$O ratios toward three protostars located in more isolated cores (Fig.~\ref{fig:HDOH2O}). The HDO/H$_2$O ratios for those sources are remarkably consistent and higher by factors 2--4 compared to those for the ``classical'' hot corinos located in more clustered regions. These differences could reflect the physics in the environment from which they arise, with the more isolated cores forming in colder environments and/or evolving more slowly through those. However, to fully disentangle these effects, it is clearly necessary to expand the samples of well-studied sources to other regions and compare to other tracers of the (temporal) protostellar chemical evolution.

\subsection{Complex organics}\label{ss:com_fractionation}
The characterization of the isotopic ratios of COMs also provides important insight into the astrochemistry in star forming regions. The good relative calibration across the ALMA bands, and the number of transitions any COM isotopolog typically shows in these regions, makes it possible to derive the relative abundances and thus levels of fractionation with high accuracy if the isotopologs co-exist in the high-density environment where LTE applies.

The first interstellar detections of deuterated methanol were obtained toward the Orion~KL region with the IRAM 30\,m telescope for both CH$_3$OD and CH$_2$DOH \citep[][]{mauersberger88,jacq93}. The deuterium fractionation of methanol as traced more recently with PdBI and \textit{Herschel} is 0.5--0.6\% toward the compact ridge and lower than 0.4\% toward the hot core \citep[][]{peng12,neill13}. Deuterated methyl formate was also first detected in the ISM toward Orion~KL with the 30\,m telescope, CH$_3$OCDO only tentatively but CH$_2$DOCHO robustly, both with a deuterium fractionation of $\sim$4\% with some variations across the region \citep[][]{margules10,coudert13}. The different deuteration levels of methanol and methyl formate may indicate that the former was formed at earlier stages than the latter.

A systematic study of the deuteration of COMs in Sgr~B2(N2) was performed with the EMoCA survey \citep[][]{belloche16}. The low level of deuteration found for a few COMs in Sgr~B2(N2) compared to Orion~KL and other high-mass protostellar objects surveyed in methanol by \citet{fontani15a} was interpreted as resulting either from the higher prestellar-phase temperatures that seem to characterize the Galactic center region, or from a lower elemental abundance of deuterium itself in the Galactic center region due to stellar processing, as was argued in the past on the basis of measurements performed with smaller molecules on larger scales, albeit with large uncertainties \citep[][]{jacq99,lubowich00,polehampton02}. Measurements toward the hot cores NGC6334I \citep{bogelund18} and NGC~7538-IRS1 \citep{ospinazamudio19}, both located in the Galactic disk, found low values similar to Sgr~B2(N2), which supports the temperature interpretation for Sgr~B2 rather than that of astration.

Compared to the high-mass star forming regions, the degree of deuteration is much higher toward low-mass protostars such as IRAS~16293-2422 as witnessed by early measurements \citep{vandishoeck95} as well as detections of doubly and triply deuterated variants of species such as CH$_3$OH \citep[e.g.,][]{parise02,parise04} using single-dish telescopes. Toward the two main components of this source, the PILS survey has provided systematic inventories of the D/H ratios as well as some estimates of $^{12}$C/$^{13}$C ratios for COMs \citep{jorgensen16,jorgensen18,coutens16,calcutt18a,manigand19iras16293A}. For some species the D/H ratios can be as high as 5--8\%, which translate to the abundance of the deuterated isotopologs relative to the main isotopologs in some cases being as high as 20\% due to statistics (see also discussion of water above). In fact, in the context of PILS, most deuterated isotopologs of two-carbon-containing COMs for which spectroscopy is available  have been identified.

Interestingly, some variations are seen in the deuteration between different species toward the B component of IRAS~16293-2422. Toward that source a range of the two carbon-atom containing species including CH$_2$OHCHO \citep{jorgensen16}, CH$_3$CHO, CH$_3$OCHO, CH$_3$OCH$_3$, and C$_2$H$_5$OH \citep{jorgensen18}, all show ratios of 4--8\%, while many of the smaller/simpler molecules, H$_2$CO \citep{persson18}, CH$_3$OH \citep{jorgensen18}, CH$_3$CN \citep{calcutt18a} and the S-bearing species \citep{drozdovskaya18} show lower ratios of $\approx$2\%. Toward the other source, IRAS16293A, the deuteration is similarly high for some species but because of larger uncertainties no clear trend is discernible \citep{manigand19iras16293A}. One possible explanation for this heterogeneity observed toward IRAS16293B, and possibly also causing the differences between CH$_3$OH and CH$_3$OCHO toward Orion mentioned earlier, could be that the former group of species are formed later in the evolution of the protostellar core and inherit the D/H ratio then.

The $^{12}$C/$^{13}$C ratios may also carry information. The spectral surveys performed toward Sgr~B2(N) led to the determination of the $^{12}$C/$^{13}$C isotopic ratio for a number of COMs. The EMoCA survey toward Sgr~B2(N2) yields values of 20--21 for CH$_3$CN, C$_2$H$_3$CN, and HC$_3$N \citep[][]{belloche16} and 25--27 for CH$_3$OH, C$_2$H$_5$OH, and NH$_2$CHO \citep[][]{mueller16a,belloche17}, with uncertainties likely not higher than 10--15\%. A ratio of 32 was derived from the singly $^{13}$C-substituted isotopologs of C$_2$H$_5$CN, which is surprisingly high compared to the other cyanides, all the more so as the ratio derived from the singly and doubly $^{13}$C-substituted isotopologs of C$_2$H$_5$CN is 26 \citep[][]{margules16}. The origin of this difference is still unclear.
$^{12}$C/$^{13}$C ratios were also derived from H$_2$CS, CH$_3$CCH, NH$_2$CHO, and C$_2$H$_5$CN toward Sgr~B2(N) by \citet{halfen17} with their single-dish ARO survey. They obtained values ranging from 15 to 33 with uncertainties on the order of 30\%, with an average value of $24 \pm 9$, consistent with the range of values derived from EMoCA. In IRAS16293B, some indications were seen for differences in the $^{12}$C/$^{13}$C ratio for a few of the species with higher D/H ratios, which, again, may reflect fractionation triggered by FUV irradiation by the embedded protostar \citep{jorgensen18}. 

Other important information about the formation of these organics may come from possible differences in the isotope ratios for different functional groups in individual COMs. \cite{favre14mf} studied the two $^{13}$C isotopologs of methyl formate in Orion~KL with ALMA and found no significant fractionation difference between the two variants. A similar conclusion was reached based on the EMoCA measurements of the $^{12}$C/$^{13}$C ratios mentioned above. Toward IRAS16293B, most of the complex organics do not show variations in their D/H ratios after accounting for the statistics when there are multiple hydrogen atoms  \citep{jorgensen16,jorgensen18}. This again suggests a homogeneous time of formation during the cold phase for each individual species at which point the D/H ratio is inherited, while exchange reactions later in the warm gas play a smaller role. There is one exception to this trend, namely CH$_3$CHO, for which the CHO-group shows a higher D/H ratio than the CH$_3$-group when corrected for statistics \citep{coudert19,manigand19iras16293A}. Whether the D/H ratio for this species could be influenced by other (e.g., gas-phase) processes remains an open question. 

It has been suggested that the CH$_2$DOH/CH$_3$OD ratio could vary between regions of low- and high-mass star formation \citep[e.g.,][and references therein]{ospinazamudio19}, ranging from about unity toward Orion \citep[or even lower toward NGC6334I, see][]{bogelund18}, over ratios of close to the statistical value of 3 toward high-mass protostars, to much higher ratios claimed toward low-mass star forming regions. However, there are known issues with the predicted line intensities of CH$_2$DOH that may have caused the column density estimates to be off in some cases. Also, in particular, the higher ratios for low-mass sources have suffered from problems due to extended emission seen in single-dish beams as well as optical thickness of a range of the brighter CH$_2$DOH transitions. Recent estimates for the warm gas toward solar-type protostars are consistent with the statistical ratios (\citealt{jorgensen18,taquet20}). Consequently, while the estimates toward Orion indicate a CH$_2$DOH/CH$_3$OD ratio below the statistical value, which may reflect the physical structure of the region, the conclusion that there is a dichotomy between low- and high-mass star forming regions in terms of the CH$_2$DOH/CH$_3$OD ratios is not robust at this point.

A final puzzle is posed by the detections of doubly-deuterated variants of a range of species on small scales toward IRAS16293B, including  H$_2$CO \citep{persson18},  CH$_3$CN \citep{calcutt18a}, and CH$_3$OCHO \citep{manigand19}. As seen for water, the D/H ratios derived from the ratios of the column densities for the doubly- and singly-deuterated isotopologs are significantly higher by factors of 5--10 than those derived from the singly- and non-deuterated isotopologs once corrected for statistics. However, if these species are all formed in the later stages through hydrogenation of CO-rich ices, this difference can, in contrast to water, not be attributed to differences in formation times between the various isotopologs. An alternative mechanism could be abstraction and substitution reactions, but as noted above those will to be equally efficient for different functional groups of a given molecule \citep[see discussion in][]{manigand19}.

\section{ORIGIN AND EVOLUTION OF CHEMICAL COMPLEXITY}
\label{s:origin_complexity}
As described throughout this review, one of the ultimate goals of astrochemistry is to understand the degree of chemical complexity that can arise before and during the star formation process and how much, if any, of this complexity may be inherited by 
emerging (proto)planetary systems in general. With the wide diversity of environments in which complex organic molecules are now detected (Sects.~\ref{s:inventories} and \ref{s:environments}) and their intricate link to the physical evolution of individual sources (Sect.~\ref{s:physical_evolution}), a key question is to what degree the chemistry varies. With the  systematic inventories of molecular species becoming available for objects in different evolutionary stages and environments, some clues to answering these questions may be obtained by comparing them. Held together with detailed comparison of the isotopic compositions (Sect.~\ref{s:fractionation}), the hope is that such measurements will help us to place the origin of our own Solar System and its chemistry in a more general star (and planet) formation context. In this section we pick-up this thread in a discussion of similarities and differences among measured  abundances of complex organic molecules toward different types of environments related to star formation, recent cometary values, and predictions from chemical models. 

\subsection{Similarities and differences between IRAS16293B, Sgr~B2(N2) and Comet 67P/C-G}
An example of such a comparison between the PILS IRAS16293B results and measurements toward Comet 67P/C-G obtained from the ROSINA instrument on \emph{Rosetta} is presented by \cite{drozdovskaya19}. Their results show a correlation between the protostellar and cometary measurements of the abundances of the CHO-, N-, and S-bearing species when estimated relative to CH$_3$OH, CH$_3$CN, and CH$_3$SH, although with some scatter. For the CHO- and N-bearing species it further appears that the cometary relative abundances are slightly enhanced compared to the protostellar ones. \cite{drozdovskaya19} concluded that the volatiles present at the earliest stages of the protostellar evolution are inherited by the cometary bodies, but also that some additional degree of processing could have occurred during the protoplanetary disk stage leading to further production of more complex species.

\begin{figure}[!htb]
    \includegraphics[width=\hsize]{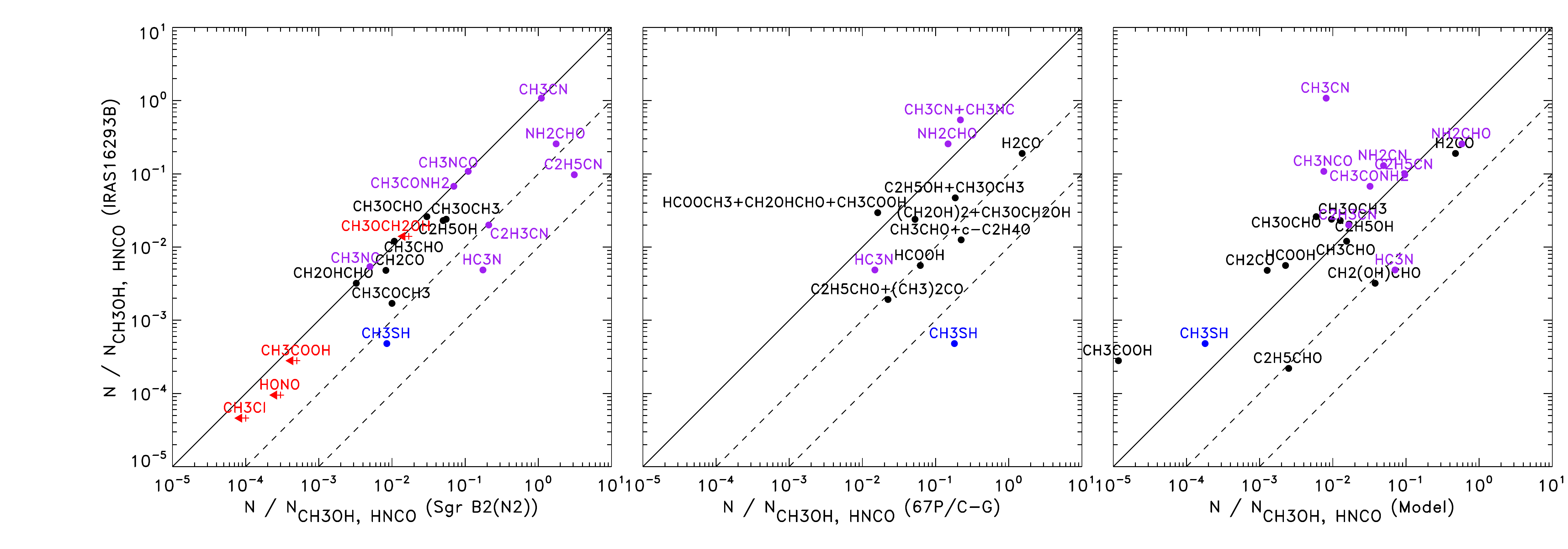}
    \caption{Comparison between the abundances of IRAS16293B from the PILS survey to those toward Sgr~B2(N2) from the EMoCA survey (left), a subset of those measured for Comet 67P/C-G from \textit{Rosetta} (middle), and in models (right). The abundances are measured relative to HNCO for the N-bearing species (data points shown with purple) and CH$_3$OH for the O-bearing species (data points shown in black and upper limits shown in red) and CH$_3$SH (data points shown in blue). The lines indicate one-to-one abundances (solid line) and factors of 10 and 100 enhancement in Sgr~B2(N2), Comet~67P/C-G, or the models compared to IRAS16293B (dashed lines). A list of the molecular abundances and references are provided in Table~\ref{t:abundances} in the Appendix.}
    \label{f:pils_emoca}
\end{figure}

Obviously, a relevant question in this comparison is how representative IRAS16293B is for the environment of the protosun -- and, more generally, how much variation is seen between different star-forming regions. To start addressing these questions, Fig.~\ref{f:pils_emoca} compares the abundances toward IRAS16293B from PILS with those toward Sgr~B2(N2) from EMoCA (\emph{left}), with the same species from the \textit{Rosetta} values for Comet 67P/C-G (\emph{middle}), as well as results from the MAGICKAL simulations presented by \cite{mueller16a} and \cite{belloche17} intended to reproduce the chemistry in Sgr~B2(N2) (\emph{right}). The full list of abundances and references are provided in Table~\ref{t:abundances} in the Appendix. For the oxygen-bearing species and CH$_3$SH, we use CH$_3$OH as a reference species, and for the nitrogen-bearing species, we use HNCO. This choice of HNCO differs from \cite{drozdovskaya19}, who normalised the nitrogen-bearing species relative to CH$_3$CN. For the observed datasets this is not significant, as the HNCO/CH$_3$CN ratios are identical toward IRAS16293B and Sgr~B2(N2). However, in the models the CH$_3$CN abundance is significantly lower (see below), while adopting HNCO makes for a more direct comparison to the observations.

The  most striking takeaway from Fig.~\ref{f:pils_emoca} is the excellent correlation between the abundances measured toward IRAS16293B and Sgr~B2(N2) directly demonstrating that CH$_3$OH and HNCO (or CH$_3$CN) serve as good proxies for the groups of oxygen- and nitrogen-bearing species toward IRAS16293B and Sgr~B2(N2). It is important to note that the abundances of HNCO relative to CH$_3$OH differ significantly. Relative to IRAS16293B the HNCO/CH$_3$OH ratio is close to a factor 14 higher toward Sgr~B2(N2), a factor 35 higher toward Comet~67P/C-G, and a factor 9 higher in the models. These differences between the observed HNCO/CH$_3$OH ratios, and implicitly the difference between the O- and N-bearing species, may reflect an underlying chemical differentiation similar to those discussed in Sect.~\ref{ss:otherhotcores} -- e.g., because the groups of species trace different physical components toward the sources.

Figure~\ref{f:pils_emoca} also indicates some other immediate trends. In terms of the overall variations, Sgr~B2(N2) and IRAS16293B appear more similar to each other than to Comet 67P/C-G or to the hot core models. In fact, while the hot core models were set up to simulate the chemistry of Sgr~B2(N2), they are actually in slightly better overall agreement with the abundances observed toward IRAS16293B once the differences between the O- and N-bearing species are taken out through the normalisations with CH$_3$OH and HNCO. Specifically, the abundances are in general slightly lower in the models compared to IRAS16293B, whereas the abundances toward Sgr~B2(N2), with very few exceptions, are slightly higher than those toward IRAS16293B, although only in a few cases by more than an order of magnitude.

Aside from NH$_2$CHO and the three nitrile group-bearing species, HC$_3$N, C$_2$H$_3$CN, and C$_2$H$_5$CN, the nitrogen-bearing species appear to agree well between Sgr~B2(N2) and IRAS16293B. The fact that the C$_2$H$_{\mathrm{X}}$CN species are similarly elevated toward Sgr B2(N2) over IRAS16293B perhaps indicates that all three of these species are behaving in concert through some shared chemistry or common dependence on physical conditions. In the case of C$_2$H$_3$CN and C$_2$H$_5$CN, their similar behavior across sources may be at least partially explained by the chemical models, which suggest that gas-phase C$_2$H$_3$CN is a direct (and major) product of the destruction of gas-phase C$_2$H$_5$CN \cite[see][]{garrod17}. Several competing theories have been offered as to the production mechanism for NH$_2$CHO in hot cores/corinos (see Sect.~\ref{ss:modeling}), including both gas and grain-surface processes. If indeed {\em all} of those suggested processes were operative, rather than one single, dominant process, then discrepancies between sources should perhaps be expected, if not easily explained.

In the chemical models (Fig.~\ref{f:pils_emoca}, \emph{right}), abundances of most N-bearing species are lower than values from the observations and cometary measurements. It is possible that some of this general discrepancy relates to the use of a ratio against HNCO, if this molecule is not well reproduced by the models. However, the modeled abundance of CH$_3$CN is especially divergent from the observed value, falling as many as two orders of magnitude short. The models with which we compare here suggest that much of the gas-phase CH$_3$CN is formed through gas-phase chemistry at high temperatures and densities, although the dominance of this process would require a sufficient time period under such conditions. Other recent models \citep{bonfand19} that use an alternative, and more detailed, physical description in conjunction with an updated binding energy for CH$_3$CN indicate a grain-surface origin for this molecule. The abundance of CH$_3$CN may therefore be especially sensitive to physical conditions and evolutionary timescales in individual sources, thus requiring a more accurate physical treatment in the chemical models to determine the origins of this molecule. Deeper integration of gas-grain chemical modeling treatments with magneto-hydrodynamical simulations of the star-formation process would be of great benefit to distinguishing between formation mechanisms for key molecules such as this, especially in the case where observations and models diverge by more than an order of magnitude.

The O-bearing species appear even more closely correlated between IRAS16293B and Sgr~B2(N2) than the N-bearing species with, in particular, a number of the CHO species in almost 1-to-1 agreement. Comparing the O- and N-bearing species between IRAS16293B and Comet 67P/C-G, the opposite appears to be the case, with better agreement for the N-bearing species while the O-bearing species are somewhat more abundant toward the comet as also noted by \cite{drozdovskaya19}. The models reproduce the column densities for the O-bearing species within about an order of magnitude but with larger scatter compared to the observational datasets. Overall, CH$_3$SH does not show any systematic trends, but a comparison to CH$_3$OH is perhaps not the most obvious for that species. 

\subsection{A wider census of oxygen- and nitrogen-bearing species in star forming environments}
\begin{figure}
\includegraphics[width=\hsize]{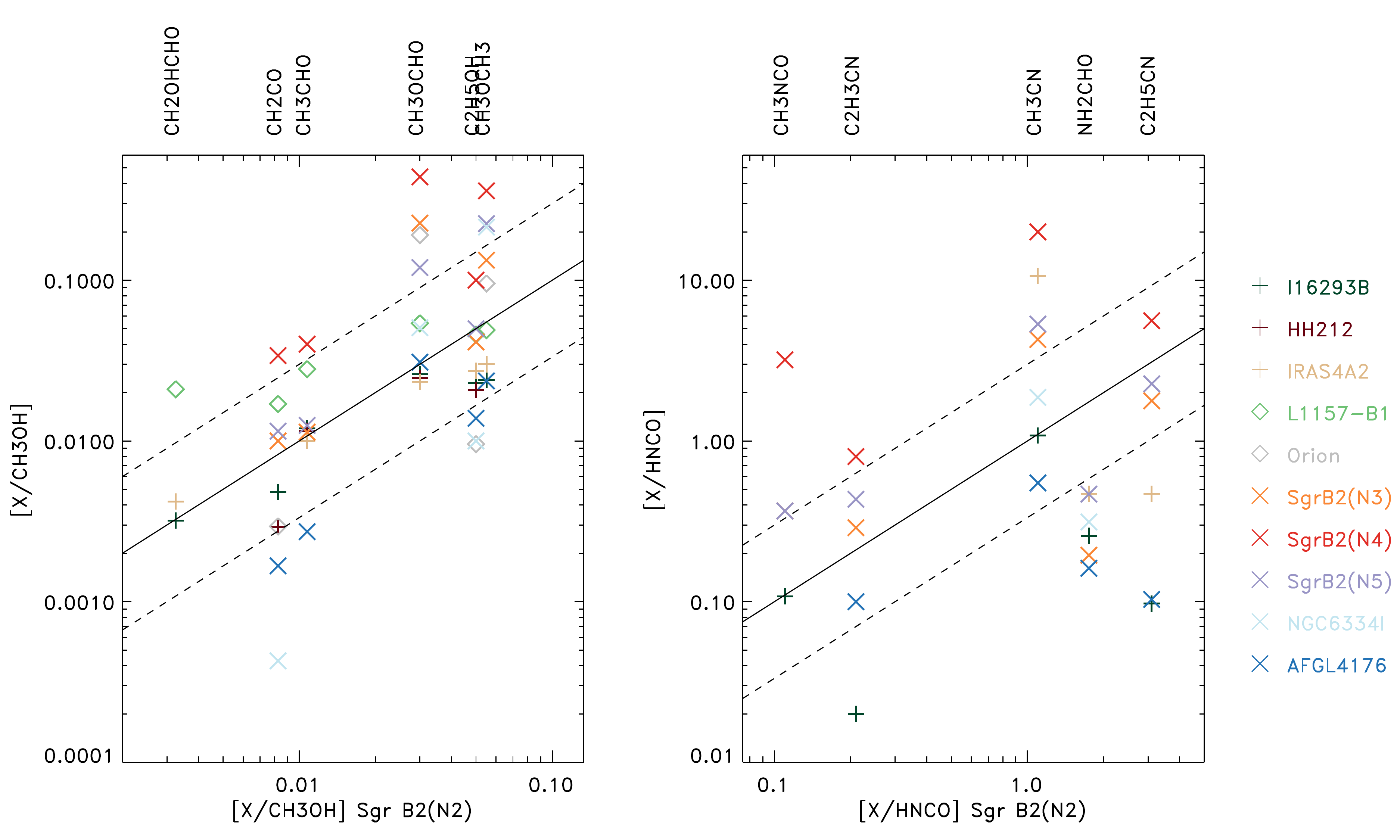}
\caption{Comparison between the abundances of O- (\emph{left}) and N-bearing species (\emph{right}) toward Sgr~B2(N2) \citep[][and A.~Belloche priv. comm.]{mueller16a,belloche16,belloche17,bonfand19} on the X-axis and a range of sources in the literature on the Y-axis (see legend on the right-hand side): the low-mass protostellar hot corinos IRAS16293B \citep{jorgensen18,calcutt18a} and NGC~1333-IRAS4A2 \citep{lopezsepulcre17}, the possible disk atmosphere toward the low-mass protostar HH212 \citep{lee19hh212}, the shocked region B1 in the L1157 outflow, the other hot cores in Sgr~B2(N), N3, N4, and N5 \citep[][and priv. comm.]{bonfand19}, the high-mass star forming region NGC6334I \citep{zernickel12}, the Orion compact ridge \citep{crockett14} and the high-mass protostar AFGL~4176 \citep{bogelund19afgl4176}. The low- and high-mass sources are shown with plus-signs and crosses, respectively, and the two shocks (Orion and L1157-B1) with diamonds. The abundances of the O-bearing species are measured as their column densities relative to those of CH$_3$OH while the N-bearing species are measured relative to HNCO. The three diagonal lines in each panel indicate equal abundances (solid) and abundances factors 3 above or below those in Sgr~B2(N2) (dashed).}\label{f:CompCom}
\end{figure}

Similar comparisons can also be made for a wider set of sources of different types for which smaller inventories are present in the literature. Figure~\ref{f:CompCom} shows the abundances of the common O-bearing species (relative to CH$_3$OH; \emph{left}) and N-bearing species (relative to HNCO; \emph{right}) toward a range of sources. Tentative correlations are also seen here with abundances in agreement for the O-bearing species in most cases within a factor of about 3--5. A few outliers are seen, but no source or species systematically stands out. 
For the N-bearing species slightly more scatter is seen and there is no clear sign of an overall correlation compared to Sgr B2(N2). The exception to this is the ratio between NH$_2$CHO and HNCO which shows a relatively constant abundance of HNCO of about 20\% compared to NH$_2$CHO -- except for Sgr~B2(N2) where NH$_2$CHO is close to a factor 2 more abundant than HNCO. The former value is in agreement with empirical correlations between the two species previously reported  \citep[e.g.,][]{bisschop07,mendoza14,lopezsepulcre15}. This correlation has previously been taken as evidence for the formation of NH$_2$CHO through the addition of H atoms to solid-phase HNCO. However, experimental evidence suggests that at least the final step in this process does not proceed with any efficiency, instead re-forming HNCO \citep{noble15}, although other work \citep{haupa19} indicates that inter-conversion of HNCO and NH$_2$CHO through the addition of H is indeed possible. A direct chemical relationship, via ice chemistry, between these two molecules therefore remains somewhat uncertain, although this does not rule out that similar chemical conditions would promote independent production of both. For example, \cite{ligterink18amides} suggest that both molecules could form contemporaneously in the ices through radical addition.

Furthermore, some astrochemical models suggest that much of the observable gas-phase HNCO could be a product of the destruction of larger molecules, such as urea, NH$_2$CONH$_2$, through ion-molecule chemistry \citep{garrod08b,tideswell10}. In this scenario, any HNCO released directly from the grains is destroyed in the gas phase much earlier on, at lower temperatures. If NH$_2$CONH$_2$ itself, a known interstellar molecule \citep{belloche19}, is predominantly a product of NH$_2$CHO, then a case for a direct chemical link between HNCO and NH$_2$CHO may be made. On the other hand, HNCO and NH$_2$CHO may simply be good independent indicators of the overall C, N, and O abundances available to the chemistry. Indeed, models by \cite{quenard18} indicate that these two molecules are linked through their similar response to temperature, not a direct chemical link. Correlations between chemically-related molecules, therefore, should not necessarily be taken to indicate a causal relationship in their interstellar production.

Clearly much work needs to be done to confirm the trends outlined above --
in particular, the relatively small scatter in abundances for many COMs in Sgr B2(N2) vs. IRAS16293B, as well as the overall agreements across different types of sources. If confirmed, this trend could have important implications, i.e., that the  conditions where the complex organics form are relatively homogeneous. A speculation could be that this reflects that many of the complex species form in the colder environments with the subsequent evolution (e.g., heating of gas and dust in low- vs. high-mass star forming regions or in shocks) playing a smaller role. In that case, the small scatter would provide a better defined target for models and laboratory experiments to shoot for. The possible direct detection of solid-phase COMs more complex than methanol in cold regions could also be highly valuable in confirming such a picture and also help addressing the link, if any, between the COMs observed in the gas phase toward cold prestellar cores and those in hot cores and corinos. 

Further, while strong similarities are seen between IRAS16293B and Sgr B2(N2) those still differ from the Comet 67P/C-G measurements. The agreement between, e.g., the IRAS16293B and AFGL4176 column densities (Fig.~\ref{f:CompCom}), suggests that this trend may apply more generally. If confirmed through future systematic chemical inventories, this would further emphasise that chemical processing must have taken place between the protostellar and cometary stages. Accounting for this will likely be critical to address the relative importance of the earliest stages of star formation on any eventual chemical diversity in protoplanetary disks and (exo)planetary systems.

\begin{summary}[SUMMARY POINTS]
\begin{enumerate}
\item Complex organic molecules are 
detected in a wide range of environments ranging from cold gas in prestellar cores to the warm gas on Solar System scales close to individual protostars (Sect.~\ref{s:inventories}).
\item The recent detections of aromatic, branched, and chiral molecules have revealed a wider range of complexity in the molecular structures available to the chemistry of star-forming regions 
and, by implication, to the 
earliest stages of protostellar and protoplanetary disk evolution (Sect.~\ref{ss:complexity}).
\item From a theoretical/modeling perspective, the specific formation mechanisms for individual species are still open for discussion with both grain-surface/ice and gas-phase processes providing possible formation pathways (Sect.~\ref{ss:modeling}). 
\item Spatial differentiation between individual molecules and groups of species are seen toward some regions, which may reflect their physical structure or evolution, and/or more microphysical effects such as variations in binding energies (Sect.~\ref{s:environments}).
\item The importance of the detailed physical evolution of protostars on their resulting chemistry is complex, with  processes such as disk formation, accretion shocks, and episodic accretion -- as well as shock events such as those seen in the Orion KL region -- all 
potentially affecting the degree of chemical processing on a source-by-source basis (Sect.~\ref{s:physical_evolution}). However, the importance of these effects on the end-result, e.g., in terms of the relative abundances of COMs, remains to be quantified.
\item The degree of fractionation varies between different types of regions, for example with the enhanced deuteration of water and complex organics in low- versus high-mass star-forming regions. Systematic variations are also seen in the fractionation between groups of molecules, perhaps reflecting the conditions under which they form (Sect.~\ref{s:fractionation}).
\item Comparisons between the relative abundances of groups of molecules show promise for providing empirical insights into the origin of chemical complexity and the link to the Solar System. Specifically an excellent agreement between the abundances measured toward two significantly different regions, namely the low-mass protostar IRAS16293B and the Galactic center/high-mass star-forming region Sgr~B2(N2), may suggest that interstellar complex chemistry is relatively robust to environmental effects (Sect.~\ref{s:origin_complexity}). Such correlations also set well-defined targets for future work on chemical modeling and laboratory experiments.
\end{enumerate}
\end{summary}

\begin{issues}[FUTURE ISSUES]
\begin{enumerate}
\item The presence of complex organic molecules toward cold regions highlights the importance of understanding the composition of dust-grain ice mantles (Sect.~\ref{ss:prestellar_coms}). A key task for upcoming observations with the James Webb Space Telescope (JWST) will be to reveal the composition of these ices in terms of complex organic molecules. Those observations will thereby help to define
the link between the ice chemistry and the large gas-phase inventories established through observations at longer wavelengths.
\item Both from a physical and chemical point of view, there is significant work to do in comparing the predictions of numerical (magneto)hydrodynamical models to the molecular signatures observed toward protostars, as well as understanding, for example, the processes leading to the formation of disks and accretion through the evolution of protostars. Deeper integration of detailed chemical models with hydrodynamical simulations, to account for the chemical effects of more realistic treatments of physical evolution, would go some way to disentangling inaccuracies in the chemical treatments from the uncertainties associated purely with source structure.
\item Significant work needs to be done on expanding the large molecular inventory surveys to wider samples of sources in a systematic way. This is recognised as one of the key scientific goals of the ALMA Development Roadmap \citep{alma2030}. Such upgrades will enable comparisons similar to those of Fig.~\ref{f:pils_emoca} to many more sources in diverse environments and stages of evolution.
\item Moving forward with the characterization of the degree of molecular complexity in the gas phase of star forming regions will require sensitive surveys that can go beyond the confusion limit reached today. Solutions may be to target lower frequencies, e.g., with ALMA and NOEMA below 80 GHz or with ngVLA -- or to perform high angular resolution observations to further resolve regions with narrow linewidths.
\item The high angular resolution achievable with ALMA may allow astrochemical studies in the time domain such as proposed in connection with studies of COMs in the Orion KL region and outflows (Sects.~\ref{ss:outflows} and \ref{ss:orion}). If indeed possible, this would be an interesting avenue to explore further.
\end{enumerate}
\end{issues}

\section*{DISCLOSURE STATEMENT}
The authors are not aware of any affiliations, memberships, funding, or financial holdings that
might be perceived as affecting the objectivity of this review.

\section*{ACKNOWLEDGMENTS}
We are grateful to Yuri Aikawa, Geoff Blake, Paola Caselli, Eric Herbst, Niels Ligterink, Laurent Margules and Ewine van Dishoeck for reading and providing comments about the manuscript. Also, we thank Per Bjerkeli, Timea Csengeri, Sigurd Jensen and Laurent Pagani for providing material used in some of the figures. JKJ acknowledges support by the European Research Council (ERC) under the European Union's Horizon 2020 research and innovation programme through ERC Consolidator Grant ``S4F'' (grant agreement No~646908). AB acknowledges support by the Deutsche Forschungsgemeinschaft (DFG) through the collaborative research grant SFB 956 (project ID 184018867), sub-project B3. RTG acknowledges support from the NASA Astrophysics Research and Analysis program (grant NNX15AG07G), the NASA Emerging Worlds program (grant NNX17AE23G), and the NASA Astrophysics Theory Program (grant 80NSSCI8K0558). 

\appendix
\section{Molecular abundances}\label{s:abundanceappendix}
\begin{table}[!htb]
\caption{Abundances relative to CH$_3$OH used in Fig.~\ref{f:pils_emoca} from measurements toward IRAS16293B, SgrB2(N2) and Comet 67P/C-G as well as predicted by models.} \label{t:abundances}
\begin{tabular}{l|c|c|c|c} \hline
Species            & IRAS16293B$^{\rm a}$ & Sgr~B2(N2)$^{\rm b}$  & Model predictions$^{\rm c}$ & Comet 67P/C-G$^{\rm d}$    \\ 
                   & (PILS)               & (EMoCA) & &     \\ \hline
CH$_3$OH           & 1.0         &       1.0     & 1.0      & 1.0      \\ \hline
H$_2$CO            & 0.19        &      $\ldots$ & 0.48     & 1.5      \\ \hline
C$_2$H$_5$OH       & 0.023       &       0.050   & 0.013    & \multirow{2}{*}{0.19$^{\rm e}$}     \\ \cline{1-4}
CH$_3$OCH$_3$      & 0.024       &       0.055   & 0.0096   &          \\ \hline
CH$_3$OCHO         & 0.026       &       0.030   & 0.0059   & \multirow{3}{*}{0.016$^{\rm e}$}         \\ \cline{1-4}
CH$_2$OHCHO        & 0.0034      &       0.0033$\ast$  & 0.038    &          \\ \cline{1-4}
CH$_3$COOH         & 0.00028     &    $<$0.0005$\ast$  & 1.2e-05  &          \\ \hline
CH$_3$CHO          & 0.012       &       0.011   & 0.015    & \multirow{2}{*}{0.22$^{\rm e}$}     \\ \cline{1-4}
c-C$_2$H$_4$O      & 0.00054     &      $\ldots$ & $\ldots$ &          \\ \hline
CH$_3$OCH$_2$OH    & 0.014       &     $<$0.017$\ast$  & 0.0044   & \multirow{2}{*}{0.052$^{\rm e}$}    \\ \cline{1-4}
(CH$_2$OH)$_2$     & 0.0099$^{\rm f}$     &      $\ldots$ & $\ldots$ &          \\ \hline
CH$_3$COCH$_3$     & 0.0017      &       0.010   & $\ldots$ & \multirow{2}{*}{0.022$^{\rm e}$}   \\ \cline{1-4}
C$_2$H$_5$CHO      & 0.00022     &      $\ldots$ & $\ldots$ &          \\ \hline
NH$_2$CHO          & 0.0010      &       0.088   & 0.020    & 0.019    \\ \hline
CH$_3$CN           & 0.004       &       0.055   & 0.00028  & \multirow{2}{*}{0.028$^{\rm e}$}    \\ \cline{1-4}
CH$_3$NC           & 2.0e-05     &       0.00025 & $\ldots$ &          \\ \hline
HNCO               & 0.0037      &       0.050   & 0.034    & 0.13     \\ \hline
HC$_3$N            & 1.8e-05     &       0.0088  & 0.0024   & 0.0019   \\ \hline
CH$_3$SH           & 0.00048     &       0.0085  & 0.00018  & 0.18     \\ \hline
CH$_3$NCO          & 0.00040     &       0.0055  & 0.00026  & $\ldots$ \\ \hline
C$_2$H$_5$CN       & 0.00036     &       0.16    & 0.0033   & $\ldots$ \\ \hline
C$_2$H$_3$CN       & 7.4e-05     &       0.011   & 0.00056  & $\ldots$ \\ \hline
CH$_3$CONH$_2$     & 0.00025     &       0.0035  & 0.0011   & $\ldots$ \\ \hline
HONO               & 9.0e-05     &    $<$0.00030$\ast$ & $\ldots$ & $\ldots$ \\ \hline
CH$_2$CO           & 0.0048      &       0.0083$\ast$  & 0.0013   & $\ldots$ \\ \hline
\end{tabular}
\begin{tabnote}
$^{\rm a}$Measurements toward IRAS16293B in connection with the PILS program \citep{jorgensen16,jorgensen18,coutens16,coutens19hono,lykke17,calcutt18a,calcutt18b,ligterink17,drozdovskaya18,persson18,manigand19iras16293A}, see also summary by \cite{drozdovskaya19}. $^{\rm b}$Measurements toward Sgr~B2(N2) in connection with the EMoCA program \citep{belloche16,belloche17,mueller16a,bonfand19,ordu19,willis19}. New measurements presented here for the first time (Belloche, priv. comm) indicated with a ``$\ast$ $^{\rm c}$''Results from the MAGICKAL simulations presented by \cite{mueller16a} and \cite{belloche17}. $^{\rm d}$Measurements from ROSINA instrument presented by \cite{drozdovskaya19}. $^{\rm e}$Species indistinguishable in the ROSINA spectrometer measurements of Comet 67P/C-G. $^{\rm f}$Refers to total abundance of the two lowest state conformers, $aGg'$ and $gGg'$, of ethylene glycol, (CH$_2$OH)$_2$.
\end{tabnote}
\end{table}

\bibliographystyle{ar-style2}

\begin{thebibliography}{}
\expandafter\ifx\csname natexlab\endcsname\relax\def\natexlab#1{#1}\fi

\bibitem[{{Acharyya} \& {Herbst}(2018)}]{acharyya18}
{Acharyya} K, {Herbst} E. 2018.
\textit{\apj} 859:51

\bibitem[{{Ag{\'u}ndez} et~al.(2008){Ag{\'u}ndez}, {Cernicharo}, {Gu{\'e}lin},
  {Gerin}, {McCarthy} \& {Thaddeus}}]{agundez08}
{Ag{\'u}ndez} M, {Cernicharo} J, {Gu{\'e}lin} M, {Gerin} M, {McCarthy} MC,
  {Thaddeus} P. 2008.
\textit{\aap} 478:L19--L22

\bibitem[{{Ag{\'u}ndez} et~al.(2019){Ag{\'u}ndez}, {Marcelino}, {Cernicharo},
  {Roueff} \& {Tafalla}}]{agundez19}
{Ag{\'u}ndez} M, {Marcelino} N, {Cernicharo} J, {Roueff} E, {Tafalla} M. 2019.
\textit{\aap} 625:A147

\bibitem[{{Aikawa} et~al.(2012){Aikawa}, {Wakelam}, {Hersant}, {Garrod} \&
  {Herbst}}]{aikawa12}
{Aikawa} Y, {Wakelam} V, {Hersant} F, {Garrod} RT, {Herbst} E. 2012.
\textit{\apj} 760:40

\bibitem[{{Alexander} et~al.(2012){Alexander}, {Bowden}, {Fogel}, {Howard},
  {Herd} \& {Nittler}}]{alexander12}
{Alexander} CMO, {Bowden} R, {Fogel} ML, {Howard} KT, {Herd} CDK, {Nittler} LR.
  2012.
\textit{Science} 337:721

\bibitem[{{Allen} \& {Burton}(1993)}]{allen93}
{Allen} DA, {Burton} MG. 1993.
\textit{\nat} 363:54--56

\bibitem[{{Allen} \& {Robinson}(1977)}]{allen77}
{Allen} M, {Robinson} GW. 1977.
\textit{\apj} 212:396--415

\bibitem[{{Allen} et~al.(2017){Allen}, {van der Tak}, {S{\'a}nchez-Monge},
  {Cesaroni} \& {Beltr{\'a}n}}]{allen17}
{Allen} V, {van der Tak} FFS, {S{\'a}nchez-Monge} {\'A}, {Cesaroni} R,
  {Beltr{\'a}n} MT. 2017.
\textit{\aap} 603:A133

\bibitem[{{Alonso} et~al.(2016){Alonso}, {Kolesnikov{\'a}}, {Tercero},
  {Cabezas}, {Alonso} et~al.}]{alonso16}
{Alonso} ER, {Kolesnikov{\'a}} L, {Tercero} B, {Cabezas} C, {Alonso} JL, et~al.
  2016.
\textit{\apj} 832:42

\bibitem[{{{\'A}lvarez-Barcia} et~al.(2018){{\'A}lvarez-Barcia}, {Russ},
  {K{\"a}stner} \& {Lamberts}}]{alvarez-barcia18}
{{\'A}lvarez-Barcia} S, {Russ} P, {K{\"a}stner} J, {Lamberts} T. 2018.
\textit{\mnras} 479:2007--2015

\bibitem[{{Anderl} et~al.(2016){Anderl}, {Maret}, {Cabrit}, {Belloche}, {Maury}
  et~al.}]{anderl16}
{Anderl} S, {Maret} S, {Cabrit} S, {Belloche} A, {Maury} AJ, et~al. 2016.
\textit{\aap} 591:A3

\bibitem[{{Arce} et~al.(2008){Arce}, {Santiago-Garc{\'{\i}}a}, {J{\o}rgensen},
  {Tafalla} \& {Bachiller}}]{arce08}
{Arce} HG, {Santiago-Garc{\'{\i}}a} J, {J{\o}rgensen} JK, {Tafalla} M,
  {Bachiller} R. 2008.
\textit{\apjl} 681:L21--L24

\bibitem[{{Artur de la Villarmois} et~al.(2019){Artur de la Villarmois},
  {J{\o}rgensen}, {Kristensen}, {Bergin}, {Harsono} et~al.}]{artur19sample}
{Artur de la Villarmois} E, {J{\o}rgensen} JK, {Kristensen} LE, {Bergin} EA,
  {Harsono} D, et~al. 2019.
\textit{\aap} 626:A71

\bibitem[{{Artur de la Villarmois} et~al.(2018){Artur de la Villarmois},
  {Kristensen}, {J{\o}rgensen}, {Bergin}, {Brinch} et~al.}]{artur18}
{Artur de la Villarmois} E, {Kristensen} LE, {J{\o}rgensen} JK, {Bergin} EA,
  {Brinch} C, et~al. 2018.
\textit{\aap} 614:A26

\bibitem[{{Audard} et~al.(2014){Audard}, {{\'A}brah{\'a}m}, {Dunham}, {Green},
  {Grosso} et~al.}]{audardppvi}
{Audard} M, {{\'A}brah{\'a}m} P, {Dunham} MM, {Green} JD, {Grosso} N, et~al.
  2014.
\textit{{Episodic Accretion in Young Stars}}. In \textit{Protostars and Planets
  VI}, eds. H~{Beuther}, RS~{Klessen}, CP~{Dullemond}, T~{Henning}

\bibitem[{{Bacmann} et~al.(2012){Bacmann}, {Taquet}, {Faure}, {Kahane} \&
  {Ceccarelli}}]{bacmann12}
{Bacmann} A, {Taquet} V, {Faure} A, {Kahane} C, {Ceccarelli} C. 2012.
\textit{\aap} 541:L12

\bibitem[{{Bally} et~al.(2017){Bally}, {Ginsburg}, {Arce}, {Eisner},
  {Youngblood} et~al.}]{bally17}
{Bally} J, {Ginsburg} A, {Arce} H, {Eisner} J, {Youngblood} A, et~al. 2017.
\textit{\apj} 837:60

\bibitem[{{Bally} \& {Zinnecker}(2005)}]{bally05}
{Bally} J, {Zinnecker} H. 2005.
\textit{\aj} 129:2281--2293

\bibitem[{{Balucani}, {Ceccarelli} \& {Taquet}(2015)}]{balucani15}
{Balucani} N, {Ceccarelli} C, {Taquet} V. 2015.
\textit{\mnras} 449:L16--L20

\bibitem[{{Barger} \& {Garrod}(2020)}]{barger19}
{Barger} CJ, {Garrod} RT. 2020.
\textit{\apj} 888:38

\bibitem[{{Barone} et~al.(2015){Barone}, {Latouche}, {Skouteris}, {Vazart},
  {Balucani} et~al.}]{barone15}
{Barone} V, {Latouche} C, {Skouteris} D, {Vazart} F, {Balucani} N, et~al. 2015.
\textit{\mnras} 453:L31--L35

\bibitem[{{Bell} et~al.(2014){Bell}, {Cernicharo}, {Viti}, {Marcelino}, {Palau}
  et~al.}]{bell14}
{Bell} TA, {Cernicharo} J, {Viti} S, {Marcelino} N, {Palau} A, et~al. 2014.
\textit{\aap} 564:A114

\bibitem[{{Belloche} et~al.(2014){Belloche}, {Garrod}, {M{\"u}ller} \&
  {Menten}}]{belloche14}
{Belloche} A, {Garrod} RT, {M{\"u}ller} HSP, {Menten} KM. 2014.
\textit{Science} 345:1584--1587

\bibitem[{{Belloche} et~al.(2019){Belloche}, {Garrod}, {M{\"u}ller}, {Menten},
  {Medvedev} et~al.}]{belloche19}
{Belloche} A, {Garrod} RT, {M{\"u}ller} HSP, {Menten} KM, {Medvedev} I, et~al.
  2019.
\textit{\aap} 628:A10

\bibitem[{{Belloche} et~al.(2017){Belloche}, {Meshcheryakov}, {Garrod},
  {Ilyushin}, {Alekseev} et~al.}]{belloche17}
{Belloche} A, {Meshcheryakov} AA, {Garrod} RT, {Ilyushin} VV, {Alekseev} EA,
  et~al. 2017.
\textit{\aap} 601:A49

\bibitem[{{Belloche} et~al.(2016){Belloche}, {M{\"u}ller}, {Garrod} \&
  {Menten}}]{belloche16}
{Belloche} A, {M{\"u}ller} HSP, {Garrod} RT, {Menten} KM. 2016.
\textit{\aap} 587:A91

\bibitem[{{Belloche} et~al.(2013){Belloche}, {M{\"u}ller}, {Menten}, {Schilke}
  \& {Comito}}]{belloche13}
{Belloche} A, {M{\"u}ller} HSP, {Menten} KM, {Schilke} P, {Comito} C. 2013.
\textit{\aap} 559:A47

\bibitem[{{Benedettini} et~al.(2012){Benedettini}, {Busquet}, {Lefloch},
  {Codella}, {Cabrit} et~al.}]{benedettini12}
{Benedettini} M, {Busquet} G, {Lefloch} B, {Codella} C, {Cabrit} S, et~al.
  2012.
\textit{\aap} 539:L3

\bibitem[{{Bennett} et~al.(2007){Bennett}, {Chen}, {Sun}, {Chang} \&
  {Kaiser}}]{bennett07}
{Bennett} CJ, {Chen} SH, {Sun} BJ, {Chang} AHH, {Kaiser} RI. 2007.
\textit{\apj} 660:1588--1608

\bibitem[{{Bennett} et~al.(2011){Bennett}, {Hama}, {Kim}, {Kawasaki} \&
  {Kaiser}}]{bennett11}
{Bennett} CJ, {Hama} T, {Kim} YS, {Kawasaki} M, {Kaiser} RI. 2011.
\textit{\apj} 727:27

\bibitem[{{Bergner} et~al.(2016){Bergner}, {{\"O}berg}, {Rajappan} \&
  {Fayolle}}]{bergner16}
{Bergner} JB, {{\"O}berg} KI, {Rajappan} M, {Fayolle} EC. 2016.
\textit{\apj} 829:85

\bibitem[{{Bertin} et~al.(2016){Bertin}, {Romanzin}, {Doronin}, {Philippe},
  {Jeseck} et~al.}]{bertin16}
{Bertin} M, {Romanzin} C, {Doronin} M, {Philippe} L, {Jeseck} P, et~al. 2016.
\textit{\apjl} 817:L12

\bibitem[{{Bisschop} et~al.(2008){Bisschop}, {J{\o}rgensen}, {Bourke},
  {Bottinelli} \& {van Dishoeck}}]{bisschop08}
{Bisschop} SE, {J{\o}rgensen} JK, {Bourke} TL, {Bottinelli} S, {van Dishoeck}
  EF. 2008.
\textit{\aap} 488:959--968

\bibitem[{{Bisschop} et~al.(2007){Bisschop}, {J{\o}rgensen}, {van Dishoeck} \&
  {de Wachter}}]{bisschop07}
{Bisschop} SE, {J{\o}rgensen} JK, {van Dishoeck} EF, {de Wachter} EBM. 2007.
\textit{\aap} 465:913--929

\bibitem[{{Bizzocchi} et~al.(2014){Bizzocchi}, {Caselli}, {Spezzano} \&
  {Leonardo}}]{bizzocchi14}
{Bizzocchi} L, {Caselli} P, {Spezzano} S, {Leonardo} E. 2014.
\textit{\aap} 569:A27

\bibitem[{{Bjerkeli} et~al.(2016){Bjerkeli}, {J{\o}rgensen}, {Bergin},
  {Frimann}, {Harsono} et~al.}]{bjerkeli16water}
{Bjerkeli} P, {J{\o}rgensen} JK, {Bergin} EA, {Frimann} S, {Harsono} D, et~al.
  2016.
\textit{\aap} 595:A39

\bibitem[{{Bjerkeli}, {J{\o}rgensen} \& {Brinch}(2016)}]{bjerkeli16outflow}
{Bjerkeli} P, {J{\o}rgensen} JK, {Brinch} C. 2016.
\textit{\aap} 587:A145

\bibitem[{{Blake} et~al.(1996){Blake}, {Mundy}, {Carlstrom}, {Padin}, {Scott}
  et~al.}]{blake96}
{Blake} GA, {Mundy} LG, {Carlstrom} JE, {Padin} S, {Scott} SL, et~al. 1996.
\textit{\apjl} 472:L49

\bibitem[{{B{\o}gelund} et~al.(2019{\natexlab{a}}){B{\o}gelund}, {Barr},
  {Taquet}, {Ligterink}, {Persson} et~al.}]{bogelund19afgl4176}
{B{\o}gelund} EG, {Barr} AG, {Taquet} V, {Ligterink} NFW, {Persson} MV, et~al.
  2019{\natexlab{a}}.
\textit{\aap} 628:A2

\bibitem[{{B{\o}gelund} et~al.(2019{\natexlab{b}}){B{\o}gelund}, {McGuire},
  {Hogerheijde}, {van Dishoeck} \& {Ligterink}}]{bogelund19methylamine}
{B{\o}gelund} EG, {McGuire} BA, {Hogerheijde} MR, {van Dishoeck} EF,
  {Ligterink} NFW. 2019{\natexlab{b}}.
\textit{\aap} 624:A82

\bibitem[{{B{\o}gelund} et~al.(2018){B{\o}gelund}, {McGuire}, {Ligterink},
  {Taquet}, {Brogan} et~al.}]{bogelund18}
{B{\o}gelund} EG, {McGuire} BA, {Ligterink} NFW, {Taquet} V, {Brogan} CL,
  et~al. 2018.
\textit{\aap} 615:A88

\bibitem[{{Bonfand} et~al.(2019){Bonfand}, {Belloche}, {Garrod}, {Menten},
  {Willis} et~al.}]{bonfand19}
{Bonfand} M, {Belloche} A, {Garrod} RT, {Menten} KM, {Willis} E, et~al. 2019.
\textit{\aap} 628:A27

\bibitem[{{Bonfand} et~al.(2017){Bonfand}, {Belloche}, {Menten}, {Garrod} \&
  {M{\"u}ller}}]{bonfand17}
{Bonfand} M, {Belloche} A, {Menten} KM, {Garrod} RT, {M{\"u}ller} HSP. 2017.
\textit{\aap} 604:A60

\bibitem[{{Boogert}, {Gerakines} \& {Whittet}(2015)}]{boogert15}
{Boogert} ACA, {Gerakines} PA, {Whittet} DCB. 2015.
\textit{\araa} 53:541--581

\bibitem[{{Brinch} \& {J{\o}rgensen}(2013)}]{brinch13}
{Brinch} C, {J{\o}rgensen} JK. 2013.
\textit{\aap} 559:A82

\bibitem[{{Brouillet} et~al.(2013){Brouillet}, {Despois}, {Baudry}, {Peng},
  {Favre} et~al.}]{brouillet13}
{Brouillet} N, {Despois} D, {Baudry} A, {Peng} TC, {Favre} C, et~al. 2013.
\textit{\aap} 550:A46

\bibitem[{{Brown}, {Charnley} \& {Millar}(1988)}]{brown88}
{Brown} PD, {Charnley} SB, {Millar} TJ. 1988.
\textit{\mnras} 231:409--417

\bibitem[{{Burkhardt} et~al.(2016){Burkhardt}, {Dollhopf}, {Corby}, {Carroll},
  {Shingledecker} et~al.}]{burkhardt16}
{Burkhardt} AM, {Dollhopf} NM, {Corby} JF, {Carroll} PB, {Shingledecker} CN,
  et~al. 2016.
\textit{\apj} 827:21

\bibitem[{{Burkhardt} et~al.(2019){Burkhardt}, {Shingledecker}, {Le Gal},
  {McGuire}, {Remijan} \& {Herbst}}]{burkhardt19}
{Burkhardt} AM, {Shingledecker} CN, {Le Gal} R, {McGuire} BA, {Remijan} AJ,
  {Herbst} E. 2019.
\textit{\apj} 881:32

\bibitem[{{Butscher} et~al.(2016){Butscher}, {Duvernay}, {Danger} \&
  {Chiavassa}}]{butscher16}
{Butscher} T, {Duvernay} F, {Danger} G, {Chiavassa} T. 2016.
\textit{\aap} 593:A60

\bibitem[{{Butscher} et~al.(2017){Butscher}, {Duvernay}, {Rimola},
  {Segado-Centellas} \& {Chiavassa}}]{butscher17}
{Butscher} T, {Duvernay} F, {Rimola} A, {Segado-Centellas} M, {Chiavassa} T.
  2017.
\textit{Physical Chemistry Chemical Physics (Incorporating Faraday
  Transactions)} 19:2857--2866

\bibitem[{{Calcutt} et~al.(2018{\natexlab{a}}){Calcutt}, {Fiechter}, {Willis},
  {M{\"u}ller}, {Garrod} et~al.}]{calcutt18b}
{Calcutt} H, {Fiechter} MR, {Willis} ER, {M{\"u}ller} HSP, {Garrod} RT, et~al.
  2018{\natexlab{a}}.
\textit{\aap} 617:A95

\bibitem[{{Calcutt} et~al.(2018{\natexlab{b}}){Calcutt}, {J{\o}rgensen},
  {M{\"u}ller}, {Kristensen}, {Coutens} et~al.}]{calcutt18a}
{Calcutt} H, {J{\o}rgensen} JK, {M{\"u}ller} HSP, {Kristensen} LE, {Coutens} A,
  et~al. 2018{\natexlab{b}}.
\textit{\aap} 616:A90

\bibitem[{{Calcutt} et~al.(2014){Calcutt}, {Viti}, {Codella}, {Beltr{\'a}n},
  {Fontani} \& {Woods}}]{calcutt14}
{Calcutt} H, {Viti} S, {Codella} C, {Beltr{\'a}n} MT, {Fontani} F, {Woods} PM.
  2014.
\textit{\mnras} 443:3157--3173

\bibitem[{{Carpenter} et~al.(2018){Carpenter}, {Iono}, {Testi}, {Whyborn},
  {Wootten} \& {Evans}}]{alma2030}
{Carpenter} J, {Iono} D, {Testi} L, {Whyborn} N, {Wootten} A, {Evans} N. 2018.
{\emph{The ALMA Development Roadmap}
  (https://almaobservatory.org/wp-content/uploads/2018/07/20180712-alma-development-roadmap.pdf)}

\bibitem[{{Caselli} \& {Ceccarelli}(2012)}]{caselli12review}
{Caselli} P, {Ceccarelli} C. 2012.
\textit{\aapr} 20:56

\bibitem[{{Caselli} et~al.(2012){Caselli}, {Keto}, {Bergin}, {Tafalla},
  {Aikawa} et~al.}]{caselli12water}
{Caselli} P, {Keto} E, {Bergin} EA, {Tafalla} M, {Aikawa} Y, et~al. 2012.
\textit{\apjl} 759:L37

\bibitem[{{Caselli} et~al.(2010){Caselli}, {Keto}, {Pagani}, {Aikawa},
  {Y{\i}ld{\i}z} et~al.}]{caselli10}
{Caselli} P, {Keto} E, {Pagani} L, {Aikawa} Y, {Y{\i}ld{\i}z} UA, et~al. 2010.
\textit{\aap} 521:L29

\bibitem[{{Cassen} \& {Moosman}(1981)}]{cassen81}
{Cassen} P, {Moosman} A. 1981.
\textit{Icarus} 48:353--376

\bibitem[{{Caswell}(1996)}]{caswell96}
{Caswell} JL. 1996.
\textit{\mnras} 283:606--612

\bibitem[{{Ceccarelli} et~al.(2014){Ceccarelli}, {Caselli},
  {Bockel{\'e}e-Morvan}, {Mousis}, {Pizzarello} et~al.}]{ceccarellippvi}
{Ceccarelli} C, {Caselli} P, {Bockel{\'e}e-Morvan} D, {Mousis} O, {Pizzarello}
  S, et~al. 2014.
\textit{{Deuterium Fractionation: The Ariadne's Thread from the Precollapse
  Phase to Meteorites and Comets Today}}. In \textit{Protostars and Planets
  VI}, eds. H~{Beuther}, RS~{Klessen}, CP~{Dullemond}, T~{Henning}

\bibitem[{{Cernicharo} et~al.(2019){Cernicharo}, {Gallego},
  {L{\'o}pez-P{\'e}rez}, {Tercero}, {Tanarro} et~al.}]{cernicharo19}
{Cernicharo} J, {Gallego} JD, {L{\'o}pez-P{\'e}rez} JA, {Tercero} F, {Tanarro}
  I, et~al. 2019.
\textit{\aap} 626:A34

\bibitem[{{Cernicharo} et~al.(2016){Cernicharo}, {Kisiel}, {Tercero},
  {Kolesnikov{\'a}}, {Medvedev} et~al.}]{cernicharo16}
{Cernicharo} J, {Kisiel} Z, {Tercero} B, {Kolesnikov{\'a}} L, {Medvedev} IR,
  et~al. 2016.
\textit{\aap} 587:L4

\bibitem[{{Charnley}, {Tielens} \& {Millar}(1992)}]{charnley92}
{Charnley} SB, {Tielens} AGGM, {Millar} TJ. 1992.
\textit{\apjl} 399:L71

\bibitem[{{Chuang} et~al.(2016){Chuang}, {Fedoseev}, {Ioppolo}, {van Dishoeck}
  \& {Linnartz}}]{chuang16}
{Chuang} KJ, {Fedoseev} G, {Ioppolo} S, {van Dishoeck} EF, {Linnartz} H. 2016.
\textit{\mnras} 455:1702--1712

\bibitem[{{Cleeves} et~al.(2014){Cleeves}, {Bergin}, {Alexander}, {Du},
  {Graninger} et~al.}]{cleeves14}
{Cleeves} LI, {Bergin} EA, {Alexander} CMOD, {Du} F, {Graninger} D, et~al.
  2014.
\textit{Science} 345:1590--1593

\bibitem[{{Codella} et~al.(2016){Codella}, {Ceccarelli}, {Bianchi}, {Podio},
  {Bachiller} et~al.}]{codella16svs13}
{Codella} C, {Ceccarelli} C, {Bianchi} E, {Podio} L, {Bachiller} R, et~al.
  2016.
\textit{\mnras} 462:L75--L79

\bibitem[{{Codella} et~al.(2017){Codella}, {Ceccarelli}, {Caselli}, {Balucani},
  {Barone} et~al.}]{codella17}
{Codella} C, {Ceccarelli} C, {Caselli} P, {Balucani} N, {Barone} V, et~al.
  2017.
\textit{\aap} 605:L3

\bibitem[{{Codella} et~al.(2012){Codella}, {Ceccarelli}, {Lefloch}, {Fontani},
  {Busquet} et~al.}]{codella12}
{Codella} C, {Ceccarelli} C, {Lefloch} B, {Fontani} F, {Busquet} G, et~al.
  2012.
\textit{\apjl} 757:L9

\bibitem[{{Codella} et~al.(2010){Codella}, {Lefloch}, {Ceccarelli},
  {Cernicharo}, {Caux} et~al.}]{codella10chess}
{Codella} C, {Lefloch} B, {Ceccarelli} C, {Cernicharo} J, {Caux} E, et~al.
  2010.
\textit{\aap} 518:L112

\bibitem[{{Corby}(2016)}]{corby16}
{Corby} J. 2016.
\textit{{Astrochemistry in The Age of Broadband Radio Astronomy}}.
Ph.D. thesis, University of Virginia

\bibitem[{{Corby} et~al.(2015){Corby}, {Jones}, {Cunningham}, {Menten},
  {Belloche} et~al.}]{corby15}
{Corby} JF, {Jones} PA, {Cunningham} MR, {Menten} KM, {Belloche} A, et~al.
  2015.
\textit{\mnras} 452:3969--3993

\bibitem[{{Cordiner} et~al.(2017){Cordiner}, {Charnley}, {Kisiel}, {McGuire} \&
  {Kuan}}]{cordiner17}
{Cordiner} MA, {Charnley} SB, {Kisiel} Z, {McGuire} BA, {Kuan} YJ. 2017.
\textit{\apj} 850:187

\bibitem[{{Coudert} et~al.(2013){Coudert}, {Drouin}, {Tercero}, {Cernicharo},
  {Guillemin} et~al.}]{coudert13}
{Coudert} LH, {Drouin} BJ, {Tercero} B, {Cernicharo} J, {Guillemin} JC, et~al.
  2013.
\textit{\apj} 779:119

\bibitem[{{Coudert} et~al.(2019){Coudert}, {Margul{\`e}s}, {Vastel},
  {Motiyenko}, {Caux} \& {Guillemin}}]{coudert19}
{Coudert} LH, {Margul{\`e}s} L, {Vastel} C, {Motiyenko} R, {Caux} E,
  {Guillemin} JC. 2019.
\textit{\aap} 624:A70

\bibitem[{{Coutens} et~al.(2014){Coutens}, {J{\o}rgensen}, {Persson}, {van
  Dishoeck}, {Vastel} \& {Taquet}}]{coutens14}
{Coutens} A, {J{\o}rgensen} JK, {Persson} MV, {van Dishoeck} EF, {Vastel} C,
  {Taquet} V. 2014.
\textit{\apjl} 792:L5

\bibitem[{{Coutens} et~al.(2016){Coutens}, {J{\o}rgensen}, {van der Wiel},
  {M{\"u}ller}, {Lykke} et~al.}]{coutens16}
{Coutens} A, {J{\o}rgensen} JK, {van der Wiel} MHD, {M{\"u}ller} HSP, {Lykke}
  JM, et~al. 2016.
\textit{\aap} 590:L6

\bibitem[{{Coutens} et~al.(2019){Coutens}, {Ligterink}, {Loison}, {Wakelam},
  {Calcutt} et~al.}]{coutens19hono}
{Coutens} A, {Ligterink} NFW, {Loison} JC, {Wakelam} V, {Calcutt} H, et~al.
  2019.
\textit{\aap} 623:L13

\bibitem[{{Coutens} et~al.(2015){Coutens}, {Persson}, {J{\o}rgensen},
  {Wampfler} \& {Lykke}}]{coutens15}
{Coutens} A, {Persson} MV, {J{\o}rgensen} JK, {Wampfler} SF, {Lykke} JM. 2015.
\textit{\aap} 576:A5

\bibitem[{{Coutens} et~al.(2012){Coutens}, {Vastel}, {Caux}, {Ceccarelli},
  {Bottinelli} et~al.}]{coutens12}
{Coutens} A, {Vastel} C, {Caux} E, {Ceccarelli} C, {Bottinelli} S, et~al. 2012.
\textit{\aap} 539:A132

\bibitem[{{Coutens} et~al.(2013){Coutens}, {Vastel}, {Cazaux}, {Bottinelli},
  {Caux} et~al.}]{coutens13}
{Coutens} A, {Vastel} C, {Cazaux} S, {Bottinelli} S, {Caux} E, et~al. 2013.
\textit{\aap} 553:A75

\bibitem[{{Crockett} et~al.(2015){Crockett}, {Bergin}, {Neill}, {Favre},
  {Blake} et~al.}]{crockett15}
{Crockett} NR, {Bergin} EA, {Neill} JL, {Favre} C, {Blake} GA, et~al. 2015.
\textit{\apj} 806:239

\bibitem[{{Crockett} et~al.(2014){Crockett}, {Bergin}, {Neill}, {Favre},
  {Schilke} et~al.}]{crockett14}
{Crockett} NR, {Bergin} EA, {Neill} JL, {Favre} C, {Schilke} P, et~al. 2014.
\textit{\apj} 787:112

\bibitem[{{Csengeri} et~al.(2019){Csengeri}, {Belloche}, {Bontemps},
  {Wyrowski}, {Menten} \& {Bouscasse}}]{csengeri19}
{Csengeri} T, {Belloche} A, {Bontemps} S, {Wyrowski} F, {Menten} KM,
  {Bouscasse} L. 2019.
\textit{\aap} 632:A57

\bibitem[{{Csengeri} et~al.(2018){Csengeri}, {Bontemps}, {Wyrowski},
  {Belloche}, {Menten} et~al.}]{csengeri18}
{Csengeri} T, {Bontemps} S, {Wyrowski} F, {Belloche} A, {Menten} KM, et~al.
  2018.
\textit{\aap} 617:A89

\bibitem[{{Cuppen} et~al.(2017){Cuppen}, {Walsh}, {Lamberts}, {Semenov},
  {Garrod} et~al.}]{cuppen17}
{Cuppen} HM, {Walsh} C, {Lamberts} T, {Semenov} D, {Garrod} RT, et~al. 2017.
\textit{\ssr} 212:1--58

\bibitem[{{Dartois} et~al.(2003){Dartois}, {Thi}, {Geballe}, {Deboffle},
  {d'Hendecourt} \& {van Dishoeck}}]{dartois03}
{Dartois} E, {Thi} W, {Geballe} TR, {Deboffle} D, {d'Hendecourt} L, {van
  Dishoeck} E. 2003.
\textit{\aap} 399:1009--1020

\bibitem[{{Degli Esposti} et~al.(2017){Degli Esposti}, {Dore}, {Melosso},
  {Kobayashi}, {Fujita} \& {Ozeki}}]{degliesposti17}
{Degli Esposti} C, {Dore} L, {Melosso} M, {Kobayashi} K, {Fujita} C, {Ozeki} H.
  2017.
\textit{\apjs} 230:26

\bibitem[{{Drozdovskaya} et~al.(2018){Drozdovskaya}, {van Dishoeck},
  {J{\o}rgensen}, {Calmonte}, {van der Wiel} et~al.}]{drozdovskaya18}
{Drozdovskaya} MN, {van Dishoeck} EF, {J{\o}rgensen} JK, {Calmonte} U, {van der
  Wiel} MHD, et~al. 2018.
\textit{\mnras} 476:4949--4964

\bibitem[{{Drozdovskaya} et~al.(2019){Drozdovskaya}, {van Dishoeck}, {Rubin},
  {J{\o}rgensen} \& {Altwegg}}]{drozdovskaya19}
{Drozdovskaya} MN, {van Dishoeck} EF, {Rubin} M, {J{\o}rgensen} JK, {Altwegg}
  K. 2019.
\textit{\mnras} 490:50--79

\bibitem[{{Dulieu} et~al.(2019){Dulieu}, {Nguyen}, {Congiu}, {Baouche} \&
  {Taquet}}]{dulieu19}
{Dulieu} F, {Nguyen} T, {Congiu} E, {Baouche} S, {Taquet} V. 2019.
\textit{\mnras} 484:L119--L123

\bibitem[{{Dunham} et~al.(2014){Dunham}, {Stutz}, {Allen}, {Evans}, {Fischer}
  et~al.}]{dunhamppvi}
{Dunham} MM, {Stutz} AM, {Allen} LE, {Evans} N.~J. I, {Fischer} WJ, et~al.
  2014.
\textit{{The Evolution of Protostars: Insights from Ten Years of Infrared
  Surveys with Spitzer and Herschel}}. In \textit{Protostars and Planets VI},
  eds. H~{Beuther}, RS~{Klessen}, CP~{Dullemond}, T~{Henning}

\bibitem[{{Ehrenfreund} et~al.(1997){Ehrenfreund}, {Boogert}, {Gerakines},
  {Tielens} \& {van Dishoeck}}]{ehrenfreund97}
{Ehrenfreund} P, {Boogert} ACA, {Gerakines} PA, {Tielens} AGGM, {van Dishoeck}
  EF. 1997.
\textit{\aap} 328:649--669

\bibitem[{{El-Abd} et~al.(2019){El-Abd}, {Brogan}, {Hunter}, {Willis}, {Garrod}
  \& {McGuire}}]{elabd19}
{El-Abd} SJ, {Brogan} CL, {Hunter} TR, {Willis} ER, {Garrod} RT, {McGuire} BA.
  2019.
\textit{\apj} 883:129

\bibitem[{{Enrique-Romero} et~al.(2016){Enrique-Romero}, {Rimola}, {Ceccarelli}
  \& {Balucani}}]{enrique-romero16}
{Enrique-Romero} J, {Rimola} A, {Ceccarelli} C, {Balucani} N. 2016.
\textit{\mnras} 459:L6--L10

\bibitem[{Faure, Lique \& Remijan(2018)}]{faure18}
Faure A, Lique F, Remijan AJ. 2018.
\textit{The Journal of Physical Chemistry Letters} 9:3199--3204

\bibitem[{{Faure} et~al.(2014){Faure}, {Remijan}, {Szalewicz} \&
  {Wiesenfeld}}]{faure14}
{Faure} A, {Remijan} AJ, {Szalewicz} K, {Wiesenfeld} L. 2014.
\textit{\apj} 783:72

\bibitem[{{Favre} et~al.(2014){Favre}, {Carvajal}, {Field}, {J{\o}rgensen},
  {Bisschop} et~al.}]{favre14mf}
{Favre} C, {Carvajal} M, {Field} D, {J{\o}rgensen} JK, {Bisschop} SE, et~al.
  2014.
\textit{\apjs} 215:25

\bibitem[{{Favre} et~al.(2017){Favre}, {Pagani}, {Goldsmith}, {Bergin},
  {Carvajal} et~al.}]{favre17}
{Favre} C, {Pagani} L, {Goldsmith} PF, {Bergin} EA, {Carvajal} M, et~al. 2017.
\textit{\aap} 604:L2

\bibitem[{{Fayolle} et~al.(2017){Fayolle}, {{\"O}berg}, {J{\o}rgensen},
  {Altwegg}, {Calcutt} et~al.}]{fayolle17}
{Fayolle} EC, {{\"O}berg} KI, {J{\o}rgensen} JK, {Altwegg} K, {Calcutt} H,
  et~al. 2017.
\textit{Nature Astronomy} 1:703--708

\bibitem[{{Fedoseev} et~al.(2016){Fedoseev}, {Chuang}, {van Dishoeck},
  {Ioppolo} \& {Linnartz}}]{fedoseev16}
{Fedoseev} G, {Chuang} KJ, {van Dishoeck} EF, {Ioppolo} S, {Linnartz} H. 2016.
\textit{\mnras} 460:4297--4309

\bibitem[{{Fedoseev} et~al.(2015){Fedoseev}, {Cuppen}, {Ioppolo}, {Lamberts} \&
  {Linnartz}}]{fedoseev15}
{Fedoseev} G, {Cuppen} HM, {Ioppolo} S, {Lamberts} T, {Linnartz} H. 2015.
\textit{\mnras} 448:1288--1297

\bibitem[{{Feng} et~al.(2015){Feng}, {Beuther}, {Henning}, {Semenov}, {Palau}
  \& {Mills}}]{feng15}
{Feng} S, {Beuther} H, {Henning} T, {Semenov} D, {Palau} A, {Mills} EAC. 2015.
\textit{\aap} 581:A71

\bibitem[{{Fontani} et~al.(2015a){Fontani}, {Busquet}, {Palau}, {Caselli},
  {S{\'a}nchez-Monge} et~al.}]{fontani15a}
{Fontani} F, {Busquet} G, {Palau} A, {Caselli} P, {S{\'a}nchez-Monge} {\'A},
  et~al. 2015a.
\textit{\aap} 575:A87

\bibitem[{{Fontani} et~al.(2014){Fontani}, {Codella}, {Ceccarelli}, {Lefloch},
  {Viti} \& {Benedettini}}]{fontani14}
{Fontani} F, {Codella} C, {Ceccarelli} C, {Lefloch} B, {Viti} S, {Benedettini}
  M. 2014.
\textit{\apjl} 788:L43

\bibitem[{{Friedel} \& {Snyder}(2008)}]{friedel08}
{Friedel} DN, {Snyder} LE. 2008.
\textit{\apj} 672:962--973

\bibitem[{{Frimann} et~al.(2017){Frimann}, {J{\o}rgensen}, {Dunham}, {Bourke},
  {Kristensen} et~al.}]{frimann17}
{Frimann} S, {J{\o}rgensen} JK, {Dunham} MM, {Bourke} TL, {Kristensen} LE,
  et~al. 2017.
\textit{\aap} 602:A120

\bibitem[{{Frimann} et~al.(2016){Frimann}, {J{\o}rgensen}, {Padoan} \&
  {Haugb{\o}lle}}]{frimann16b}
{Frimann} S, {J{\o}rgensen} JK, {Padoan} P, {Haugb{\o}lle} T. 2016.
\textit{\aap} 587:A60

\bibitem[{{Furuya} \& {Aikawa}(2018)}]{furuya18}
{Furuya} K, {Aikawa} Y. 2018.
\textit{\apj} 857:105

\bibitem[{{Furuya} et~al.(2015){Furuya}, {Aikawa}, {Hincelin}, {Hassel},
  {Bergin} et~al.}]{furuya15}
{Furuya} K, {Aikawa} Y, {Hincelin} U, {Hassel} GE, {Bergin} EA, et~al. 2015.
\textit{\aap} 584:A124

\bibitem[{{Furuya} et~al.(2017){Furuya}, {Drozdovskaya}, {Visser}, {van
  Dishoeck}, {Walsh} et~al.}]{furuya17}
{Furuya} K, {Drozdovskaya} MN, {Visser} R, {van Dishoeck} EF, {Walsh} C, et~al.
  2017.
\textit{\aap} 599:A40

\bibitem[{{Furuya}, {van Dishoeck} \& {Aikawa}(2016)}]{furuya16}
{Furuya} K, {van Dishoeck} EF, {Aikawa} Y. 2016.
\textit{\aap} 586:A127

\bibitem[{{Gaches} \& {Offner}(2018)}]{gaches18}
{Gaches} BAL, {Offner} SSR. 2018.
\textit{\apj} 861:87

\bibitem[{{Garrod}(2008)}]{garrod08a}
{Garrod} RT. 2008.
\textit{\aap} 491:239--251

\bibitem[{{Garrod}(2013)}]{garrod13}
{Garrod} RT. 2013.
\textit{\apj} 765:60

\bibitem[{{Garrod}(2019)}]{garrod19}
{Garrod} RT. 2019.
\textit{\apj} 884:69

\bibitem[{{Garrod} et~al.(2017){Garrod}, {Belloche}, {M{\"u}ller} \&
  {Menten}}]{garrod17}
{Garrod} RT, {Belloche} A, {M{\"u}ller} HSP, {Menten} KM. 2017.
\textit{\aap} 601:A48

\bibitem[{{Garrod} \& {Herbst}(2006)}]{garrod06}
{Garrod} RT, {Herbst} E. 2006.
\textit{\aap} 457:927--936

\bibitem[{{Garrod} \& {Pauly}(2011)}]{garrod11}
{Garrod} RT, {Pauly} T. 2011.
\textit{\apj} 735:15

\bibitem[{{Garrod} et~al.(2009){Garrod}, {Vasyunin}, {Semenov}, {Wiebe} \&
  {Henning}}]{garrod09}
{Garrod} RT, {Vasyunin} AI, {Semenov} DA, {Wiebe} DS, {Henning} T. 2009.
\textit{\apjl} 700:L43--L46

\bibitem[{{Garrod}, {Wakelam} \& {Herbst}(2007)}]{garrod07}
{Garrod} RT, {Wakelam} V, {Herbst} E. 2007.
\textit{\aap} 467:1103--1115

\bibitem[{{Garrod}, {Widicus Weaver} \& {Herbst}(2008)}]{garrod08b}
{Garrod} RT, {Widicus Weaver} SL, {Herbst} E. 2008.
\textit{\apj} 682:283--302

\bibitem[{{Garrod} et~al.(2005){Garrod}, {Williams}, {Hartquist}, {Rawlings} \&
  {Viti}}]{garrod05}
{Garrod} RT, {Williams} DA, {Hartquist} TW, {Rawlings} JMC, {Viti} S. 2005.
\textit{\mnras} 356:654--664

\bibitem[{{Gaume} et~al.(1995){Gaume}, {Claussen}, {de Pree}, {Goss} \&
  {Mehringer}}]{gaume95}
{Gaume} RA, {Claussen} MJ, {de Pree} CG, {Goss} WM, {Mehringer} DM. 1995.
\textit{\apj} 449:663

\bibitem[{{Gerakines} et~al.(1999){Gerakines}, {Whittet}, {Ehrenfreund},
  {Boogert}, {Tielens} et~al.}]{gerakines99}
{Gerakines} PA, {Whittet} DCB, {Ehrenfreund} P, {Boogert} ACA, {Tielens} AGGM,
  et~al. 1999.
\textit{\apj} 522:357--377

\bibitem[{{Ginsburg} et~al.(2018){Ginsburg}, {Bally}, {Barnes}, {Bastian},
  {Battersby} et~al.}]{ginsburg18}
{Ginsburg} A, {Bally} J, {Barnes} A, {Bastian} N, {Battersby} C, et~al. 2018.
\textit{\apj} 853:171

\bibitem[{{Goddi} et~al.(2011){Goddi}, {Greenhill}, {Humphreys}, {Chandler} \&
  {Matthews}}]{goddi11}
{Goddi} C, {Greenhill} LJ, {Humphreys} EML, {Chandler} CJ, {Matthews} LD. 2011.
\textit{\apjl} 739:L13

\bibitem[{{Goldsmith} \& {Langer}(1999)}]{goldsmith99}
{Goldsmith} PF, {Langer} WD. 1999.
\textit{\apj} 517:209--225

\bibitem[{{G{\'o}mez} et~al.(2005){G{\'o}mez}, {Rodr{\'I}guez}, {Loinard},
  {Lizano}, {Poveda} \& {Allen}}]{gomez05}
{G{\'o}mez} L, {Rodr{\'I}guez} LF, {Loinard} L, {Lizano} S, {Poveda} A, {Allen}
  C. 2005.
\textit{\apj} 635:1166--1172

\bibitem[{{Guzm{\'a}n} et~al.(2015){Guzm{\'a}n}, {Sanhueza}, {Contreras},
  {Smith}, {Jackson} et~al.}]{guzman15}
{Guzm{\'a}n} AE, {Sanhueza} P, {Contreras} Y, {Smith} HA, {Jackson} JM, et~al.
  2015.
\textit{\apj} 815:130

\bibitem[{{Halfen} et~al.(2006){Halfen}, {Apponi}, {Woolf}, {Polt} \&
  {Ziurys}}]{halfen06}
{Halfen} DT, {Apponi} AJ, {Woolf} N, {Polt} R, {Ziurys} LM. 2006.
\textit{\apj} 639:237--245

\bibitem[{{Halfen}, {Woolf} \& {Ziurys}(2017)}]{halfen17}
{Halfen} DT, {Woolf} NJ, {Ziurys} LM. 2017.
\textit{\apj} 845:158

\bibitem[{{Hamberg} et~al.(2010){Hamberg}, {{\"O}sterdahl}, {Thomas},
  {Zhaunerchyk}, {Vigren} et~al.}]{hamberg10}
{Hamberg} M, {{\"O}sterdahl} F, {Thomas} RD, {Zhaunerchyk} V, {Vigren} E,
  et~al. 2010.
\textit{\aap} 514:A83

\bibitem[{{Harsono} et~al.(2018){Harsono}, {Bjerkeli}, {van der Wiel},
  {Ramsey}, {Maud} et~al.}]{harsono18}
{Harsono} D, {Bjerkeli} P, {van der Wiel} MHD, {Ramsey} JP, {Maud} LT, et~al.
  2018.
\textit{Nature Astronomy} 2:646--651

\bibitem[{{Harsono} et~al.(2014){Harsono}, {J{\o}rgensen}, {van Dishoeck},
  {Hogerheijde}, {Bruderer} et~al.}]{harsono14}
{Harsono} D, {J{\o}rgensen} JK, {van Dishoeck} EF, {Hogerheijde} MR, {Bruderer}
  S, et~al. 2014.
\textit{\aap} 562:A77

\bibitem[{{Hasegawa} \& {Herbst}(1993)}]{hasegawa93}
{Hasegawa} TI, {Herbst} E. 1993.
\textit{\mnras} 263:589

\bibitem[{{Haupa}, {Tarczay} \& {Lee}(2019)}]{haupa19}
{Haupa} KA, {Tarczay} G, {Lee} YP. 2019.
\textit{J. Am. Chem. Soc.} 141:11614--11620

\bibitem[{{Heays} et~al.(2014){Heays}, {Visser}, {Gredel}, {Ubachs}, {Lewis}
  et~al.}]{heays14}
{Heays} AN, {Visser} R, {Gredel} R, {Ubachs} W, {Lewis} BR, et~al. 2014.
\textit{\aap} 562:A61

\bibitem[{{Henderson} \& {Gudipati}(2015)}]{henderson15}
{Henderson} BL, {Gudipati} MS. 2015.
\textit{\apj} 800:66

\bibitem[{{Henkel} et~al.(1987){Henkel}, {Jacq}, {Mauersberger}, {Menten} \&
  {Steppe}}]{henkel87}
{Henkel} C, {Jacq} T, {Mauersberger} R, {Menten} KM, {Steppe} H. 1987.
\textit{\aap} 188:L1--L4

\bibitem[{{Herbst} \& {van Dishoeck}(2009)}]{herbst09}
{Herbst} E, {van Dishoeck} EF. 2009.
\textit{\araa} 47:427--480

\bibitem[{{Herczeg} et~al.(2012){Herczeg}, {Karska}, {Bruderer}, {Kristensen},
  {van Dishoeck} et~al.}]{herczeg12}
{Herczeg} GJ, {Karska} A, {Bruderer} S, {Kristensen} LE, {van Dishoeck} EF,
  et~al. 2012.
\textit{\aap} 540:A84

\bibitem[{{Hern{\'a}ndez-Hern{\'a}ndez}
  et~al.(2014){Hern{\'a}ndez-Hern{\'a}ndez}, {Zapata}, {Kurtz} \&
  {Garay}}]{hernandez-hernandez14}
{Hern{\'a}ndez-Hern{\'a}ndez} V, {Zapata} L, {Kurtz} S, {Garay} G. 2014.
\textit{\apj} 786:38

\bibitem[{{Hollis}(2005)}]{hollis05}
{Hollis} JM. 2005.
\textit{{Complex Molecules and the GBT: Is Isomerism the Key?}} In
  \textit{Astrochemistry: Recent Successes and Current Challenges}, eds.
  DC~{Lis}, GA~{Blake}, E~{Herbst}, vol. 231 of \textit{IAU Symposium}

\bibitem[{{Hollis} et~al.(2004b){Hollis}, {Jewell}, {Lovas} \&
  {Remijan}}]{hollis04b}
{Hollis} JM, {Jewell} PR, {Lovas} FJ, {Remijan} A. 2004b.
\textit{\apjl} 613:L45--L48

\bibitem[{{Hollis} et~al.(2004a){Hollis}, {Jewell}, {Lovas}, {Remijan} \&
  {M{\o}llendal}}]{hollis04a}
{Hollis} JM, {Jewell} PR, {Lovas} FJ, {Remijan} A, {M{\o}llendal} H. 2004a.
\textit{\apj} 610:L21--L24

\bibitem[{{Hollis} et~al.(2006b){Hollis}, {Lovas}, {Remijan}, {Jewell},
  {Ilyushin} \& {Kleiner}}]{hollis06b}
{Hollis} JM, {Lovas} FJ, {Remijan} AJ, {Jewell} PR, {Ilyushin} VV, {Kleiner} I.
  2006b.
\textit{\apjl} 643:L25--L28

\bibitem[{{Hollis} et~al.(2006a){Hollis}, {Remijan}, {Jewell} \&
  {Lovas}}]{hollis06a}
{Hollis} JM, {Remijan} AJ, {Jewell} PR, {Lovas} FJ. 2006a.
\textit{\apj} 642:933--939

\bibitem[{{Hsieh} et~al.(2018){Hsieh}, {Murillo}, {Belloche}, {Hirano}, {Walsh}
  et~al.}]{hsieh18}
{Hsieh} TH, {Murillo} NM, {Belloche} A, {Hirano} N, {Walsh} C, et~al. 2018.
\textit{\apj} 854:15

\bibitem[{{Hsieh} et~al.(2019){Hsieh}, {Murillo}, {Belloche}, {Hirano}, {Walsh}
  et~al.}]{hsieh19}
{Hsieh} TH, {Murillo} NM, {Belloche} A, {Hirano} N, {Walsh} C, et~al. 2019.
\textit{\apj} 884:149

\bibitem[{{Indriolo} et~al.(2015){Indriolo}, {Neufeld}, {Gerin}, {Schilke},
  {Benz} et~al.}]{indriolo15}
{Indriolo} N, {Neufeld} DA, {Gerin} M, {Schilke} P, {Benz} AO, et~al. 2015.
\textit{\apj} 800:40

\bibitem[{{Jacobsen} et~al.(2019){Jacobsen}, {J{\o}rgensen}, {Di Francesco},
  {Evans}, {Choi} \& {Lee}}]{jacobsen19}
{Jacobsen} SK, {J{\o}rgensen} JK, {Di Francesco} J, {Evans} NJI, {Choi} M,
  {Lee} JE. 2019.
\textit{\aap} 629

\bibitem[{{Jacq} et~al.(1999){Jacq}, {Baudry}, {Walmsley} \&
  {Caselli}}]{jacq99}
{Jacq} T, {Baudry} A, {Walmsley} CM, {Caselli} P. 1999.
\textit{\aap} 347:957--966

\bibitem[{{Jacq} et~al.(1993){Jacq}, {Walmsley}, {Mauersberger}, {Anderson},
  {Herbst} \& {De Lucia}}]{jacq93}
{Jacq} T, {Walmsley} CM, {Mauersberger} R, {Anderson} T, {Herbst} E, {De Lucia}
  FC. 1993.
\textit{\aap} 271:276

\bibitem[{{Jensen} et~al.(2019){Jensen}, {J{\o}rgensen}, {Kristensen},
  {Furuya}, {Coutens} et~al.}]{jensen19}
{Jensen} SS, {J{\o}rgensen} JK, {Kristensen} LE, {Furuya} K, {Coutens} A,
  et~al. 2019.
\textit{\aap} 631:A25

\bibitem[{{Jim{\'e}nez-Serra} et~al.(2016){Jim{\'e}nez-Serra}, {Vasyunin},
  {Caselli}, {Marcelino}, {Billot} et~al.}]{jimenezserra16}
{Jim{\'e}nez-Serra} I, {Vasyunin} AI, {Caselli} P, {Marcelino} N, {Billot} N,
  et~al. 2016.
\textit{\apjl} 830:L6

\bibitem[{{Jin} \& {Garrod}(2020)}]{jin20}
{Jin} M, {Garrod} RT. 2020.
\textit{\apj} {submitted}

\bibitem[{{Jones} et~al.(2008){Jones}, {Burton}, {Cunningham}, {Menten},
  {Schilke} et~al.}]{jones08}
{Jones} PA, {Burton} MG, {Cunningham} MR, {Menten} KM, {Schilke} P, et~al.
  2008.
\textit{\mnras} 386:117--137

\bibitem[{{Jones} et~al.(2011){Jones}, {Burton}, {Tothill} \&
  {Cunningham}}]{jones11}
{Jones} PA, {Burton} MG, {Tothill} NFH, {Cunningham} MR. 2011.
\textit{\mnras} 411:2293--2310

\bibitem[{{J{\o}rgensen} et~al.(2012){J{\o}rgensen}, {Favre}, {Bisschop},
  {Bourke}, {van Dishoeck} \& {Schmalzl}}]{jorgensen12}
{J{\o}rgensen} JK, {Favre} C, {Bisschop} SE, {Bourke} TL, {van Dishoeck} EF,
  {Schmalzl} M. 2012.
\textit{\apjl} 757:L4

\bibitem[{{J{\o}rgensen} et~al.(2006){J{\o}rgensen}, {Johnstone}, {van
  Dishoeck} \& {Doty}}]{jorgensen06}
{J{\o}rgensen} JK, {Johnstone} D, {van Dishoeck} EF, {Doty} SD. 2006.
\textit{\aap} 449:609--619

\bibitem[{{J{\o}rgensen} et~al.(2018){J{\o}rgensen}, {M{\"u}ller}, {Calcutt},
  {Coutens}, {Drozdovskaya} et~al.}]{jorgensen18}
{J{\o}rgensen} JK, {M{\"u}ller} HSP, {Calcutt} H, {Coutens} A, {Drozdovskaya}
  MN, et~al. 2018.
\textit{\aap} 620:A170

\bibitem[{{J{\o}rgensen} et~al.(2016){J{\o}rgensen}, {van der Wiel}, {Coutens},
  {Lykke}, {M{\"u}ller} et~al.}]{jorgensen16}
{J{\o}rgensen} JK, {van der Wiel} MHD, {Coutens} A, {Lykke} JM, {M{\"u}ller}
  HSP, et~al. 2016.
\textit{\aap} 595:A117

\bibitem[{{J{\o}rgensen} \& {van Dishoeck}(2010{\natexlab{a}})}]{iras4b_hdo}
{J{\o}rgensen} JK, {van Dishoeck} EF. 2010{\natexlab{a}}.
\textit{\apjl} 725:L172--L175

\bibitem[{{J{\o}rgensen} \& {van Dishoeck}(2010{\natexlab{b}})}]{iras4b_h2o}
{J{\o}rgensen} JK, {van Dishoeck} EF. 2010{\natexlab{b}}.
\textit{\apjl} 710:L72--L76

\bibitem[{{J{\o}rgensen} et~al.(2009){J{\o}rgensen}, {van Dishoeck}, {Visser},
  {Bourke}, {Wilner} et~al.}]{jorgensen09}
{J{\o}rgensen} JK, {van Dishoeck} EF, {Visser} R, {Bourke} TL, {Wilner} DJ,
  et~al. 2009.
\textit{\aap} 507:861--879

\bibitem[{{J{\o}rgensen} et~al.(2013){J{\o}rgensen}, {Visser}, {Sakai},
  {Bergin}, {Brinch} et~al.}]{jorgensen13}
{J{\o}rgensen} JK, {Visser} R, {Sakai} N, {Bergin} EA, {Brinch} C, et~al. 2013.
\textit{\apj} 779:L22

\bibitem[{{J{\o}rgensen} et~al.(2015){J{\o}rgensen}, {Visser}, {Williams} \&
  {Bergin}}]{jorgensen15}
{J{\o}rgensen} JK, {Visser} R, {Williams} JP, {Bergin} EA. 2015.
\textit{\aap} 579:A23

\bibitem[{{Kahane} et~al.(2013){Kahane}, {Ceccarelli}, {Faure} \&
  {Caux}}]{kahane13}
{Kahane} C, {Ceccarelli} C, {Faure} A, {Caux} E. 2013.
\textit{\apjl} 763:L38

\bibitem[{{Kaifu} et~al.(1974){Kaifu}, {Morimoto}, {Nagane}, {Akabane},
  {Iguchi} \& {Takagi}}]{kaifu74}
{Kaifu} N, {Morimoto} M, {Nagane} K, {Akabane} K, {Iguchi} T, {Takagi} K. 1974.
\textit{\apjl} 191:L135--L137

\bibitem[{{Kalenskii} \& {Johansson}(2010)}]{kalenskii10}
{Kalenskii} SV, {Johansson} LEB. 2010.
\textit{Astronomy Reports} 54:1084--1104

\bibitem[{{Kalv{\={a}}ns}(2018)}]{kalvans18}
{Kalv{\={a}}ns} J. 2018.
\textit{\mnras} 478:2753--2765

\bibitem[{{Kim} et~al.(2012){Kim}, {Evans}, {Dunham}, {Lee} \&
  {Pontoppidan}}]{kim12}
{Kim} HJ, {Evans} II NJ, {Dunham} MM, {Lee} JE, {Pontoppidan} KM. 2012.
\textit{\apj} 758:38

\bibitem[{{Kolesnikov{\'a}} et~al.(2018){Kolesnikov{\'a}}, {Tercero}, {Alonso},
  {Guillemin}, {Cernicharo} \& {Alonso}}]{kolesnikova18}
{Kolesnikov{\'a}} L, {Tercero} B, {Alonso} ER, {Guillemin} JC, {Cernicharo} J,
  {Alonso} JL. 2018.
\textit{\aap} 609:A24

\bibitem[{{Krim} et~al.(2018){Krim}, {Jonusas}, {Guillemin}, {Y\'{a}\~{n}ez} \&
  {Lamsabhi}}]{krim18}
{Krim} L, {Jonusas} M, {Guillemin} JC, {Y\'{a}\~{n}ez} M, {Lamsabhi} AM. 2018.
\textit{Phys. Chem. Chem. Phys.} 20:19971

\bibitem[{{Le Petit} et~al.(2016){Le Petit}, {Ruaud}, {Bron}, {Godard},
  {Roueff} et~al.}]{lepetit16}
{Le Petit} F, {Ruaud} M, {Bron} E, {Godard} B, {Roueff} E, et~al. 2016.
\textit{\aap} 585:A105

\bibitem[{{Lee} et~al.(2019{\natexlab{a}}){Lee}, {Codella}, {Li} \&
  {Liu}}]{lee19hh212}
{Lee} CF, {Codella} C, {Li} ZY, {Liu} SY. 2019{\natexlab{a}}.
\textit{\apj} 876:63

\bibitem[{{Lee} et~al.(2017){Lee}, {Li}, {Ho}, {Hirano}, {Zhang} \&
  {Shang}}]{lee17coms}
{Lee} CF, {Li} ZY, {Ho} PTP, {Hirano} N, {Zhang} Q, {Shang} H. 2017.
\textit{\apj} 843:27

\bibitem[{{Lee}(2007)}]{lee07}
{Lee} JE. 2007.
\textit{Journal of Korean Astronomical Society} 40:83--89

\bibitem[{{Lee} et~al.(2019{\natexlab{b}}){Lee}, {Lee}, {Baek}, {Aikawa},
  {Cieza} et~al.}]{lee19v883ori}
{Lee} JE, {Lee} S, {Baek} G, {Aikawa} Y, {Cieza} L, et~al. 2019{\natexlab{b}}.
\textit{Nature Astronomy} 3:314--319

\bibitem[{{Lefloch} et~al.(2017){Lefloch}, {Ceccarelli}, {Codella}, {Favre},
  {Podio} et~al.}]{lefloch17}
{Lefloch} B, {Ceccarelli} C, {Codella} C, {Favre} C, {Podio} L, et~al. 2017.
\textit{\mnras} 469:L73--L77

\bibitem[{{Li} et~al.(2017){Li}, {Shen}, {Wang}, {Chen}, {Li} et~al.}]{li17}
{Li} J, {Shen} Z, {Wang} J, {Chen} X, {Li} D, et~al. 2017.
\textit{\apj} 849:115

\bibitem[{{Ligterink} et~al.(2018{\natexlab{a}}){Ligterink}, {Calcutt},
  {Coutens}, {Kristensen}, {Bourke} et~al.}]{ligterink18amines}
{Ligterink} NFW, {Calcutt} H, {Coutens} A, {Kristensen} LE, {Bourke} TL, et~al.
  2018{\natexlab{a}}.
\textit{\aap} 619:A28

\bibitem[{{Ligterink} et~al.(2017){Ligterink}, {Coutens}, {Kofman},
  {M{\"u}ller}, {Garrod} et~al.}]{ligterink17}
{Ligterink} NFW, {Coutens} A, {Kofman} V, {M{\"u}ller} HSP, {Garrod} RT, et~al.
  2017.
\textit{\mnras} 469:2219--2229

\bibitem[{{Ligterink}, {Tenenbaum} \& {van Dishoeck}(2015)}]{ligterink15}
{Ligterink} NFW, {Tenenbaum} ED, {van Dishoeck} EF. 2015.
\textit{\aap} 576:A35

\bibitem[{{Ligterink} et~al.(2018{\natexlab{b}}){Ligterink}, {Terwisscha van
  Scheltinga}, {Taquet}, {J{\o}rgensen}, {Cazaux} et~al.}]{ligterink18amides}
{Ligterink} NFW, {Terwisscha van Scheltinga} J, {Taquet} V, {J{\o}rgensen} JK,
  {Cazaux} S, et~al. 2018{\natexlab{b}}.
\textit{\mnras} 480:3628--3643

\bibitem[{{Lindberg} et~al.(2014){Lindberg}, {J{\o}rgensen}, {Brinch},
  {Haugb{\o}lle}, {Bergin} et~al.}]{lindberg14alma}
{Lindberg} JE, {J{\o}rgensen} JK, {Brinch} C, {Haugb{\o}lle} T, {Bergin} EA,
  et~al. 2014.
\textit{\aap} 566:A74

\bibitem[{{Linsky}(2003)}]{linsky03}
{Linsky} JL. 2003.
\textit{Space Science Reviews} 106:49--60

\bibitem[{{Lis} et~al.(2013){Lis}, {Biver}, {Bockel{\'e}e-Morvan}, {Hartogh},
  {Bergin} et~al.}]{lis13}
{Lis} DC, {Biver} N, {Bockel{\'e}e-Morvan} D, {Hartogh} P, {Bergin} EA, et~al.
  2013.
\textit{\apjl} 774:L3

\bibitem[{{Liu} et~al.(2011){Liu}, {Parise}, {Kristensen}, {Visser}, {van
  Dishoeck} \& {G{\"u}sten}}]{liu11}
{Liu} F, {Parise} B, {Kristensen} L, {Visser} R, {van Dishoeck} EF,
  {G{\"u}sten} R. 2011.
\textit{\aap} 527:A19

\bibitem[{{Longmore} et~al.(2012){Longmore}, {Rathborne}, {Bastian}, {Alves},
  {Ascenso} et~al.}]{longmore12}
{Longmore} SN, {Rathborne} J, {Bastian} N, {Alves} J, {Ascenso} J, et~al. 2012.
\textit{\apj} 746:117

\bibitem[{{Loomis} et~al.(2013){Loomis}, {Zaleski}, {Steber}, {Neill}, {Muckle}
  et~al.}]{loomis13}
{Loomis} RA, {Zaleski} DP, {Steber} AL, {Neill} JL, {Muckle} MT, et~al. 2013.
\textit{\apjl} 765:L9

\bibitem[{{L{\'o}pez-Sepulcre} et~al.(2015){L{\'o}pez-Sepulcre}, {Jaber},
  {Mendoza}, {Lefloch}, {Ceccarelli} et~al.}]{lopezsepulcre15}
{L{\'o}pez-Sepulcre} A, {Jaber} AA, {Mendoza} E, {Lefloch} B, {Ceccarelli} C,
  et~al. 2015.
\textit{\mnras} 449:2438--2458

\bibitem[{{L{\'o}pez-Sepulcre} et~al.(2017){L{\'o}pez-Sepulcre}, {Sakai},
  {Neri}, {Imai}, {Oya} et~al.}]{lopezsepulcre17}
{L{\'o}pez-Sepulcre} A, {Sakai} N, {Neri} R, {Imai} M, {Oya} Y, et~al. 2017.
\textit{\aap} 606:A121

\bibitem[{{Lu}, {Chang} \& {Aikawa}(2018)}]{lu18}
{Lu} Y, {Chang} Q, {Aikawa} Y. 2018.
\textit{\apj} 869:165

\bibitem[{{Lubowich} et~al.(2000){Lubowich}, {Pasachoff}, {Balonek}, {Millar},
  {Tremonti} et~al.}]{lubowich00}
{Lubowich} DA, {Pasachoff} JM, {Balonek} TJ, {Millar} TJ, {Tremonti} C, et~al.
  2000.
\textit{\nat} 405:1025--1027

\bibitem[{{Luhman} et~al.(2017){Luhman}, {Robberto}, {Tan}, {Andersen}, {Giulia
  Ubeira Gabellini} et~al.}]{luhman17}
{Luhman} KL, {Robberto} M, {Tan} JC, {Andersen} M, {Giulia Ubeira Gabellini} M,
  et~al. 2017.
\textit{\apjl} 838:L3

\bibitem[{{Lykke} et~al.(2017){Lykke}, {Coutens}, {J{\o}rgensen}, {van der
  Wiel}, {Garrod} et~al.}]{lykke17}
{Lykke} JM, {Coutens} A, {J{\o}rgensen} JK, {van der Wiel} MHD, {Garrod} RT,
  et~al. 2017.
\textit{\aap} 597:A53

\bibitem[{{Lyons} \& {Young}(2005)}]{lyons05}
{Lyons} JR, {Young} ED. 2005.
\textit{\nat} 435:317--320

\bibitem[{{Mangum} \& {Shirley}(2015)}]{mangum15}
{Mangum} JG, {Shirley} YL. 2015.
\textit{\pasp} 127:266

\bibitem[{{Manigand} et~al.(2019){Manigand}, {Calcutt}, {J{\o}rgensen},
  {Taquet}, {M{\"u}ller} et~al.}]{manigand19}
{Manigand} S, {Calcutt} H, {J{\o}rgensen} JK, {Taquet} V, {M{\"u}ller} HSP,
  et~al. 2019.
\textit{\aap} 623:A69

\bibitem[{{Manigand} et~al.(2020){Manigand}, {J{\o}rgensen}, {Calcutt},
  {M{\"u}ller}, {Ligterink} et~al.}]{manigand19iras16293A}
{Manigand} S, {J{\o}rgensen} JK, {Calcutt} H, {M{\"u}ller} HSP, {Ligterink}
  NFW, et~al. 2020.
\textit{\aap} 635:A48

\bibitem[{{Margul{\`e}s} et~al.(2016){Margul{\`e}s}, {Belloche}, {M{\"u}ller},
  {Motiyenko}, {Guillemin} et~al.}]{margules16}
{Margul{\`e}s} L, {Belloche} A, {M{\"u}ller} HSP, {Motiyenko} RA, {Guillemin}
  JC, et~al. 2016.
\textit{\aap} 590:A93

\bibitem[{{Margul{\`e}s} et~al.(2010){Margul{\`e}s}, {Huet}, {Demaison},
  {Carvajal}, {Kleiner} et~al.}]{margules10}
{Margul{\`e}s} L, {Huet} TR, {Demaison} J, {Carvajal} M, {Kleiner} I, et~al.
  2010.
\textit{\apj} 714:1120--1132

\bibitem[{{Margul{\`e}s} et~al.(2017){Margul{\`e}s}, {McGuire}, {Senent},
  {Motiyenko}, {Remijan} \& {Guillemin}}]{margules17}
{Margul{\`e}s} L, {McGuire} BA, {Senent} ML, {Motiyenko} RA, {Remijan} A,
  {Guillemin} JC. 2017.
\textit{\aap} 601:A50

\bibitem[{{Mart{\'\i}n-Dom{\'e}nech}, {Mu{\~n}oz Caro} \&
  {Cruz-D{\'\i}az}(2016)}]{martin-domenech16}
{Mart{\'\i}n-Dom{\'e}nech} R, {Mu{\~n}oz Caro} GM, {Cruz-D{\'\i}az} GA. 2016.
\textit{\aap} 589:A107

\bibitem[{{Martin-Drumel} et~al.(2019){Martin-Drumel}, {Lee}, {Belloche},
  {Zingsheim}, {Thorwirth} et~al.}]{martinDrumel19}
{Martin-Drumel} MA, {Lee} KLK, {Belloche} A, {Zingsheim} O, {Thorwirth} S,
  et~al. 2019.
\textit{\aap} 623:A167

\bibitem[{{Mauersberger} et~al.(1988){Mauersberger}, {Henkel}, {Jacq} \&
  {Walmsley}}]{mauersberger88}
{Mauersberger} R, {Henkel} C, {Jacq} T, {Walmsley} CM. 1988.
\textit{\aap} 194:L1--L4

\bibitem[{{Mauersberger} et~al.(1991){Mauersberger}, {Henkel}, {Walmsley},
  {Sage} \& {Wiklind}}]{mauersberger91}
{Mauersberger} R, {Henkel} C, {Walmsley} CM, {Sage} LJ, {Wiklind} T. 1991.
\textit{\aap} 247:307

\bibitem[{{McGuire} et~al.(2018){McGuire}, {Burkhardt}, {Kalenskii},
  {Shingledecker}, {Remijan} et~al.}]{mcguire18}
{McGuire} BA, {Burkhardt} AM, {Kalenskii} S, {Shingledecker} CN, {Remijan} AJ,
  et~al. 2018.
\textit{Science} 359:202--205

\bibitem[{{McGuire} et~al.(2015){McGuire}, {Carroll}, {Dollhopf}, {Crockett},
  {Corby} et~al.}]{mcguire15}
{McGuire} BA, {Carroll} PB, {Dollhopf} NM, {Crockett} NR, {Corby} JF, et~al.
  2015.
\textit{\apj} 812:76

\bibitem[{{McGuire} et~al.(2016){McGuire}, {Carroll}, {Loomis}, {Finneran},
  {Jewell} et~al.}]{mcguire16}
{McGuire} BA, {Carroll} PB, {Loomis} RA, {Finneran} IA, {Jewell} PR, et~al.
  2016.
\textit{Science} 352:1449--1452

\bibitem[{{McGuire} et~al.(2017){McGuire}, {Shingledecker}, {Willis},
  {Burkhardt}, {El-Abd} et~al.}]{mcguire17}
{McGuire} BA, {Shingledecker} CN, {Willis} ER, {Burkhardt} AM, {El-Abd} S,
  et~al. 2017.
\textit{\apj} 851:L46

\bibitem[{{Mendoza} et~al.(2014){Mendoza}, {Lefloch}, {L{\'o}pez-Sepulcre},
  {Ceccarelli}, {Codella} et~al.}]{mendoza14}
{Mendoza} E, {Lefloch} B, {L{\'o}pez-Sepulcre} A, {Ceccarelli} C, {Codella} C,
  et~al. 2014.
\textit{\mnras} 445:151--161

\bibitem[{{Minier} et~al.(2003){Minier}, {Ellingsen}, {Norris} \&
  {Booth}}]{minier03}
{Minier} V, {Ellingsen} SP, {Norris} RP, {Booth} RS. 2003.
\textit{\aap} 403:1095--1100

\bibitem[{{Molet} et~al.(2019){Molet}, {Brouillet}, {Nony}, {Gusdorf}, {Motte}
  et~al.}]{molet19}
{Molet} J, {Brouillet} N, {Nony} T, {Gusdorf} A, {Motte} F, et~al. 2019.
\textit{\aap} 626:A132

\bibitem[{{M{\"o}ller}, {Endres} \& {Schilke}(2017)}]{moller17}
{M{\"o}ller} T, {Endres} C, {Schilke} P. 2017.
\textit{\aap} 598:A7

\bibitem[{{Motiyenko} et~al.(2019){Motiyenko}, {Armieieva}, {Margul{\`e}s},
  {Alekseev} \& {Guillemin}}]{motiyenko19}
{Motiyenko} RA, {Armieieva} IA, {Margul{\`e}s} L, {Alekseev} EA, {Guillemin}
  JC. 2019.
\textit{\aap} 623:A162

\bibitem[{{Mottram} et~al.(2013){Mottram}, {van Dishoeck}, {Schmalzl},
  {Kristensen}, {Visser} et~al.}]{mottram13}
{Mottram} JC, {van Dishoeck} EF, {Schmalzl} M, {Kristensen} LE, {Visser} R,
  et~al. 2013.
\textit{\aap} 558:A126

\bibitem[{{M{\"u}ller} et~al.(2016a){M{\"u}ller}, {Belloche}, {Xu}, {Lees},
  {Garrod} et~al.}]{mueller16a}
{M{\"u}ller} HSP, {Belloche} A, {Xu} LH, {Lees} RM, {Garrod} RT, et~al. 2016a.
\textit{\aap} 587:A92

\bibitem[{{M{\"u}ller} et~al.(2005){M{\"u}ller}, {Schl{\"o}der}, {Stutzki} \&
  {Winnewisser}}]{cdms2}
{M{\"u}ller} HSP, {Schl{\"o}der} F, {Stutzki} J, {Winnewisser} G. 2005.
\textit{Journal of Molecular Structure} 742:215--227

\bibitem[{{M{\"u}ller} et~al.(2001){M{\"u}ller}, {Thorwirth}, {Roth} \&
  {Winnewisser}}]{cdms1}
{M{\"u}ller} HSP, {Thorwirth} S, {Roth} DA, {Winnewisser} G. 2001.
\textit{\aap} 370:L49--L52

\bibitem[{{M{\"u}ller} et~al.(2016b){M{\"u}ller}, {Walters}, {Wehres},
  {Belloche}, {Wilkins} et~al.}]{mueller16b}
{M{\"u}ller} HSP, {Walters} A, {Wehres} N, {Belloche} A, {Wilkins} OH, et~al.
  2016b.
\textit{\aap} 595:A87

\bibitem[{{Muller} et~al.(2013){Muller}, {Beelen}, {Black}, {Curran},
  {Horellou} et~al.}]{muller13}
{Muller} S, {Beelen} A, {Black} JH, {Curran} SJ, {Horellou} C, et~al. 2013.
\textit{\aap} 551:A109

\bibitem[{{Muller} et~al.(2011){Muller}, {Beelen}, {Gu{\'e}lin}, {Aalto},
  {Black} et~al.}]{muller11}
{Muller} S, {Beelen} A, {Gu{\'e}lin} M, {Aalto} S, {Black} JH, et~al. 2011.
\textit{\aap} 535:A103

\bibitem[{{Muller} et~al.(2014){Muller}, {Combes}, {Gu{\'e}lin}, {G{\'e}rin},
  {Aalto} et~al.}]{muller14}
{Muller} S, {Combes} F, {Gu{\'e}lin} M, {G{\'e}rin} M, {Aalto} S, et~al. 2014.
\textit{\aap} 566:A112

\bibitem[{{Murillo} et~al.(2013){Murillo}, {Lai}, {Bruderer}, {Harsono} \& {van
  Dishoeck}}]{murillo13}
{Murillo} NM, {Lai} SP, {Bruderer} S, {Harsono} D, {van Dishoeck} EF. 2013.
\textit{\aap} 560:A103

\bibitem[{{Neill} et~al.(2014){Neill}, {Bergin}, {Lis}, {Schilke}, {Crockett}
  et~al.}]{neill14}
{Neill} JL, {Bergin} EA, {Lis} DC, {Schilke} P, {Crockett} NR, et~al. 2014.
\textit{\apj} 789:8

\bibitem[{{Neill} et~al.(2013){Neill}, {Crockett}, {Bergin}, {Pearson} \&
  {Xu}}]{neill13}
{Neill} JL, {Crockett} NR, {Bergin} EA, {Pearson} JC, {Xu} LH. 2013.
\textit{\apj} 777:85

\bibitem[{{Neill} et~al.(2012){Neill}, {Muckle}, {Zaleski}, {Steber}, {Pate}
  et~al.}]{neill12}
{Neill} JL, {Muckle} MT, {Zaleski} DP, {Steber} AL, {Pate} BH, et~al. 2012.
\textit{\apj} 755:153

\bibitem[{{Neill} et~al.(2011){Neill}, {Steber}, {Muckle}, {Zaleski},
  {Lattanzi} et~al.}]{neill11}
{Neill} JL, {Steber} AL, {Muckle} MT, {Zaleski} DP, {Lattanzi} V, et~al. 2011.
\textit{Journal of Physical Chemistry A} 115:6472--6480

\bibitem[{{Neufeld} \& {Hollenbach}(1994)}]{neufeld94}
{Neufeld} DA, {Hollenbach} DJ. 1994.
\textit{\apj} 428:170--185

\bibitem[{{Noble} et~al.(2015){Noble}, {Theule}, {Congiu}, {Dulieu}, {Bonnin}
  et~al.}]{noble15}
{Noble} JA, {Theule} P, {Congiu} E, {Dulieu} F, {Bonnin} M, et~al. 2015.
\textit{\aap} 576:A91

\bibitem[{{{\"O}berg}(2016)}]{oberg16}
{{\"O}berg} KI. 2016.
\textit{Chemical Reviews} 116:9631--9663

\bibitem[{{{\"O}berg} et~al.(2013){{\"O}berg}, {Boamah}, {Fayolle}, {Garrod},
  {Cyganowski} \& {van der Tak}}]{oberg13}
{{\"O}berg} KI, {Boamah} MD, {Fayolle} EC, {Garrod} RT, {Cyganowski} CJ, {van
  der Tak} F. 2013.
\textit{\apj} 771:95

\bibitem[{{{\"O}berg} et~al.(2010){{\"O}berg}, {Bottinelli}, {J{\o}rgensen} \&
  {van Dishoeck}}]{oberg10}
{{\"O}berg} KI, {Bottinelli} S, {J{\o}rgensen} JK, {van Dishoeck} EF. 2010.
\textit{\apj} 716:825--834

\bibitem[{{{\"O}berg} et~al.(2014){{\"O}berg}, {Fayolle}, {Reiter} \&
  {Cyganowski}}]{oberg14}
{{\"O}berg} KI, {Fayolle} EC, {Reiter} JB, {Cyganowski} C. 2014.
\textit{Faraday Discussions} 168:81--101

\bibitem[{{{\"O}berg} et~al.(2009){{\"O}berg}, {Garrod}, {van Dishoeck} \&
  {Linnartz}}]{oberg09}
{{\"O}berg} KI, {Garrod} RT, {van Dishoeck} EF, {Linnartz} H. 2009.
\textit{\aap} 504:891--913

\bibitem[{{{\"O}berg} et~al.(2011){{\"O}berg}, {van der Marel}, {Kristensen} \&
  {van Dishoeck}}]{oberg11serpens}
{{\"O}berg} KI, {van der Marel} N, {Kristensen} LE, {van Dishoeck} EF. 2011.
\textit{\apj} 740:14

\bibitem[{{Ohishi} et~al.(2019){Ohishi}, {Suzuki}, {Hirota}, {Saito} \&
  {Kaifu}}]{ohishi19}
{Ohishi} M, {Suzuki} T, {Hirota} T, {Saito} M, {Kaifu} N. 2019.
\textit{\pasj} 71:86

\bibitem[{{Ordu} et~al.(2012){Ordu}, {M{\"u}ller}, {Walters}, {Nu{\~n}ez},
  {Lewen} et~al.}]{ordu12}
{Ordu} MH, {M{\"u}ller} HSP, {Walters} A, {Nu{\~n}ez} M, {Lewen} F, et~al.
  2012.
\textit{\aap} 541:A121

\bibitem[{{Ordu} et~al.(2019){Ordu}, {Zingsheim}, {Belloche}, {Lewen}, {Garrod}
  et~al.}]{ordu19}
{Ordu} MH, {Zingsheim} O, {Belloche} A, {Lewen} F, {Garrod} RT, et~al. 2019.
\textit{\aap} 629:A72

\bibitem[{{Ospina-Zamudio} et~al.(2019){Ospina-Zamudio}, {Favre}, {Kounkel},
  {Xu}, {Neill} et~al.}]{ospinazamudio19}
{Ospina-Zamudio} J, {Favre} C, {Kounkel} M, {Xu} LH, {Neill} J, et~al. 2019.
\textit{\aap} 627:A80

\bibitem[{{Oya} et~al.(2016){Oya}, {Sakai}, {L{\'o}pez-Sepulcre}, {Watanabe},
  {Ceccarelli} et~al.}]{oya16}
{Oya} Y, {Sakai} N, {L{\'o}pez-Sepulcre} A, {Watanabe} Y, {Ceccarelli} C,
  et~al. 2016.
\textit{\apj} 824:88

\bibitem[{{Oya} et~al.(2017){Oya}, {Sakai}, {Watanabe}, {Higuchi}, {Hirota}
  et~al.}]{oya17}
{Oya} Y, {Sakai} N, {Watanabe} Y, {Higuchi} AE, {Hirota} T, et~al. 2017.
\textit{\apj} 837:174

\bibitem[{{Pagani} et~al.(2019){Pagani}, {Bergin}, {Goldsmith}, {Melnick},
  {Snell} \& {Favre}}]{pagani19}
{Pagani} L, {Bergin} E, {Goldsmith} PF, {Melnick} G, {Snell} R, {Favre} C.
  2019.
\textit{\aap} 624:L5

\bibitem[{{Pagani} et~al.(2017){Pagani}, {Favre}, {Goldsmith}, {Bergin},
  {Snell} \& {Melnick}}]{pagani17}
{Pagani} L, {Favre} C, {Goldsmith} PF, {Bergin} EA, {Snell} R, {Melnick} G.
  2017.
\textit{\aap} 604:A32

\bibitem[{{Palau} et~al.(2017){Palau}, {Walsh}, {S{\'a}nchez-Monge}, {Girart},
  {Cesaroni} et~al.}]{palau17}
{Palau} A, {Walsh} C, {S{\'a}nchez-Monge} {\'A}, {Girart} JM, {Cesaroni} R,
  et~al. 2017.
\textit{\mnras} 467:2723--2752

\bibitem[{{Parise} et~al.(2004){Parise}, {Castets}, {Herbst}, {Caux},
  {Ceccarelli} et~al.}]{parise04}
{Parise} B, {Castets} A, {Herbst} E, {Caux} E, {Ceccarelli} C, et~al. 2004.
\textit{\aap} 416:159--163

\bibitem[{{Parise} et~al.(2002){Parise}, {Ceccarelli}, {Tielens}, {Herbst},
  {Lefloch} et~al.}]{parise02}
{Parise} B, {Ceccarelli} C, {Tielens} AGGM, {Herbst} E, {Lefloch} B, et~al.
  2002.
\textit{\aap} 393:L49--L53

\bibitem[{{Peng} et~al.(2013){Peng}, {Despois}, {Brouillet}, {Baudry}, {Favre}
  et~al.}]{peng13}
{Peng} TC, {Despois} D, {Brouillet} N, {Baudry} A, {Favre} C, et~al. 2013.
\textit{\aap} 554:A78

\bibitem[{{Peng} et~al.(2012){Peng}, {Despois}, {Brouillet}, {Parise} \&
  {Baudry}}]{peng12}
{Peng} TC, {Despois} D, {Brouillet} N, {Parise} B, {Baudry} A. 2012.
\textit{\aap} 543:A152

\bibitem[{{Persson} et~al.(2018){Persson}, {J{\o}rgensen}, {M{\"u}ller},
  {Coutens}, {van Dishoeck} et~al.}]{persson18}
{Persson} MV, {J{\o}rgensen} JK, {M{\"u}ller} HSP, {Coutens} A, {van Dishoeck}
  EF, et~al. 2018.
\textit{\aap} 610:A54

\bibitem[{{Persson}, {J{\o}rgensen} \& {van Dishoeck}(2013)}]{persson13}
{Persson} MV, {J{\o}rgensen} JK, {van Dishoeck} EF. 2013.
\textit{\aap} 549:L3

\bibitem[{{Persson} et~al.(2014){Persson}, {J{\o}rgensen}, {van Dishoeck} \&
  {Harsono}}]{persson14}
{Persson} MV, {J{\o}rgensen} JK, {van Dishoeck} EF, {Harsono} D. 2014.
\textit{\aap} 563:A74

\bibitem[{{Pickett} et~al.(1998){Pickett}, {Poynter}, {Cohen}, {Delitsky},
  {Pearson} \& {Muller}}]{jpl}
{Pickett} HM, {Poynter} IRL, {Cohen} EA, {Delitsky} ML, {Pearson} JC, {Muller}
  HSP. 1998.
\textit{\jqsrt} 60:883--890

\bibitem[{{Pizzarello} \& {Groy}(2011)}]{pizzarello11}
{Pizzarello} S, {Groy} TL. 2011.
\textit{\gca} 75:645--656

\bibitem[{{Plambeck} et~al.(2009){Plambeck}, {Wright}, {Friedel}, {Widicus
  Weaver}, {Bolatto} et~al.}]{plambeck09}
{Plambeck} RL, {Wright} MCH, {Friedel} DN, {Widicus Weaver} SL, {Bolatto} AD,
  et~al. 2009.
\textit{\apjl} 704:L25--L28

\bibitem[{{Polehampton} et~al.(2002){Polehampton}, {Baluteau}, {Ceccarelli},
  {Swinyard} \& {Caux}}]{polehampton02}
{Polehampton} ET, {Baluteau} JP, {Ceccarelli} C, {Swinyard} BM, {Caux} E. 2002.
\textit{\aap} 388:L44--L47

\bibitem[{{Potapov} et~al.(2017){Potapov}, {J{\"a}ger}, {Henning}, {Jonusas} \&
  {Krim}}]{potapov17}
{Potapov} A, {J{\"a}ger} C, {Henning} T, {Jonusas} M, {Krim} L. 2017.
\textit{\apj} 846:131

\bibitem[{{Poteet} et~al.(2013){Poteet}, {Pontoppidan}, {Megeath}, {Watson},
  {Isokoski} et~al.}]{poteet13}
{Poteet} CA, {Pontoppidan} KM, {Megeath} ST, {Watson} DM, {Isokoski} K, et~al.
  2013.
\textit{\apj} 766:117

\bibitem[{{Prodanovi{\'c}}, {Steigman} \& {Fields}(2010)}]{prodanovic10}
{Prodanovi{\'c}} T, {Steigman} G, {Fields} BD. 2010.
\textit{\mnras} 406:1108--1115

\bibitem[{{Pulliam}, {McGuire} \& {Remijan}(2012)}]{pulliam12}
{Pulliam} RL, {McGuire} BA, {Remijan} AJ. 2012.
\textit{\apj} 751:1

\bibitem[{{Qasim} et~al.(2019){Qasim}, {Fedoseev}, {Chuang}, {Taquet},
  {Lamberts} et~al.}]{qasim19}
{Qasim} D, {Fedoseev} G, {Chuang} KJ, {Taquet} V, {Lamberts} T, et~al. 2019.
\textit{\aap} 627:A1

\bibitem[{{Qin} et~al.(2015){Qin}, {Schilke}, {Wu}, {Wu}, {Liu} et~al.}]{qin15}
{Qin} SL, {Schilke} P, {Wu} J, {Wu} Y, {Liu} T, et~al. 2015.
\textit{\apj} 803:39

\bibitem[{{Qin} et~al.(2010){Qin}, {Wu}, {Huang}, {Zhao}, {Li} et~al.}]{qin10}
{Qin} SL, {Wu} Y, {Huang} M, {Zhao} G, {Li} D, et~al. 2010.
\textit{\apj} 711:399--416

\bibitem[{{Qu{\'e}nard} et~al.(2018){Qu{\'e}nard}, {Jim{\'e}nez-Serra}, {Viti},
  {Holdship} \& {Coutens}}]{quenard18}
{Qu{\'e}nard} D, {Jim{\'e}nez-Serra} I, {Viti} S, {Holdship} J, {Coutens} A.
  2018.
\textit{\mnras} 474:2796--2812

\bibitem[{{Rab} et~al.(2017){Rab}, {Elbakyan}, {Vorobyov}, {G{\"u}del},
  {Dionatos} et~al.}]{rab17}
{Rab} C, {Elbakyan} V, {Vorobyov} E, {G{\"u}del} M, {Dionatos} O, et~al. 2017.
\textit{\aap} 604:A15

\bibitem[{{Remijan} et~al.(2004){Remijan}, {Shiao}, {Friedel}, {Meier} \&
  {Snyder}}]{remijan04}
{Remijan} A, {Shiao} YS, {Friedel} DN, {Meier} DS, {Snyder} LE. 2004.
\textit{\apj} 617:384--398

\bibitem[{{Remijan} et~al.(2008){Remijan}, {Hollis}, {Lovas}, {Stork}, {Jewell}
  \& {Meier}}]{remijan08}
{Remijan} AJ, {Hollis} JM, {Lovas} FJ, {Stork} WD, {Jewell} PR, {Meier} DS.
  2008.
\textit{\apjl} 675:L85

\bibitem[{{Requena-Torres} et~al.(2006){Requena-Torres}, {Mart{\'\i}n-Pintado},
  {Rodr{\'\i}guez-Franco}, {Mart{\'\i}n}, {Rodr{\'\i}guez-Fern{\'a}ndez} \& {de
  Vicente}}]{requena-torres06}
{Requena-Torres} MA, {Mart{\'\i}n-Pintado} J, {Rodr{\'\i}guez-Franco} A,
  {Mart{\'\i}n} S, {Rodr{\'\i}guez-Fern{\'a}ndez} NJ, {de Vicente} P. 2006.
\textit{\aap} 455:971--985

\bibitem[{{Rivilla} et~al.(2017){Rivilla}, {Beltr{\'a}n}, {Cesaroni},
  {Fontani}, {Codella} \& {Zhang}}]{rivilla17}
{Rivilla} VM, {Beltr{\'a}n} MT, {Cesaroni} R, {Fontani} F, {Codella} C, {Zhang}
  Q. 2017.
\textit{\aap} 598:A59

\bibitem[{{Robert}, {Gautier} \& {Dubrulle}(2000)}]{robert00}
{Robert} F, {Gautier} D, {Dubrulle} B. 2000.
\textit{Space Science Reviews} 92:201--224

\bibitem[{{Safron} et~al.(2015){Safron}, {Fischer}, {Megeath}, {Furlan},
  {Stutz} et~al.}]{safron15}
{Safron} EJ, {Fischer} WJ, {Megeath} ST, {Furlan} E, {Stutz} AM, et~al. 2015.
\textit{\apjl} 800:L5

\bibitem[{{Sakai} et~al.(2016){Sakai}, {Oya}, {L{\'o}pez-Sepulcre}, {Watanabe},
  {Sakai} et~al.}]{sakai16}
{Sakai} N, {Oya} Y, {L{\'o}pez-Sepulcre} A, {Watanabe} Y, {Sakai} T, et~al.
  2016.
\textit{\apjl} 820:L34

\bibitem[{{Sakai} et~al.(2009){Sakai}, {Sakai}, {Hirota}, {Burton} \&
  {Yamamoto}}]{sakai09}
{Sakai} N, {Sakai} T, {Hirota} T, {Burton} M, {Yamamoto} S. 2009.
\textit{\apj} 697:769--786

\bibitem[{{Sakai} et~al.(2014){Sakai}, {Sakai}, {Hirota}, {Watanabe},
  {Ceccarelli} et~al.}]{sakai14}
{Sakai} N, {Sakai} T, {Hirota} T, {Watanabe} Y, {Ceccarelli} C, et~al. 2014.
\textit{\nat} 507:78--80

\bibitem[{{S{\'a}nchez-Monge} et~al.(2018){S{\'a}nchez-Monge}, {Schilke},
  {Ginsburg}, {Cesaroni} \& {Schmiedeke}}]{sanchezmonge18}
{S{\'a}nchez-Monge} {\'A}, {Schilke} P, {Ginsburg} A, {Cesaroni} R,
  {Schmiedeke} A. 2018.
\textit{\aap} 609:A101

\bibitem[{{S{\'a}nchez-Monge} et~al.(2017){S{\'a}nchez-Monge}, {Schilke},
  {Schmiedeke}, {Ginsburg}, {Cesaroni} et~al.}]{sanchezmonge17}
{S{\'a}nchez-Monge} {\'A}, {Schilke} P, {Schmiedeke} A, {Ginsburg} A,
  {Cesaroni} R, et~al. 2017.
\textit{\aap} 604:A6

\bibitem[{{Schw{\"o}rer} et~al.(2019){Schw{\"o}rer}, {S{\'a}nchez-Monge},
  {Schilke}, {M{\"o}ller}, {Ginsburg} et~al.}]{schwoerer19}
{Schw{\"o}rer} A, {S{\'a}nchez-Monge} {\'A}, {Schilke} P, {M{\"o}ller} T,
  {Ginsburg} A, et~al. 2019.
\textit{\aap} 628:A6

\bibitem[{{Sewi{\l}o} et~al.(2018){Sewi{\l}o}, {Indebetouw}, {Charnley},
  {Zahorecz}, {Oliveira} et~al.}]{sewilo18}
{Sewi{\l}o} M, {Indebetouw} R, {Charnley} SB, {Zahorecz} S, {Oliveira} JM,
  et~al. 2018.
\textit{\apjl} 853:L19

\bibitem[{{Shimonishi} et~al.(2016){Shimonishi}, {Onaka}, {Kawamura} \&
  {Aikawa}}]{shimonishi16}
{Shimonishi} T, {Onaka} T, {Kawamura} A, {Aikawa} Y. 2016.
\textit{\apj} 827:72

\bibitem[{{Shimonishi} et~al.(2018){Shimonishi}, {Watanabe}, {Nishimura},
  {Aikawa}, {Yamamoto} et~al.}]{shimonishi18}
{Shimonishi} T, {Watanabe} Y, {Nishimura} Y, {Aikawa} Y, {Yamamoto} S, et~al.
  2018.
\textit{\apj} 862:102

\bibitem[{{Shingledecker} et~al.(2018){Shingledecker}, {Tennis}, {Le Gal} \&
  {Herbst}}]{shingledecker18}
{Shingledecker} CN, {Tennis} J, {Le Gal} R, {Herbst} E. 2018.
\textit{\apj} 861:20

\bibitem[{{Sipil{\"a}}, {Caselli} \& {Harju}(2015)}]{sipilae15benchmark}
{Sipil{\"a}} O, {Caselli} P, {Harju} J. 2015.
\textit{\aap} 578:A55

\bibitem[{{Skouteris} et~al.(2018){Skouteris}, {Balucani}, {Ceccarelli},
  {Vazart}, {Puzzarini} et~al.}]{skouteris18}
{Skouteris} D, {Balucani} N, {Ceccarelli} C, {Vazart} F, {Puzzarini} C, et~al.
  2018.
\textit{\apj} 854:135

\bibitem[{{Skouteris} et~al.(2017){Skouteris}, {Vazart}, {Ceccarelli},
  {Balucani}, {Puzzarini} \& {Barone}}]{skouteris17}
{Skouteris} D, {Vazart} F, {Ceccarelli} C, {Balucani} N, {Puzzarini} C,
  {Barone} V. 2017.
\textit{\mnras} 468:L1--L5

\bibitem[{{Soma} et~al.(2015){Soma}, {Sakai}, {Watanabe} \&
  {Yamamoto}}]{soma15}
{Soma} T, {Sakai} N, {Watanabe} Y, {Yamamoto} S. 2015.
\textit{\apj} 802:74

\bibitem[{{Song} \& {K{\"a}stner}(2016)}]{song16}
{Song} L, {K{\"a}stner} J. 2016.
\textit{Physical Chemistry Chemical Physics (Incorporating Faraday
  Transactions)} 18:29278--29285

\bibitem[{{Sugimura} et~al.(2011){Sugimura}, {Yamaguchi}, {Sakai}, {Umemoto},
  {Sakai} et~al.}]{sugimura11}
{Sugimura} M, {Yamaguchi} T, {Sakai} T, {Umemoto} T, {Sakai} N, et~al. 2011.
\textit{\pasj} 63:459--472

\bibitem[{{Suzuki} et~al.(2018){Suzuki}, {Ohishi}, {Saito}, {Hirota},
  {Majumdar} \& {Wakelam}}]{suzuki18}
{Suzuki} T, {Ohishi} M, {Saito} M, {Hirota} T, {Majumdar} L, {Wakelam} V. 2018.
\textit{\apjs} 237:3

\bibitem[{{Taniguchi} et~al.(2018){Taniguchi}, {Saito}, {Majumdar},
  {Shimoikura}, {Dobashi} et~al.}]{taniguchi18}
{Taniguchi} K, {Saito} M, {Majumdar} L, {Shimoikura} T, {Dobashi} K, et~al.
  2018.
\textit{\apj} 866:150

\bibitem[{{Taquet} et~al.(2019){Taquet}, {Bianchi}, {Codella}, {Persson},
  {Ceccarelli} et~al.}]{taquet20}
{Taquet} V, {Bianchi} E, {Codella} C, {Persson} MV, {Ceccarelli} C, et~al.
  2019.
\textit{\aap} 632:A19

\bibitem[{{Taquet}, {Charnley} \& {Sipil{\"a}}(2014)}]{taquet14}
{Taquet} V, {Charnley} SB, {Sipil{\"a}} O. 2014.
\textit{\apj} 791:1

\bibitem[{{Taquet} et~al.(2015){Taquet}, {L{\'o}pez-Sepulcre}, {Ceccarelli},
  {Neri}, {Kahane} \& {Charnley}}]{taquet15}
{Taquet} V, {L{\'o}pez-Sepulcre} A, {Ceccarelli} C, {Neri} R, {Kahane} C,
  {Charnley} SB. 2015.
\textit{\apj} 804:81

\bibitem[{{Taquet} et~al.(2013){Taquet}, {L{\'o}pez-Sepulcre}, {Ceccarelli},
  {Neri}, {Kahane} et~al.}]{taquet13}
{Taquet} V, {L{\'o}pez-Sepulcre} A, {Ceccarelli} C, {Neri} R, {Kahane} C,
  et~al. 2013.
\textit{\apjl} 768:L29

\bibitem[{{Taquet}, {Wirstr{\"o}m} \& {Charnley}(2016)}]{taquet16}
{Taquet} V, {Wirstr{\"o}m} ES, {Charnley} SB. 2016.
\textit{\apj} 821:46

\bibitem[{{Taquet} et~al.(2017){Taquet}, {Wirstr{\"o}m}, {Charnley}, {Faure},
  {L{\'o}pez-Sepulcre} \& {Persson}}]{taquet17}
{Taquet} V, {Wirstr{\"o}m} ES, {Charnley} SB, {Faure} A, {L{\'o}pez-Sepulcre}
  A, {Persson} CM. 2017.
\textit{\aap} 607:A20

\bibitem[{{Taylor} et~al.(1984){Taylor}, {Storey}, {Sandell}, {Williams} \&
  {Zealey}}]{taylor84}
{Taylor} KNR, {Storey} JWV, {Sandell} G, {Williams} PM, {Zealey} WJ. 1984.
\textit{\nat} 311:236--237

\bibitem[{{Tercero} et~al.(2015){Tercero}, {Cernicharo}, {L{\'o}pez},
  {Brouillet}, {Kolesnikov{\'a}} et~al.}]{tercero15}
{Tercero} B, {Cernicharo} J, {L{\'o}pez} A, {Brouillet} N, {Kolesnikov{\'a}} L,
  et~al. 2015.
\textit{\aap} 582:L1

\bibitem[{{Tercero} et~al.(2018){Tercero}, {Cuadrado}, {L{\'o}pez},
  {Brouillet}, {Despois} \& {Cernicharo}}]{tercero18}
{Tercero} B, {Cuadrado} S, {L{\'o}pez} A, {Brouillet} N, {Despois} D,
  {Cernicharo} J. 2018.
\textit{\aap} 620:L6

\bibitem[{{Thiel} et~al.(2017){Thiel}, {Belloche}, {Menten}, {Garrod} \&
  {M{\"u}ller}}]{thiel17}
{Thiel} V, {Belloche} A, {Menten} KM, {Garrod} RT, {M{\"u}ller} HSP. 2017.
\textit{\aap} 605:L6

\bibitem[{{Thiel} et~al.(2019){Thiel}, {Belloche}, {Menten}, {Giannetti},
  {Wiesemeyer} et~al.}]{thiel19}
{Thiel} V, {Belloche} A, {Menten} KM, {Giannetti} A, {Wiesemeyer} H, et~al.
  2019.
\textit{\aap} 623:A68

\bibitem[{{Tideswell} et~al.(2010){Tideswell}, {Fuller}, {Millar} \&
  {Markwick}}]{tideswell10}
{Tideswell} DM, {Fuller} GA, {Millar} TJ, {Markwick} AJ. 2010.
\textit{\aap} 510:A85

\bibitem[{{Tielens}(1983)}]{tielens83}
{Tielens} AGGM. 1983.
\textit{\aap} 119:177--184

\bibitem[{{Tielens}(2013)}]{tielens13}
{Tielens} AGGM. 2013.
\textit{Reviews of Modern Physics} 85:1021--1081

\bibitem[{{Tielens} \& {Hagen}(1982)}]{tielens82}
{Tielens} AGGM, {Hagen} W. 1982.
\textit{\aap} 114:245--260

\bibitem[{{Tobin} et~al.(2012){Tobin}, Hartmann, Chiang, Wilner, Looney
  et~al.}]{tobin12l1527}
{Tobin} JJ, Hartmann L, Chiang HF, Wilner DJ, Looney LW, et~al. 2012.
\textit{Nature} 492:83--85

\bibitem[{{Tychoniec} et~al.(2018){Tychoniec}, {Tobin}, {Karska}, {Chand ler},
  {Dunham} et~al.}]{tychoniec18}
{Tychoniec} {\L}, {Tobin} JJ, {Karska} A, {Chand ler} C, {Dunham} MM, et~al.
  2018.
\textit{\apjs} 238:19

\bibitem[{{van Dishoeck} et~al.(2014){van Dishoeck}, {Bergin}, {Lis} \&
  {Lunine}}]{vandishoeckppvi}
{van Dishoeck} EF, {Bergin} EA, {Lis} DC, {Lunine} JI. 2014.
\textit{Protostars and Planets VI} :835--858

\bibitem[{{van Dishoeck} \& {Blake}(1998)}]{vandishoeck98}
{van Dishoeck} EF, {Blake} GA. 1998.
\textit{\araa} 36:317--368

\bibitem[{{van Dishoeck} et~al.(1995){van Dishoeck}, {Blake}, {Jansen} \&
  {Groesbeck}}]{vandishoeck95}
{van Dishoeck} EF, {Blake} GA, {Jansen} DJ, {Groesbeck} TD. 1995.
\textit{\apj} 447:760

\bibitem[{{van 't Hoff} et~al.(2018{\natexlab{a}}){van 't Hoff}, {Persson},
  {Harsono}, {Taquet}, {J{\o}rgensen} et~al.}]{vanthoff18snowline}
{van 't Hoff} MLR, {Persson} MV, {Harsono} D, {Taquet} V, {J{\o}rgensen} JK,
  et~al. 2018{\natexlab{a}}.
\textit{\aap} 613:A29

\bibitem[{{van 't Hoff} et~al.(2018{\natexlab{b}}){van 't Hoff}, {Tobin},
  {Harsono} \& {van Dishoeck}}]{vanthoff18l1527}
{van 't Hoff} MLR, {Tobin} JJ, {Harsono} D, {van Dishoeck} EF.
  2018{\natexlab{b}}.
\textit{\aap} 615:A83

\bibitem[{{van 't Hoff} et~al.(2018{\natexlab{c}}){van 't Hoff}, {Tobin},
  {Trapman}, {Harsono}, {Sheehan} et~al.}]{vanthoff18v883ori}
{van 't Hoff} MLR, {Tobin} JJ, {Trapman} L, {Harsono} D, {Sheehan} PD, et~al.
  2018{\natexlab{c}}.
\textit{\apjl} 864:L23

\bibitem[{{van't Hoff} et~al.(2020){van't Hoff}, {van Dishoeck}, {J{\o}rgensen}
  \& {Calcutt}}]{vanthoff19}
{van't Hoff} MLR, {van Dishoeck} EF, {J{\o}rgensen} JK, {Calcutt} H. 2020.
\textit{\aap} 633:A7

\bibitem[{{Vastel} et~al.(2014){Vastel}, {Ceccarelli}, {Lefloch} \&
  {Bachiller}}]{vastel14}
{Vastel} C, {Ceccarelli} C, {Lefloch} B, {Bachiller} R. 2014.
\textit{\apjl} 795:L2

\bibitem[{{Vasyunin} \& {Herbst}(2013{\natexlab{a}})}]{vasyunin13}
{Vasyunin} AI, {Herbst} E. 2013{\natexlab{a}}.
\textit{\apj} 762:86

\bibitem[{{Vasyunin} \& {Herbst}(2013{\natexlab{b}})}]{vasyunin13b}
{Vasyunin} AI, {Herbst} E. 2013{\natexlab{b}}.
\textit{\apj} 769:34

\bibitem[{{Vasyunina} et~al.(2014){Vasyunina}, {Vasyunin}, {Herbst}, {Linz},
  {Voronkov} et~al.}]{vasyunina14}
{Vasyunina} T, {Vasyunin} AI, {Herbst} E, {Linz} H, {Voronkov} M, et~al. 2014.
\textit{\apj} 780:85

\bibitem[{{Visser} \& {Bergin}(2012)}]{visser12}
{Visser} R, {Bergin} EA. 2012.
\textit{\apjl} 754:L18

\bibitem[{{Visser}, {Bergin} \& {J{\o}rgensen}(2015)}]{visser15}
{Visser} R, {Bergin} EA, {J{\o}rgensen} JK. 2015.
\textit{\aap} 577:A102

\bibitem[{{Visser} et~al.(2013){Visser}, {J{\o}rgensen}, {Kristensen}, {van
  Dishoeck} \& {Bergin}}]{visser13}
{Visser} R, {J{\o}rgensen} JK, {Kristensen} LE, {van Dishoeck} EF, {Bergin} EA.
  2013.
\textit{\apj} 769:19

\bibitem[{{Visser}, {van Dishoeck} \&
  {Black}(2009)}]{visser09photodissociation}
{Visser} R, {van Dishoeck} EF, {Black} JH. 2009.
\textit{\aap} 503:323--343

\bibitem[{{Viti} et~al.(2004){Viti}, {Collings}, {Dever}, {McCoustra} \&
  {Williams}}]{viti04}
{Viti} S, {Collings} MP, {Dever} JW, {McCoustra} MRS, {Williams} DA. 2004.
\textit{\mnras} 354:1141--1145

\bibitem[{{Viti} \& {Williams}(1999)}]{viti99}
{Viti} S, {Williams} DA. 1999.
\textit{\mnras} 305:755--762

\bibitem[{{Vuitton} et~al.(2012){Vuitton}, {Yelle}, {Lavvas} \&
  {Klippenstein}}]{vuitton12}
{Vuitton} V, {Yelle} RV, {Lavvas} P, {Klippenstein} SJ. 2012.
\textit{\apj} 744:11

\bibitem[{{Watson} et~al.(2007){Watson}, {Bohac}, {Hull}, {Forrest}, {Furlan}
  et~al.}]{watson07}
{Watson} DM, {Bohac} CJ, {Hull} C, {Forrest} WJ, {Furlan} E, et~al. 2007.
\textit{\nat} 448:1026--1028

\bibitem[{{Wehres} et~al.(2018b){Wehres}, {Hermanns}, {Wilkins}, {Borisov},
  {Lewen} et~al.}]{wehres18b}
{Wehres} N, {Hermanns} M, {Wilkins} OH, {Borisov} K, {Lewen} F, et~al. 2018b.
\textit{\aap} 615:A140

\bibitem[{{Wehres} et~al.(2018a){Wehres}, {Ma{\ss}en}, {Borisov}, {Schmidt},
  {Lewen} et~al.}]{wehres18a}
{Wehres} N, {Ma{\ss}en} J, {Borisov} K, {Schmidt} B, {Lewen} F, et~al. 2018a.
\textit{Physical Chemistry Chemical Physics (Incorporating Faraday
  Transactions)} 20:5530--5544

\bibitem[{{Widicus Weaver}(2019)}]{widicus-weaver19}
{Widicus Weaver} SL. 2019.
\textit{\araa} 57:79--112

\bibitem[{{Willis} et~al.(2019){Willis}, {Garrod}, {Belloche}, {M{\"u}ller},
  {Barger} et~al.}]{willis19}
{Willis} ER, {Garrod} RT, {Belloche} A, {M{\"u}ller} HSP, {Barger} CJ, et~al.
  2019.
\textit{\aap}

\bibitem[{{Wyrowski} et~al.(1999){Wyrowski}, {Schilke}, {Walmsley} \&
  {Menten}}]{wyrowski99}
{Wyrowski} F, {Schilke} P, {Walmsley} CM, {Menten} KM. 1999.
\textit{\apjl} 514:L43--L46

\bibitem[{{Xu} et~al.(2008){Xu}, {Li}, {Hachisuka}, {Pandian}, {Menten} \&
  {Henkel}}]{xu08}
{Xu} Y, {Li} JJ, {Hachisuka} K, {Pandian} JD, {Menten} KM, {Henkel} C. 2008.
\textit{\aap} 485:729--734

\bibitem[{{Yamaguchi} et~al.(2011){Yamaguchi}, {Takano}, {Sakai}, {Sakai},
  {Liu} et~al.}]{yamaguchi11}
{Yamaguchi} T, {Takano} S, {Sakai} N, {Sakai} T, {Liu} SY, et~al. 2011.
\textit{\pasj} 63:L37--L41

\bibitem[{{Yamaguchi} et~al.(2012){Yamaguchi}, {Takano}, {Watanabe}, {Sakai},
  {Sakai} et~al.}]{yamaguchi12}
{Yamaguchi} T, {Takano} S, {Watanabe} Y, {Sakai} N, {Sakai} T, et~al. 2012.
\textit{\pasj} 64:105

\bibitem[{{Yen} et~al.(2015{\natexlab{a}}){Yen}, {Koch}, {Takakuwa}, {Ho},
  {Ohashi} \& {Tang}}]{yen15sample}
{Yen} HW, {Koch} PM, {Takakuwa} S, {Ho} PTP, {Ohashi} N, {Tang} YW.
  2015{\natexlab{a}}.
\textit{\apj} 799:193

\bibitem[{{Yen} et~al.(2017){Yen}, {Koch}, {Takakuwa}, {Krasnopolsky}, {Ohashi}
  \& {Aso}}]{yen17}
{Yen} HW, {Koch} PM, {Takakuwa} S, {Krasnopolsky} R, {Ohashi} N, {Aso} Y. 2017.
\textit{\apj} 834:178

\bibitem[{{Yen} et~al.(2015{\natexlab{b}}){Yen}, {Takakuwa}, {Koch}, {Aso},
  {Koyamatsu} et~al.}]{yen15b335}
{Yen} HW, {Takakuwa} S, {Koch} PM, {Aso} Y, {Koyamatsu} S, et~al.
  2015{\natexlab{b}}.
\textit{\apj} 812:129

\bibitem[{{Zakharenko} et~al.(2019){Zakharenko}, {Lewen}, {Ilyushin},
  {Drozdovskaya}, {J{\o}rgensen} et~al.}]{zakharenko19}
{Zakharenko} O, {Lewen} F, {Ilyushin} VV, {Drozdovskaya} MN, {J{\o}rgensen} JK,
  et~al. 2019.
\textit{\aap} 621:A114

\bibitem[{{Zaleski} et~al.(2013){Zaleski}, {Seifert}, {Steber}, {Muckle},
  {Loomis} et~al.}]{zaleski13}
{Zaleski} DP, {Seifert} NA, {Steber} AL, {Muckle} MT, {Loomis} RA, et~al. 2013.
\textit{\apjl} 765:L10

\bibitem[{{Zapata} et~al.(2009){Zapata}, {Schmid-Burgk}, {Ho}, {Rodr{\'\i}guez}
  \& {Menten}}]{zapata09}
{Zapata} LA, {Schmid-Burgk} J, {Ho} PTP, {Rodr{\'\i}guez} LF, {Menten} KM.
  2009.
\textit{\apjl} 704:L45--L48

\bibitem[{{Zapata}, {Schmid-Burgk} \& {Menten}(2011)}]{zapata11}
{Zapata} LA, {Schmid-Burgk} J, {Menten} KM. 2011.
\textit{\aap} 529:A24

\bibitem[{{Zeng} et~al.(2019){Zeng}, {Qu{\'e}nard}, {Jim{\'e}nez-Serra},
  {Mart{\'\i}n-Pintado}, {Rivilla} et~al.}]{zeng19}
{Zeng} S, {Qu{\'e}nard} D, {Jim{\'e}nez-Serra} I, {Mart{\'\i}n-Pintado} J,
  {Rivilla} VM, et~al. 2019.
\textit{\mnras} 484:L43--L48

\bibitem[{{Zernickel} et~al.(2012){Zernickel}, {Schilke}, {Schmiedeke}, {Lis},
  {Brogan} et~al.}]{zernickel12}
{Zernickel} A, {Schilke} P, {Schmiedeke} A, {Lis} DC, {Brogan} CL, et~al. 2012.
\textit{\aap} 546:A87

\end{thebibliography}

\end{document}